\documentclass[11pt,a4paper,notitlepage]{article}
\usepackage[utf8]{inputenc}
\usepackage[english]{babel}
\usepackage[T1]{fontenc}
\usepackage{hyperref}
\usepackage{amsmath}
\usepackage{stmaryrd}
\usepackage{amsfonts}
\usepackage{amssymb}
\usepackage{color} 
\usepackage{caption}
\usepackage{amsmath}
\usepackage{relsize,exscale}
\usepackage{array,multirow,makecell}

\usepackage[toc,page]{appendix}

\setcellgapes{1pt}
\makegapedcells
\newcolumntype{R}[1]{>{\raggedleft\arraybackslash }b{#1}}
\newcolumntype{L}[1]{>{\raggedright\arraybackslash }b{#1}}
\newcolumntype{C}[1]{>{\centering\arraybackslash }b{#1}}
\newcommand{\Tr}{\mathrm{Tr}}

\newtheorem{theorem}{Theorem}
\newtheorem{definition}{Definition}

\newtheorem{proposition}{Proposition}
\newtheorem{remark}{Remark}
\newtheorem{claim}{Claim}
\newtheorem{corollary}{Corollary}

\newcommand{\perm}{\mathrm{perm}}
\newtheorem{lemma}{Lemma}
\usepackage{enumerate}
\usepackage{subfig}
\usepackage{graphicx}

\newcommand{\U}{{\mathcal U}}

\newcommand{\SU}{\mathrm{SU}}

\newcommand{\beq}{\begin{equation}}
\newcommand{\eeq}{\end{equation}}
\newcommand{\bea}{\begin{eqnarray}}
\newcommand{\eea}{\end{eqnarray}}
\definecolor{mygray}{gray}{0.3}

\usepackage[left=2cm,right=2cm,top=3cm,bottom=3cm]{geometry}

\begin{document}

%\sffamily
%\begin{flushleft}
%Coarse Graining in Quantum Gravity: \\
%Bridging the Gap between Microscopic 
%\\
%Models and Spacetime-Physics \\
%%\today
%\end{flushleft}
\begin{center}
\textbf{\LARGE{Flowing in discrete gravity models and Ward identities: \\
A review }}\\
\bigskip

{Dine Ousmane Samary$^{a,b}$\footnote{dine.ousmanesamary@cipma.uac.bj},\,\, Vincent Lahoche$^b$\footnote{vincent.lahoche@cea.fr}, \, and \,
Ezinvi Balo\"{\i}tcha$^a$\footnote{ezinvi.baloitcha@cipma.uac.bj}
 }
 \end{center}
\vspace{15pt}
\begin{center}
a)\, Facult\'e des Sciences et Techniques (ICMPA-UNESCO Chair),
Universit\'e d'Abomey-
Calavi, 072 BP 50, Benin

b)\,  Commissariat à l'\'Energie Atomique (CEA, LIST),
 8 Avenue de la Vauve, 91120 Palaiseau, France
\end{center}

\date{\today}
%\date{\currenttime, \today}

\begin{abstract}
Ward-Takahashi identities are nonperturbative relations  between correlation functions and arising from symmetries in quantum and statistical fields theories, as Noether currents conservation for classical theories. Since their historical origin, these identities were considered to prove the exact relation between counter-terms to all order of the perturbative expansion. Recently they have been considered in relation with nonperturbative renormalization group aspects for some classes  of quantum field theories namely tensorial group field theories and matrix models, both characterized by a specific non-locality in their interactions, and expected to provide discrete models for quantum gravity. In this review, we summarize the state of the art, focusing on the conceptual aspects rather than technical subtleties, and provide a unified reflection on this novel and promising way of investigation. We attached great importance to the pedagogy and the self-consistency of the presentation. \\

\end{abstract}

%\pagebreak
%\tableofcontents

\section{Introduction: field theoretical frameworks for quantum gravity}
\begin{figure}[h!]
\begin{center}
$\vcenter{\hbox{\includegraphics[scale=0.7]{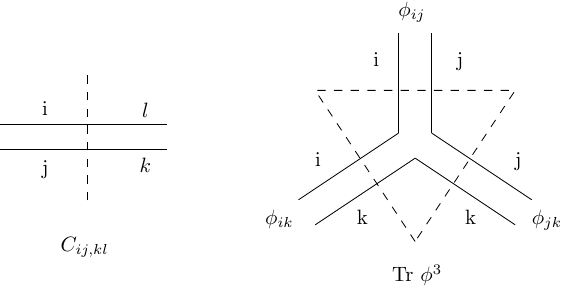} }}$
\end{center}
\caption{Propagator and vertex of the trivalent matrix model. The dual representation is pictured with dotted edges: they correspond to an edge for propagators, and a triangle for a vertex.}\label{fig1}
\end{figure}

\begin{figure}
\begin{center}
$\vcenter{\hbox{\includegraphics[scale=0.5]{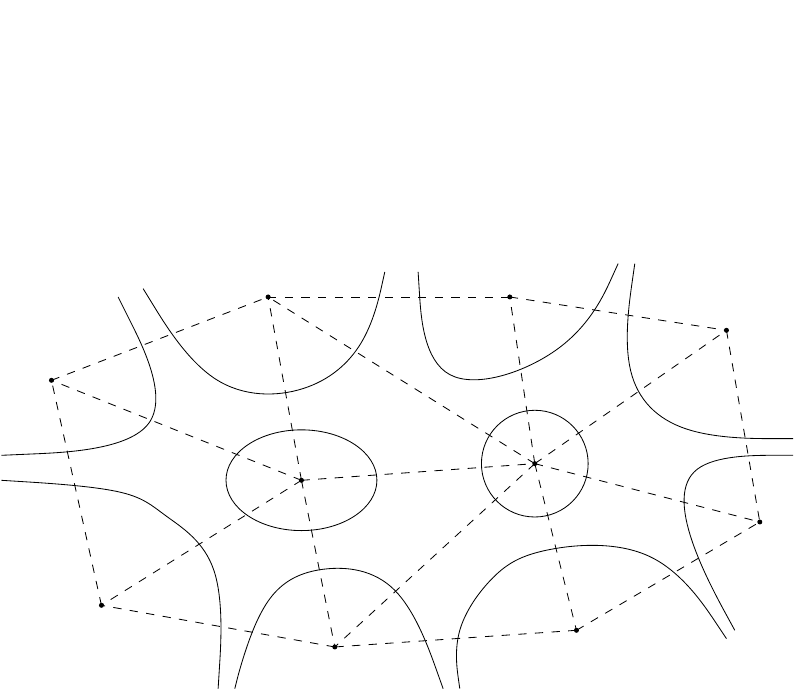} }}$
\end{center}
\caption{An example of a ribbon graph $\mathcal{G}$ and its corresponding dual triangulation $\Delta_{\mathcal{G}}$.}\label{fig22}
\end{figure}

The limitations in the possible accuracy of localization of space-time events should be a feature of a quantum theory incorporating gravitation and remains an open challenge for contemporary physics. Quantum aspects of gravity being so far from the current experimental energy scales of particle accelerators and therefore the investigations in this area are only supported by theoretical and mathematical explorations. The main difficulty at the origin of the theoretical investigation was the discovery that classical gravity does not provide a renormalizable quantum field theory over flat space-time \cite{Rovelli:2008brv}-\cite{Reuter:1996cp}. All the current tentative are then more and fewer solutions to circumvent this difficulty. However, without experimental support, the number of concurrent approaches have considerably proliferated since the pioneer works of Fierz, Pauli, Dirac, Feynman, Wheeler-DeWitt and so on between 1950-1970 (we mention a few \cite{Feynman:1963uxa}-\cite{DeWitt:1964dp}). Among the most popular nowadays, on can mention superstring theory \cite{Schwarz:1982jn}, loop quantum gravity \cite{Rovelli:1996ti} and group field theory
\cite{Oriti:2009nd}-\cite{Oriti:2009wn}, twistor theory \cite{Penrose:1972ia}, causal dynamical triangulations \cite{Ambjorn:1992eh}-\cite{Ambjorn:2013tki}, asymptotic safety scenario \cite{Eichhorn:2018yfc}, and so on, the list being so far to be exhaustive see \cite{Rovelli:1995ac}-\cite{Colosi:2004vw} and references therein. Schematically, there are two mainstreams. The background-dependent approach, assuming (at least at the first stage) the existence of a rigid extra space-time structure as a fundamental state, as it is the case for string theory; and in the opposite, the background independent approaches, as loop quantum gravity, spin foams and group field theory, where the ordinary space-time is not defined see \cite{Chirco:2019dlx}-\cite{Gielen:2018xph} and references therein. There is no solid insight showing the equivalence between these two approaches. This review is dedicated to the second kind of theories i.e. the background-independent ones; aiming to emerge the large space-time structure from the collective behavior of a very large number of elementary degrees of freedom. More precisely, we focus specifically on the subset of these approaches embedded into a quantum field theoretical formalism.
The basic scenario may be stressed in two dimensions with matrix models \cite{Brezin:1992yc}-\cite{Stanford:2019vob}. To be more precise let us consider a $N\times N$ random matrix, whose statistical properties are described by the partition function:
\begin{equation}
\mathcal{Z}(N,g):=\int d\phi\, e^{-\frac{1}{2} \Tr \phi^2-\frac{g}{\sqrt{N}} \Tr\phi^3}\,, \label{defmodel}
\end{equation}
where $d\phi$ is the invariant Haar measure on the $N\times N$ Hermitian matrices. Note that the classical action $S[\phi]:=\frac{1}{2} \Tr \phi^2+\frac{g}{\sqrt{N}} \Tr\phi^3$ admits a natural $U(N)$ symmetry due to the global trace structure. The link with quantum gravity appears with the perturbative expansion. Indeed, expanding the right-hand side perturbatively in $g$ with the propagator $C_{ij,kl}=\delta_{jk}\delta_{il}$, we generate a sum over Feynman amplitudes. Due to the structure of the interaction vertex, the amplitudes are not labelled with ordinary graphs as in standard field theory, but by \textit{ribbon graphs}, i.e. a set of vertices, edges, and faces. Let us investigate more precisely the structure of these graphs. The interaction vertex has three external points, and the global trace structure identifies the six external indices pairwise. The propagator, on the other hand, is represented by the strand which identifies the pairs of indices. The edges of the graphs are then ‘‘ribbons", and due to the pairwise identification at the vertex level, the closed trace of size $N$ occurs, and we call these the \textit{faces}. This enrichment concerning ordinary graphs allows building a one to one duality between such a ribbon graph and a two-dimensional triangulation, each vertex being in correspondence with a topological ‘‘triangle", as pictured in Figure \ref{fig1}. In Figure \ref{fig22} we provide an example of a ribbon graph and show explicitly how the correspondence with two-dimensional triangulation works.\\

From Feynman rules, the partition function expands as a sum of Feynman amplitudes
\begin{equation}
\mathcal{Z}=\sum_{\mathcal{G}}\frac{1}{s(\mathcal{G})}g^{n(\mathcal{G})}\mathcal{A}_{\mathcal{G}}\ , \label{eqmat}
\end{equation}
and it is not hard to check that, up to the rescaling $\phi\to\sqrt{N}\phi$, the amplitude $\mathcal{A}_{\mathcal{G}}$ depends only on $N$ and on the genus $h$ of the dual representation $\Delta_{\mathcal{G}}$ of $\mathcal{G}$: $\mathcal{A}_{\mathcal{G}}=N^{2-2h(\Delta_{\mathcal{G}})}$.
At this step we stressed that matrix models are statistical models for discrete surfaces -- the type of discretization, triangulation, quadrangulation, etc, depending on the number of fields in the interactions. This link with quantum gravity can be heuristically traced in the following way. Including cosmological constant $\Lambda$, classical gravity in dimension two is described by the Hilbert action:
\begin{equation}
\mathcal{S}_{2d}=\frac{1}{G}\int _{\mathcal{M}} d^2x\sqrt{-g}(-R(g)+\Lambda)=-\frac{4\pi}{G}\chi(\mathcal{M})+\frac{\Lambda}{G}A_{\mathcal{M}}\ , \label{2daction}
\end{equation}
where we used the Gauss-Bonnet theorem to compute the integral in terms of the Euler characteristic $\chi(\mathcal{M})$, and where we have denoted by $A_{\mathcal{M}}$ the area of the surface $\mathcal{M}$. Then, the classical theory only depends on two parameters, and we generally assume that only these two parameters are relevant to define this discretization. As a basic example, introducing an equilateral triangulation $\Delta_{\mathcal{M}}$ of $\mathcal{M}$, such that each triangle has a fixed area $a$, the action \eqref{2daction} can be discretized as \cite{Zinn-Justin:2014wva}-\cite{Marino:2004eq}:
\begin{equation}
\mathcal{S}_{2d}(\Delta_{\mathcal{M}}):=-\frac{4\pi}{G}\chi(\Delta_{\mathcal{M}})+\frac{\Lambda a}{G}A_{\mathcal{M}}(\Delta_{\mathcal{M}})\ .
\end{equation}
The quantum theory is then expected to be well described as a sum over triangulations, by the partition function:
\begin{equation}
\mathcal{Z}_{2d}=\sum_{\Delta} e^{\frac{4\pi}{G}\chi(\Delta_{\mathcal{M}})-\frac{\Lambda a}{G}A_{\mathcal{M}}(\Delta_{\mathcal{M}})} \label{discretized}
\end{equation}
matching with the partition function \eqref{eqmat}, up to the identification: $g \leftrightarrow e^{-\Lambda a/G}$ and $N\leftrightarrow e^{4\pi/G}$.
As a result, and following this heuristic argument, the large $N$ limit of matrix models (involving a lot of ``microscopic'' degrees of freedom) matches with the weak coupling regime of two-dimensional topological gravity. Note that, despite this argument, some strong formal results showed the equivalence between matrix models and other quantum gravity approaches. In particular, equivalence with Liouville theory at fixed topology has been stressed from the agreement with KPZ relation, see \cite{Duplantier:2009np}-\cite{Duplantier:2009np2}. Moreover, it is interesting to note that the sum \eqref{discretized} over topologies, following the heuristic path integral for gravity \cite{Brezin:1992yc}-\cite{DiFrancesco:1993cyw}.
\begin{equation}
Z_G= \sum_{\text{topology}} \int dg \,e^{-\mathcal{S}_{EH}[g]}\,,
\end{equation}
where $\mathcal{S}_{EH}$ denotes the standard Einstein-Hilbert action. In higher dimensions, group field theories (GFTs) and random tensor models (RTMs) provide a good extension of this strategy to build interacting quantum gravity degrees of freedom from a field theoretical formalism. \\

GFTs are field theories defined on $d$ copies of a group manifold $\mathbf{G}$, usually chosen for its physical relevance in the point of view of ordinary (classical) space-time. A field $\psi$ is then defined as a $L^2(\mathbf{G}^{\times d})$ function $\psi: \mathbf{G}^{\times d}\to \mathbb{C}$. As the matrix models, GFTs are characterized by the specific non-locality of their interactions, usually defined to provide the higher dimensional generalization of the triangle as elementary simplex \cite{Oriti:2009nd}. GFTs
were primarily introduced as an operational field theoretical formalism to generate spin foam amplitudes with canonical relative weights \cite{Perez:2002vg} (the number $d$ being the dimension of the corresponding simplicial decomposition labelling the Feynman amplitudes), and turned out to provide a second quantized version of loop quantum gravity \cite{Rovelli:1997yv}-\cite{Rovelli:2010bf}. Since a decade, GFTs have been the theater of important development (see \cite{Oriti:2005jr}-\cite{Oriti:2006se} and references therein), especially in direction of quantum cosmology see
\cite{deCesare:2016rsf}-\cite{Gielen:2018xph}. The strategy follows the geometrogenesis scenario, where semi-classical space-time emergence is understood as a phase transition toward a condensate phase, like Bose-Einstein condensation in many-body physics. Particularly the authors showed how ordinary quantum fluid techniques can be successfully used to extract the dynamic of a GFT condensate; the effective dynamic providing a good extension of standard quantum cosmology. In particular, investigating the semi-classical limit of the quantum equation for a GFT with Laplacian kinetic action, they show it reduces to Friedmann equation up to an isotropic ansatz. This first computational paper illustrating how the geometrogenesis scenario work has been continued through a series of the paper describing isolated horizon entropy \cite{Mandrysz:2018sle}-\cite{Gibbons:1977mu}, and corrections to the exact homogeneous model which couples gravity with scalar matter fields \cite{Gielen:2018xph}. All these results recover and extend some results obtained from another approach to quantum cosmology, especially regarding the presence of the ‘‘bounce" solving the classical singularities. \\

Other interesting aspects regarding the understanding of the physical content of GFTs, and aiming to support the condensation scenario concern the renormalization group flow investigations. Despite their physical relevance for cosmological perspectives, the interest of GFT with Laplacian propagator is appeared earlier, as a consequence of the divergences structure of the GFT \cite{Geloun:2011cy}. To cancel these divergences and then ensure the theory is well defined, radiative correction showed that counter-terms involving Laplacian kinetic terms are required at the first order in the perturbation theory. Including such a term is then appeared as a good manner to break the exact tensorial invariance (which is an additional input of tensor models discussed in the next point), providing a canonical path from ultraviolet to infrared scales and the minimal ingredient to start a renormalization, in complete analogy with what happens in the Gross-Wulkenhaar model. Moreover, in a purely mathematical point of view, Laplacian kinetic action showed to be relevant concerning an appropriate version of Osterwalder-Schrader positivity, usually considered in constructive field theory \cite{Rivasseau:2013uca}-\cite{Rivasseau:2014ima}. The renormalization program was successfully done. Rigorous renormalization theorems have been proved, allowing to classify of potentially relevant theories for their ultraviolet contain \cite{Freidel:2009hd}-\cite{Lahoche:2015ola}. Constructive aspects are discussed as well \cite{Rivasseau:2017xbk}-\cite{Lionni:2016ush}, ensuring the robustness of the field theoretical framework. Finally, a nonperturbative renormalization group program has been started, to support the condensate assumption. Models with increasing complexity have been considered, showing the existence of fixed points reminiscent of the phase transitions expected in the geometrogenesis scenario. \\

RTM random tensor models \cite{Gurau:2009tw}-\cite{Gurau:2011xq} were considered for a long time as a promising candidate to extend the success of matrix models \cite{Brezin:1992yc}-\cite{DiFrancesco:1993cyw} in higher dimensions. However, without a solid $1/N$ expansion, the initial program could not go as far as for matrix models. The situation has changed for a decade, with the works of Gurau, Rivasseau and their collaborators \cite{Gurau:2011aq}-\cite{Gurau:2011xq} and references therein, and the discovery of the colored random tensors. It appeared that building interactions for a complex model (involving two kinds of tensors $\mathbf{T}$ and $\bar{\mathbf{T}}$ of size $N$) to ensure a tensorial invariance $U(N)^{\times d}$ provides a $1/N$ expansion, organized following a generalization of the genus known as Gurau degree. Gurau degree reduces to the genus in dimension two, but is not a topological invariant for higher dimensions, and is not suitable for a topological expansion. The properties, of the leading order graphs, the so-called melons are well understood. In particular, it has been shown that their critical continuous behavior corresponds to a polymer branched phase, with Hausdorff dimension $2$ and spectral dimension $4/3$. In the generally accepted scenario, the melonic phase is then understood as the first phase transition among a series, increasing the dimension toward the one of the large scale classical space-time \cite{Calcagni:2013dna}. As for matrix models, the double scaling limit has been investigated as well, providing a deeper understanding of the continuum limit \cite{Dartois:2013sra}.  Finally, tensor models and their power counting are expected to be applied in a very large set of physical problems, so far from their quantum gravity, origins see \cite{Dartois:2018kfy}. Note also that connection between tensor theory and Euclidean gravity at the classical level is established in \cite{Diaz:2020wtr}. In the quantum gravity context, tensor models provided an important improvement to ordinary GFT; giving a different rule to build interactions as the ordinary simplicial decomposition. Building interactions to ensure a formal tensorial invariance allowed to benefit from the existence of a well-defined power counting, essential to renormalization \cite{Freidel:2009hd}-\cite{Lahoche:2015ola}. The new class of GFTs inherited from this tensorial improvement are known as tensorial group field theories (TGFTs) \cite{Freidel:2009hd}-\cite{Carrozza:2014rya}, and a large part of this review is devoted to them. Finally, for a few years, another invariant group has been considered, especially $O(N)$ invariance, allowing to generalize the success of tensors and TGFT to reals and symmetric tensors fields see also \cite{Benedetti:2019rja}-\cite{Benedetti:2019eyl} and references therein. With the renormalization group perspective, an interesting program has been initiated in \cite{Eichhorn:2014xaa}-\cite{Eichhorn:2018phj} for matrix models, and extended in \cite{Eichhorn:2017xhy} for random tensors, using renormalization group technique to provide a complementary point of view on double scaling limit, allowing, in particular, to recover critical exponent in great agreement with exact calculation.   Note that some new results concerning the FRG applied to matrix models are given in \cite{Eichhorn:2020sla}-\cite{Lahoche:2019ocf} and references therein. The final part of this review is devoted to these aspects, with the Ward identities. \\

This review takes place in the context of renormalization for quantum gravity models based on field theory, and a large part is devoted to TGFTs, and summarize a series of paper \cite{Lahoche:2019ocf}-\cite{Lahoche:2020pjo},\cite{Lahoche:2018ggd}-\cite{Lahoche:2018vun} exploring new methods of approximation to build nonperturbative solutions for the renormalization group equations, in relation with Ward-Takahashi identities. After a brief recalling about the nonperturbative formalism used in this context and Ward Takahashi identities in section \ref{sec2}, we summarize the results of \cite{Lahoche:2018oeo} for a just-renormalizable quartic model in section \eqref{sec3}. In the same section, we extend the discussion to six valence interactions and sectors beyond the melonic one. In section \ref{sec4} we review a similar investigation for matrix models, based on \cite{Eichhorn:2014xaa}-\cite{Eichhorn:2019hsa}, and highlighting the role of Ward identities in the fixed point structure provided by the renormalization group flow proposed in \cite{Lahoche:2019ocf}. Finally, we may add a short remark about the presentation. Aiming to increase the clarity of the presentation, we use the mathematical terminology "Theorem, Proposition, lemma" to bring out our main statement from the rest of the text and summarize the validity domain of the corresponding statements. Section \eqref{sec5} is devoted to our conclusion and remark.

\section{Nonperturbative renormalization group for TGFTs} \label{sec2}
As explained in our introduction the purpose of this review paper is to highlight some of our recent works, which aims to combine Ward identities with flow equations and to deduce the consequences on the existence or not of fixed points and on the asymptotic behaviors. We consider particularly the matrix and tensor models, and in the case where the models are solvable (the models with the trivial propagator), the comparison of our results with the exact computations well known in the literature is also given. Note that in the case of tensor models as well as matrix models with a trivial propagator the exact result of the flow equation is given by the solution of the closed equations of the correlation functions. On the other hand in the case of the nontrivial propagator, the solution of the closed equation becomes a challenge question \cite{Samary:2014tja}-\cite{Samary:2014oya} and just approximation is required to treat this issue. Despite this difficulty, important progress is nevertheless given in the FRG, by using a new approach called the effective vertex expansion (EVE) which allows summing all the melonic contributions, as well as all the graphs having a tree structure \, cite{Lahoche:2018ggd}-\cite{Lahoche:2019vzy}. The EVE goes beyond the usual crude truncation which is extensively used in the literature (see \cite{Geloun:2016qyb}-\cite{Carrozza:2016tih} and references therein).
In this first part of the review, we focus on nonperturbative renormalization group aspects for TGFTs. Recall that a GFT is a field theory over a group manifold, and a TGFT is characterized by the choice of the interactions, which follows the contraction rule of tensor models. We focus on the Abelian version of such a theory, setting $\mathrm{G}=\U(1)$. For this choice, the field may be equivalently described on the Fourier dual group $\mathbb{Z}^d$ by a \textit{tensor field} $T:\mathbb{Z}^d\to \mathbb{C}$. We consider a theory for two complexes fields $\varphi$ and $\bar\varphi$, requiring two complex tensors fields $T$ and $\bar{T}$. The allowed configurations are then constrained by the choice of a specific action, completing the definition of the GFT. At the classical level, for free fields we choose a Laplacian-like kinetic action:
\begin{equation}
S_{\text{kin}}[T,\bar{T}]:=\sum_{\vec{p}\in\mathbb{Z}^d} \bar{T}_{p_1\cdots p_d}\left(\vec{p}\,^2+m^2\right) {T}_{p_1\cdots p_d}\,,\label{kinfin}
\end{equation}
with the standard notation $\vec{p}\,^2:=\sum_i p_i^2$, $\vec{p}:=(p_1,\cdots,p_d)$. For the rest of this review we use the short notation $T_{\vec{p}}\equiv T_{p_1\cdots p_d}$. The equation \eqref{kinfin} defines the bare propagator $C^{-1}(\vec{p}\,):=\vec{p}\,^2+m^2$. We briefly outline the receipt to build tensorial interactions. By direct inspiration of tensor models, denoting by $N$ the \textit{size} of the tensor field, restricting the domain of the indices $p_i$ into the window $[\![-N,N ]\!]$, we require invariance with respect to independent transformations along each of the $d$ indices of the tensors:
\begin{equation}
T^\prime_{p_1\cdots p_d}=\sum_{\vec{q}\in[\![-N,N ]\!]^d} \left[\prod_{i=1}^d U^{(i)}_{p_iq_i}\right] \,T_{q_1\cdots q_d}\,,\label{unit}
\end{equation}
with $U^{(i)}(U^{(i)})^\dagger=\mathrm{id}$. The index $i$ is usually referred as \textit{color index}, and invariant interactions are then obtained contracting pairwise indices of the same colors of $T$ and $\bar{T}$ fields. Define $\mathbb{U}(N)$ as the set of unitary symmetries of size $N$, a transformation for tensors is then a set of $d$ independent elements of $\mathbb{U}(N)$, $\mathcal{U}:=(U_1,\cdots, U_d)\in\mathbb{U}(N)^d$, one per index of the tensor fields. The unitary symmetries admitting an inductive limit for arbitrary large $N$, we will implicitly consider the limit $N\to \infty$ in the rest of this review. \\

\noindent
The invariant interactions which cannot be factorized into two or smaller connected components are usually called \textit{bubble}. Remark that because the transformations are independent, the bubbles are not local in the usual sense over the group manifold $\mathrm{G}^d$. However, the locality does not make sense without physical content. In standard field theory for instance, or physics in general, the locality is defined by the way following which the fields or particles interact together, and as for tensors, this choice reflects invariance for some transformations like translations and rotations. With this respect, the transformation rule \eqref{unit} define both the nature of the field (a tensor) and the corresponding locality principle. To summarize:
\begin{definition}
Any interaction bubble is said to be local. By extension, any functions expanding as a sum of the bubble will be said local.
\end{definition}
This locality principle called \textit{traciality} in the literature has some good properties of the usual ones. In particular, it allows to define local counter-terms and to follow the standard renormalization procedure for interacting quantum fields with UV divergences. For this review, we focus on the quartic melonic model in rank $d=5$ with a single coupling, described by the classical interaction:
\begin{equation}
S_{\text{int}}[T,\bar T]= g \sum_{i=1}^d \vcenter{\hbox{\includegraphics[scale=1]{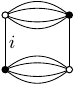} }}\,,\label{int}
\end{equation}
$g$ denoting the coupling constant and where we adopted the standard graphical convention \cite{Gurau:2011aq}-\cite{Gurau:2011xq} to picture the interaction bubble as $d$-colored bipartite regular connected graphs. The black (resp. white) nodes corresponding to $T$ (resp. $\bar{T}$) fields, and the colored edges fixing the contractions of their indices. The model that we consider is \textit{just renormalizable} in the usual sense, that is to say, all the UV divergences can be subtracted with a finite set of counter-terms, for mass, coupling and field strength. From now on, we will consider $m^2$ and $g$ as the bare couplings, sharing their counter-terms, and we introduce explicitly the wave function renormalization $Z$ replacing the propagator $C^{-1}$ by
\begin{equation}
C^{-1}(\vec{p}\,)=Z\vec{p}\,^2+m^2\,.
\end{equation}

\noindent
The equations \eqref{kin} and \eqref{int} define the classical model, without fluctuations. We quantize using path integral formulation, and define the partition function integrating over all configurations, weighted by $e^{-S}$:
\begin{equation}
\mathcal{Z}(J,\bar J):=\int dT d\bar T \,e^{-S[T,\bar T]+\langle \bar{J},T\rangle+\langle \bar{T},J\rangle}\,,\label{quantum}
\end{equation}
the sources being tensor fields themselves $J,\bar J:\mathbb{Z}^d\to \mathbb{C}$ and $\langle \bar{J},T\rangle:=\sum_{\vec{p}}\bar{J}_{\vec{p}}\, T_{\vec{p}}$. Note that the quantization procedure provide a canonical definition of what is UV and what is IR. The UV theory corresponding to the classical action $S=S_{\text{kin}}+S_{\text{int}}$ whereas the IR theory corresponds to the standard effective action defined as the Legendre transform of the free energy $\mathcal{W}:=\ln(\mathcal{Z}(J,\bar J))$. \\

\noindent
Renormalization in standard field theory allows to subtract divergences, and it has been shown that quantum GFT can be renormalized in the usual sense \cite{BenGeloun:2012pu}-\cite{BenGeloun:2011rc}. To the quantization procedure moreover, the renormalization group allows describing quantum effects \textit{scale by scale}, through more and more effective models, defining a path from UV to IR by integrating out the fluctuation of increasing size. \\

\noindent
Recognizing this path from UV to IR as an element of the quantization procedure itself, we substitute to the global quantum description \eqref{quantum} a set of models $\{\mathcal{Z}_k\}$ indexed by a referent scale $k$. This scale define what is UV, and integrated out and what is IR, and frozen out from the long distance physics. The set of scales may be discrete or continuous, and in this paper we choose a continuous description $k\in [0,\Lambda]$ for some fundamental UV cut-off $\Lambda$. There are several ways to build the functional renormalization group . We focus on the Wetterich-Morris approach \cite{Wetterich:1991be}-\cite{Wetterich:1992yh}, $ \mathcal{Z}_k(J,\bar J)$ being defined as:
\begin{equation}
\mathcal{Z}_k(J,\bar J):=\int dT d\bar T \,e^{-S_k[T,\bar T]+\langle \bar{J},T\rangle+\langle \bar{T},J\rangle}\,,\label{quantum2}
\end{equation}
with: $S_k[T,\bar T]:=S[T,\bar T]+\sum_{\vec{p}}\,\bar{T}_{\vec{p}}\, r_k(\vec{p}\,^2)T_{\vec{p}}$. The momentum dependent mass term $r_k(\vec{p}\,^2)$ called \textit{regulator} vanish for UV fluctuations $\vec{p}\,^2\gg k^2$ and becomes very large for the IR ones $\vec{p}\,^2\ll k^2$. Some additional properties for $r_k(\vec{p}\,^2)$ may be found in standard references \cite{Litim:2000ci}-\cite{Litim:2001dt}. With more details, among the standard properties of the regulator, we recall the following (for more explanations, the reader may be consult the standard reviews \cite{Delamotte:2007pf}):
\begin{enumerate}
\item $r_k(\vec{p}\,^2)$ has to have a non-vanishing ‘‘infrared" limit, i.e. $r_k(\vec{p}\,^2)\sim k^2$ for $\vec{p}\,^2/k^2\to 0$. \,
\item $r_{k}(\vec{p}\,^2) \to 0$ in the ‘‘ultraviolet'' limit , i.e. $\vec{p}\,^2/k^2\to \infty$. \,
\item $r_{k}(\vec{p}\,^2)$ has to vanish in the limit $k\to 0$, allowing to recover the original partition function.\,.
\item $r_{k}(\vec{p}\,^2)$ has to be of order $\Lambda$ in the limit $k\to \Lambda$, for some UV cut-off $\Lambda$.
\end{enumerate}
We focus essentially on the Litim's modified regulator:
\begin{equation}
r_k(\vec{p}\,^2):=Z(k)(k^2-\vec{p}\,^2)\theta(k^2-\vec{p}\,^2)\,,\label{regulator}
\end{equation}
where $\theta$ designates the Heaviside step function and $Z(k)$ is the running wave function strength.  The main advantage to use this regulator is that it allows to make analytic computations. However, it imposes a strong constraint on the anomalous dimension with respect to the UV boundary conditions. Indeed, the regulator must behaves as $k^r$ with $r>0$ in the deep UV. With the definition \eqref{regulator}: $R_k \approx Z k^2 = k^{2+\eta}$, and we must have:
\begin{equation}
\eta > \eta_c:= -2\,.\label{physicalboundeta}
\end{equation}
The renormalization group flow equation, describing the trajectory of the RG flow into the full theory space is the so called Wetterich equation \cite{Wetterich:1991be}-\cite{Wetterich:1992yh}, which for our model writes as:
\begin{equation}
\frac{\partial}{\partial k} \Gamma_k= \sum_{\vec{p}} \frac{\partial r_k}{\partial k}(\vec{p}\,) \left( \Gamma_k^{(2)}+r_k \right)^{-1}_{\vec{p}\,\vec{p}}\,,\label{Wett}
\end{equation}
where $(\Gamma_k^{(2)})_{\vec{p}\,\vec{p}\,^\prime}$ is the second derivative of the \textit{average effective action} $\Gamma_k$ with respect to the classical fields $M$ and $\bar{M}$:
\begin{equation}
\left(\Gamma_k^{(2)}\right)_{\vec{p}\,\vec{p}\,^\prime}=\frac{\partial^2\Gamma_k}{\partial M_{\vec{p}}\,\partial \bar{M}_{\vec{p}\,^\prime}}\,,
\end{equation}
where $M_{\vec{p}}=\partial \mathcal{W}_k/\partial\bar{J}_{\vec{p}}$, $\bar M_{\vec{p}}=\partial \mathcal{W}_k/\partial{J}_{\vec{p}}$ and:
\begin{align}
\Gamma_k[M,\bar M]+\sum_{\vec{p}}\,\bar{M}_{\vec{p}}\, r_k(\vec{p}\,^2)M_{\vec{p}}&:=\langle \bar{M},J\rangle+\langle \bar J,M\rangle-\mathcal{W}_k(M,\bar M)\,, \label{legendre}
\end{align}
with $\mathcal{W}_k=\ln(\mathcal{Z}_k)$. The flow equation \eqref{Wett} is a consequence of the variation of the propagator, indeed
\begin{equation}
\frac{\partial r_k}{\partial k} =\frac{\partial C^{-1}_k}{\partial k} \,,
\end{equation}
for the \textit{effective covariance} $C^{-1}_k:=C^{-1}+r_k$. But the propagator has other source of variability. In particular, it is not invariant with respect to the unitary symmetry of the classical interactions \eqref{int}. Focusing on an infinitesimal transformation : $\delta_1:=(\mathrm{id}+\epsilon,\mathrm{id},\cdots,\mathrm{id})$ acting non-trivially only on the color $1$ for some infinitesimal anti-hermitian transformations $\epsilon$, the transformation rule for the propagator follows the Lie bracket:
\begin{equation}
\mathcal{L}_{\delta_1} C^{-1}_k= [C^{-1}_k,\epsilon]\,.
\end{equation}
The source terms are non invariant as well. However, due to the translation invariance of the Lebesgue measure $dT d\bar T$ involved in the path integral \eqref{quantum2}, we must have $\mathcal{L}_{\delta_1} \mathcal{Z}_k=0$. Translating this invariance at the first order in $\epsilon$ provide a non-trivial \textit{Ward-Takahashi identity} for the quantum model \cite{Lahoche:2018oeo} and \cite{Samary:2014oya}:
\begin{theorem} \textbf{(Ward identity.)}
The non-invariance of the kinetic action with respect to unitary symmetry induce non-trivial relations between $\Gamma^{(n)}$ and $\Gamma^{(n+2)}$ for all $n$, summarized as:
\begin{align}
\sum_{\vec{p}_\bot, \vec{p}_\bot^{\prime}}{\vphantom{\sum}}'
&\bigg\{\big[C_k^{-1}(\vec{p})-C_k^{-1}(\vec{p}\,^{\prime})\big]\left[\frac{\partial^2 \mathcal{W}_k}{\partial \bar{J}_{\vec{p}\,^\prime}\,\partial {J}_{\vec{p}}}+\bar{M}_{\vec{p}}M_{\vec{p}\,^\prime}\right]-\bar{J} _{\vec{p}}\,M_{\vec{p}\,^\prime}+{J} _{\vec{p}\,^\prime}\bar{M}_{\vec{p}}\bigg\}=0\,.\label{Ward0}
\end{align}
where $\sum_{\vec{p}_\bot, \vec{p}_\bot^{\prime}}' :=\sum_{\vec{p}_\bot, \vec{p}_\bot^{\prime}} \delta_{\vec{p}\,\vec{p}_\bot^\prime}$.
\end{theorem}
In this statement we introduced the notations $\vec{p}_\bot:=(p_2,\cdots,p_d)\in\mathbb{Z}^{d-1}$ and $\delta_{\vec{p}\,\vec{p}_\bot^\prime}=\prod_{j\neq 1}\, \delta_{p_j\,p_j^\prime}$. Equations \eqref{Wett} and \eqref{Ward0} are two formal consequences of the path integral \eqref{quantum2}, coming both from the non-trivial variations of the propagator. Therefore, there are no reason to treat these two equations separately. This formal proximity is highlighted in their expanded forms, comparing equations \eqref{florence}--\eqref{florence2} and \eqref{W1}--\eqref{W2}. Instead of a set of partition function, the quantum model may be alternatively defined as an (infinite) set of effective vertices $\mathcal{Z}_k\sim \{\Gamma^{(n)}_k\}=:\mathfrak{h}_k$. RG equations dictate how to move from $\mathfrak{h}_k\underset{\text{RG}}{ \to}\mathfrak{h}_{k+\delta k}$ whereas Ward identities dictate how to move in the momentum space, along $\mathfrak{h}_k$.

\section{Effective vertex expansion and Ward identities in the melonic sector}\label{sec3}

\subsection{Effective vertex expansion}

This section essentially summarizes the state of the art in \cite{Lahoche:2018hou}-\cite{Lahoche:2018vun}. The exact RG equation cannot be solved except for very special cases. The main difficulty is that the Wetterich equation \eqref{Wett} split as an infinite hierarchical system, the derivative of $\Gamma^{(n)}$ involving $\Gamma^{(n+2)}$, and so on. Appropriate approximation schemes are then required to extract a piece of information on the exact solutions. The effective vertex expansion (EVE) is a recent technique allowing to build an approximation considering infinite sectors rather than crude truncations on the full theory space. We focus on the \textit{melonic sector}, sharing all the divergences of the model and then dominating the flow in the UV. One recalls that melonic diagrams are defined as the diagram with an optimal degree of divergence. Fixing some fundamental cut-off $\Lambda$, we consider the domain $\Lambda \gg k \gg 1$, so far from the deep UV and the deep IR regime. At this time, the flow is dominated by the renormalized couplings, which have positive or zero \textit{flow dimension} (see \cite{Lahoche:2018oeo}). We recall that the flow dimension reflects the behavior of the RG flow of the corresponding quantity, and discriminate between essential, marginal and inessential couplings just like standard dimension in quantum field theory. Because our theory is just-renormalizable, one has necessarily $[m^2]=2$ and $[g]=0$, denoted by $[X]$ the flow dimension of $X$. \\

\noindent

Note that we focus on the strictly local potential approximation, in which the EVE method works well. As recalled in the previous section, locality for TGFT means that we can be expanded as a sum (eventually infinite) of interaction bubbles. To make contact with the effective vertex formalism of this section, we have to recall the notion of boundary graphs. Indeed, effective vertex expand generally as a sum of connected diagrams, but what is relevant for the locality is boundary locality, and the boundary map $\mathcal{B}$ is defined as follow:
\begin{definition}
Let $\mathfrak{G}_p$ the set of bubbles with at most $p$ black nodes. The boundary map $\mathcal{B} : \mathfrak{G}_p\to (\mathfrak{G}_p)^{\times p}$ is defined as follows. For any connected Feynman graph $\mathcal{G}$, $\mathcal{B}(\mathcal{G})$ is the subset of external nodes, linking together with colored edges, according to their connectivity path in the graph.
\end{definition}

\begin{figure}[h!]
\begin{center}
\includegraphics[scale=0.9]{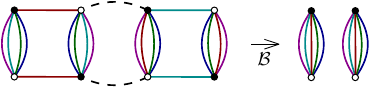}
\end{center}
\caption{Illustration of the manner that the boundary map work on a connected Feynman graph. The dotted edges correspond to Wick contractions. }\label{figboundary}
\end{figure}

\noindent
We recall that external nodes are nodes hooked to external edges. Figure \ref{figboundary} provide an illustration of the definition. Any Feynman graph $\mathcal{G}$ contributing to a local effective vertex are then such that $\mathcal{B}(\mathcal{G})$ is an interaction bubble. \\

\noindent
The basic strategy of the EVE is to close the hierarchical system coming from \eqref{Wett} using the analytic properties of the effective vertex functions and the rigid structure of the melonic diagrams. More precisely, the EVE express all the melonic effective vertices $\Gamma^{(n)}$ having negative flow dimension (that is for $n>4$) in terms of effective vertices with positive or null flow dimension, that is $\Gamma^{(2)}$ and $\Gamma^{(4)}$, and their flow is entirely governed by just-renormalizable couplings. As recalled, in this way we keep the entirety of the melonic sector and the full momentum dependence of the effective vertices. \\

\noindent
We work into the \textit{symmetric phase}, i.e. in the interior of the domain where the mean fields $M=\bar{M}=0$ make sense. This condition ensures that effective vertices with an odd number of external points, or not the same number of black and white external nodes have to be discarded from the analysis. These being called \textit{assorted functions}. Moreover, due to the momentum conservation along the boundaries of faces, $\Gamma^{(2)}_k$ must be diagonal:
\begin{equation}
\Gamma^{(2)}_{k,\,\vec{p}\,\vec{q}}=\Gamma_k^{(2)}(\vec{p}\,)\delta_{\vec{p}\,\vec{q}}\,.
\end{equation}
We denote as $G_k$ the effective $2$-point function $G^{-1}_k:=\Gamma_k^{(2)}+r_k$. \\

\noindent
The main assumption of the EVE approach is the existence of a finite analyticity domain for the leading order effective vertex functions, in which they may be identified with the resumed perturbative series. For the melonic vertex function, the existence of such an analytic domain is ensured, melons can be mapped as trees and easily summed. Moreover, these resumed functions satisfy the Ward-Takahashi identities, written without additional assumption than a cancellation of odd and assorted effective vertices. One then expect that the symmetric phase entirely covers the perturbative domain. \\

\begin{figure}
\begin{center}
\includegraphics[scale=0.5]{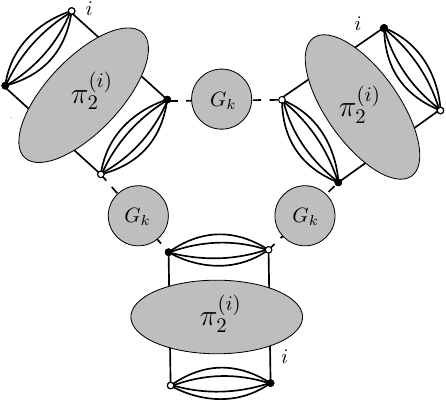}
\end{center}
\caption{Internal structure of the 1PI $6$-points graphs. } \label{fig6point}
\end{figure}

\begin{figure}
\begin{center}
\includegraphics[scale=0.3]{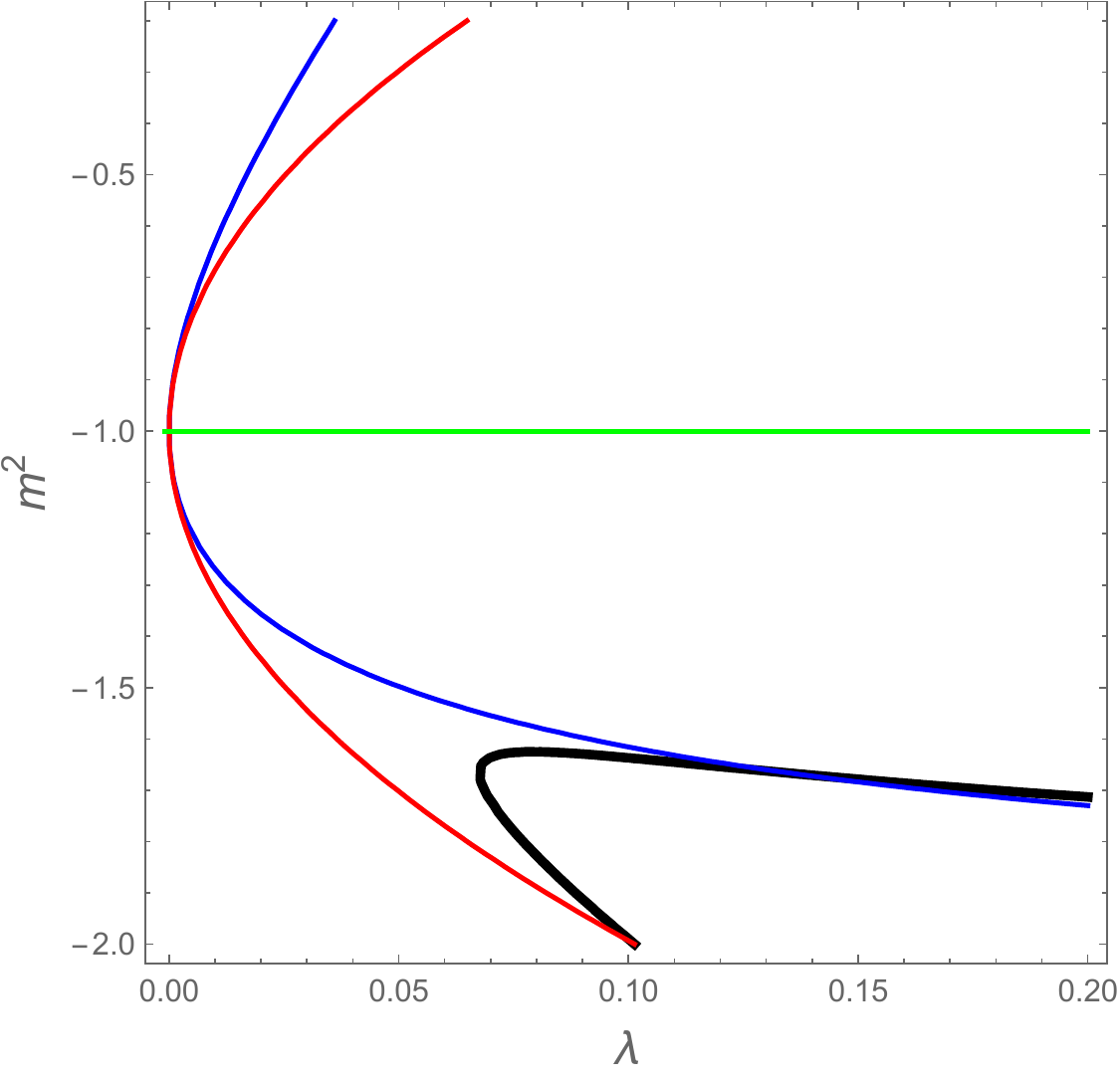}
\end{center}
\caption{Some relevant regions of the phase space. The curves $\eta=0$ (in blue), $(\eta)^{-1}=0$ (in black), and the curve $(\eta)^{-1}=0$ without took into account the momentum dependence of the vertex \eqref{vertexderiv} (in red).}\label{figattractor}
\end{figure}

\noindent
Among the properties of the melonic diagrams, we recall the following statement:
\begin{proposition}\label{propmelons}
Let $\mathcal{G}_N$ be a $2N$-point 1PI melonic diagram build with more than one vertices for a purely quartic melonic model. We call external vertices the vertices hooked to at least one external edge of $\mathcal{G}_N$ has :
\begin{itemize}
\item Two external edges per external vertices, sharing $d-1$ external faces of length one.
\item $N$ external faces of the same color running through the interior of the diagram.
\end{itemize}
\end{proposition}
Due to this proposition, the melonic effective vertex $\Gamma^{(n)}_k$ decompose as $d$ functions $\Gamma^{(n),i}_k$, labeled with a color index $i$:
\begin{equation}
\Gamma^{(n)}_{k,\,\vec{p}_1,\cdots,\vec{p}_n}=\sum_{i=1}^d \Gamma^{(n),i}_{k,\,\vec{p}_1,\cdots,\vec{p}_n}\,.
\end{equation}
The Feynman diagrams contributing to the perturbative expansion of $\Gamma^{(n,i)}_{k,\,\vec{p}_1,\cdots,\vec{p}_n}$ fix the relations between the different indices. For $n=4$ for instance, we get, from proposition \ref{propmelons}:
\begin{equation}
\Gamma_{\vec{p}_1,\vec{p}_2,\vec{p}_3,\vec{p}_4}^{(4),i} = \vcenter{\hbox{\includegraphics[scale=0.5]{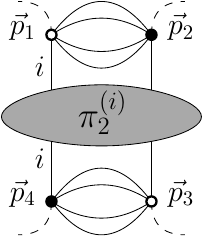} }}+\, \vcenter{\hbox{\includegraphics[scale=0.5]{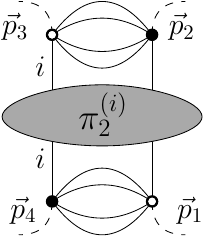} }}\,,\label{decomp4}
\end{equation}
Where the half dotted edges correspond to the amputated external propagators, and the reduced vertex functions $\pi_2^{(i)}:\mathbb{Z}^2\to \mathbb{R}$ denotes the sum of the interiors of the graphs, excluding the external nodes and the colored edges hooked to them. In the same way, one expect that the melonic effective vertex $\Gamma_{\text{melo\,}\vec{p}_1,\vec{p}_2,\vec{p}_3,\vec{p}_4,\vec{p}_5,\vec{p}_6}^{(6),i}$ is completely determined by a reduced effective vertex $\pi_3^{(i)}:\mathbb{Z}^3\to\mathbb{R}$ hooked to a boundary configuration such as:
\begin{equation}
\Gamma_{\vec{p}_1,\vec{p}_2,\vec{p}_3,\vec{p}_4,\vec{p}_5,\vec{p}_6}^{(6),i}=\vcenter{\hbox{\includegraphics[scale=0.4]{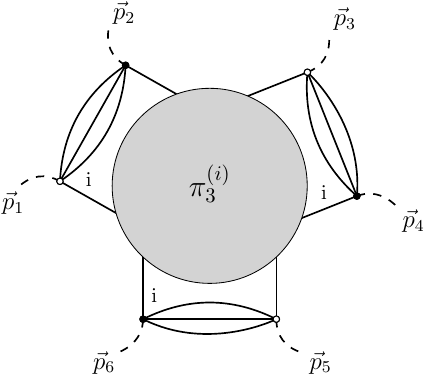} }}\,\,\, +\,\perm\,,
\end{equation}
and so one for $\Gamma^{(n),i}_{k,\,\vec{p}_1,\cdots,\vec{p}_n}$, involving the reduced vertex $\pi_n^{(i)}:\mathbb{Z}^n\to\mathbb{R}$. In the last expression, $\perm$ denote the permutation of the external edges like in \eqref{decomp4}. The reduced vertices $\pi_2^{(i)}$ can be formally resumed as a geometric series \cite{Lahoche:2018oeo}-\cite{Lahoche:2018hou}:
\begin{align}
\pi_{2,pp}^{(1)}&= 2\left(g-2g^2 \mathcal{A}_{2,p}+4g^3(\mathcal{A}_{2,p})^2-\cdots\right)=\frac{2g}{1+2g\mathcal{A}_{2,p}}\,,\label{eff4}
\end{align}
where $ \pi_{2,pp}^{(1)}$ is the diagonal element of the matrix $ \pi_{2}^{(1)}$ and :
\begin{equation}
\mathcal{A}_{n,p}:=\sum_{\vec{p}} \,G_k^n(\vec{p}\,)\delta_{p\,p_1}\,.\label{sumA}
\end{equation}
The reduced vertex $\pi_{2,pp}^{(1)}$ depends implicitly on $k$, and the renormalization conditions defining the \textit{renormalized coupling} $g_r$ are such that:
\begin{equation}
\pi_{2,00}^{(i)}\vert_{k=0}=:2g_r\,. \label{rencond}
\end{equation}
For arbitrary $k$, the zero momentum value of the reduced vertex define the effective coupling for the local quartic melonic interaction: $\pi_{2,00}^{(i)}=:2g(k)$. The explicit expression for $\pi_3^{(1)}$ may be investigated from the proposition\ref{propmelons}. The constraint over the boundaries and the recursive definition of melonic diagram enforce the internal structure pictured on Figure \ref{fig6point} below \cite{Lahoche:2018ggd}-\cite{Lahoche:2018vun}. Explicitly:
\begin{equation}
\pi_{3,ppp}^{(i)}=2(\pi_{2,pp}^{(i)})^3\,\mathcal{A}_{3,p}\,, \label{6pp}
\end{equation}
The two orientations of the external effective vertices being took into account in the definition of $\pi_{2,pp}^{(i)}$.
Expanding the exact flow equation \eqref{Wett}, and keeping only the relevant contraction for large $k$, one get the following relevant contributions for $\dot\Gamma_{k}^{(2)}$ and $\dot\Gamma_{k}^{(4)}$ :
\begin{equation}
\dot\Gamma_{k}^{(2)}(\vec{p})= -\,\sum_{i=1}^d\,\vcenter{\hbox{\includegraphics[scale=0.85]{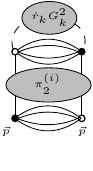} }}\,,\label{florence}
\end{equation}
and
\begin{equation}
\dot\Gamma_{k}^{(4),i}=-\,\,\vcenter{\hbox{\includegraphics[scale=0.65]{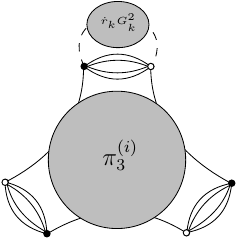} }}+\,\,4\,\vcenter{\hbox{\includegraphics[scale=0.65]{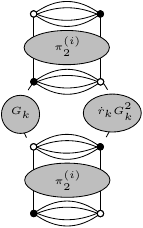} }}\,,\label{florence2}
\end{equation}
where $\dot X:=k \partial X/\partial k$. The computation require the explicit expression of $\Gamma^{(2)}_k$. In the melonic sector, the self energy obey to a closed equation, reputed difficult to solve. We approximate the exact solution by considering only the first term in the derivative expansion in the interior of the windows of momenta allowed by $\dot r_k$:
\begin{equation}
\Gamma_{k}^{(2)}(\vec{p}\,):=Z(k)\vec{p}\,^2+m^2(k)\,,\label{derivexp}
\end{equation}
where $Z(k):=\partial\Gamma_{k}^{(2)}/\partial p_1^2(\vec{0}\,)$ and $m^2(k):=\Gamma_{k}^{(2)}(\vec{0}\,)$ are both renormalized and effective field strength and mass. From the definition \eqref{regulator}, and with some calculation (see \cite{Lahoche:2018oeo}), we obtain the following statement:
\begin{proposition}
In the UV domain $\Lambda\gg k \gg 1$ and in the symmetric phase, the leading order flow equations for essential and marginal local couplings are given by:

\begin{align}
\left\{
\begin{array}{ll}
\beta_m&=-(2+\eta)\bar{m}^{2}-10 \bar{g}\,\frac{\pi^2}{(1+\bar{m}^{2})^2}\,\left(1+\frac{\eta}{6}\right)\,, \\
\beta_{g}&=-2\eta \bar{g}+4\bar{g}^2 \,\frac{\pi^2}{(1+\bar{m}^{2})^3}\,\left(1+\frac{\eta}{6}\right)\Big[1\\
&\quad-6\pi^2\bar{g}\left(\frac{1}{(1+\bar{m}^{2})^2}+\left(1+\frac{1}{1+\bar{m}^{2}}\right)\right)\Big]\,. \label{syst3}
\end{array}
\right.
\end{align}
With:
\begin{equation}
\eta=4\bar{g}\pi^2\frac{(1+\bar{m}^{2})^2-\bar{g}\pi^2(2+\bar{m}^{2})}{(1+\bar{m}^{2})^2\Omega(\bar{g},\bar{m}^{2})+2\frac{(2+\bar{m}^{2})}{3}\bar{g}^2\pi^4}\,,\label{eta1}
\end{equation}
and
\begin{equation}
\Omega(\bar m^2,\bar g):=(\bar m^2+1)^2-\pi^2\bar g\,.
\end{equation}
\end{proposition}
Where in this proposition $\beta_g:=\dot{\bar g}$, $\beta_m:=\dot{\bar{m}}^2$, $\eta=\dot{Z}/Z$, and the effective-renormalized mass and couplings are defined as: $\bar{g}:=Z^{-2}(k)g(k)$ and $\bar{m}^2:=Z^{-1}(k)k^{-2}m^2(k)$. For the computation, note that we made use of the approximation \eqref{derivexp} only for absolutely convergent quantities, and into the windows of momenta allowed by $\dot{r}_k$. As pointed out in \cite{Lahoche:2018hou}-\cite{Lahoche:2018oeo}, taking into account the full momentum dependence of the effective vertex $\pi_2^{(i)}$ in \eqref{eff4} drastically modify the expression of the anomalous dimension $\eta$ with respect to crude truncation. In particular, the singularity line discussed in \cite{Lahoche:2018ggd} disappears below the singularity $\bar{m}^2=-1$. Moreover, because all the effective melonic vertices only depend on $\bar{m}^2$ and $\bar{g}$, any fixed point for the system \eqref{syst3} is a global fixed point for the melonic sector. Note that to compute $\eta$, we required the knowledge of the derivative of the effective vertex with respect to the external momenta:
\begin{equation}
\frac{d}{dp^2}\pi_{2,pp}^{(i)}\bigg\vert_{p=0}=4g^2(k) \frac{d}{dp^2}\mathcal{A}_{2,p}\bigg\vert_{p=0}\,. \label{vertexderiv}
\end{equation}
The last sum is a superficially convergent quantity. Following the observation of the authors of \cite{Lahoche:2018oeo}, it can be computed from \eqref{6pp} and Ward identity \eqref{W2}, or directly from the truncation \eqref{derivexp}. We have\footnote{Note that we missed a factor $2$ in the computation of \cite{Lahoche:2018vun,Lahoche:2018oeo,Lahoche:2018ggd}.
\begin{equation}
\frac{d}{dp^2}\mathcal{A}_{2,p}\bigg\vert_{p=0}=-\frac{1}{Z^2k^2}\frac{\pi^2}{1+\bar{m}^2}\left(1+\frac{1}{1+\bar{m}^2} \right)\,.\
\end{equation}}
The system\footnote{Note that we missed a factor $2$ in the computation of \cite{Lahoche:2018vun,Lahoche:2018oeo,Lahoche:2018ggd}.
\begin{equation}
 \frac{d}{dp^2}\mathcal{A}_{2,p}\bigg\vert_{p=0}=-\frac{1}{Z^2k^2}\frac{\pi^2}{1+\bar{m}^2}\left(1+\frac{1}{1+\bar{m}^2} \right)\,.\
\end{equation} \eqref{syst3} admits two real fixed points solutions for $p_1:=(\bar{g}_*;\bar{m}^2_*)\approx(0.34;0.18)$ and $p_2\approx (-0.02,0.53)$. These fixed points however are expected unphysical. Indeed the first one $p_1$ has anomalous dimension $\eta_1 \approx -5.8$, under the physical bound $\eta_c=-2$ (see \eqref{physicalboundeta}) and have to be discarded. For the second fixed point $\eta_2\approx -0.4$, but it has the wrong sign with respect to the stability requirement. This result contrast with the results announced in \cite{Lahoche:2018vun,Lahoche:2018oeo,Lahoche:2018ggd} (due to the missed factor 6 in the flow equations), and with the predictions of crude truncations \cite{Lahoche:2018vun}, a disagreement that can be interpreted as an artifact of the truncation approaches. Hence, we have the following statement:

\begin{claim}\label{claim0}
Considering only the non-branching melonic sector, the effective vertex expansion allows to close the hierarchy of RG equations, and no physically relevant fixed point solutions are found.
\end{claim}

We will see in the rest of this paper that this conclusion agrees with constraints coming from Ward identities. Note with this respect that our approximations used to compute integrals withing EVE are compatible with Ward identities, see \cite{Lahoche:2018vun,Lahoche:2018oeo,Lahoche:2018ggd} and next section. Figure \ref{figattractor} summarize the improvement coming from EVE with respect to standard vertex expansion. In particular, the singularity line (in red) of the anomalous dimension without took into account the momentum dependence of the vertex is moved to the solid black line, below the singularity $\bar{m}^2=-1$ (in green).}

\subsection{The constrained non-branching melonic sector}

To close the hierarchical system derived from \eqref{Wett} and obtain the autonomous set \eqref{syst3}, we made use of the explicit expressions \eqref{eff4} and \eqref{6pp}. In this derivation, we mentioned the Ward identity but they do not contribute explicitly. In this section, we take into account their contribution and show that they introduce a strong constraint over the RG trajectories. \\

\noindent
Deriving successively the Ward identity \eqref{Ward0} with respect to external sources, and setting $J=\bar{J}=0$ at the end of the computation, we get the two following relations involving $\Gamma_k^{(4)}$ and $\Gamma_k^{(6)}$ see \cite{Lahoche:2018ggd}:
\begin{equation}
\pi_{2,00}^{(1)}\,\mathcal{L}_{2,k}=-\frac{\partial}{\partial p_1^2}\left(\Gamma_k^{(2)}(\vec{p}\,)-Z\vec{p}\,^2\right)\big\vert_{\vec{p}=\vec{0}}\,, \label{W1}
\end{equation}
and
\begin{equation}
\left(\pi_{3,00}^{(1)}\,\mathcal{L}_{2,k}-2(\pi_{2,00}^{(1)})^2\,\mathcal{L}_{3,k}\right)=-\frac{d}{dp_1^2}\pi_{2,p_1p_1}^{(1)}\big\vert_{p_1=0}\,,\label{W2}
\end{equation}
where:
\begin{equation}
\mathcal{L}_{n,k}:= \sum_{\vec{p}_\bot}\left(Z+\frac{\partial r_k}{\partial p_1^2} (\vec{p}_\bot)\right)G^n_k(\vec{p}_\bot)\,. \label{Lsum}
\end{equation}
Note that, to investigate the local potential approximation, we kept only the contributions with connected boundaries. We will return to this point later. \\

It can be easily checked that the structure equations \eqref{eff4} and \eqref{6pp} satisfy the second Ward identity \eqref{W2} see \cite{Lahoche:2018oeo}-\cite{Lahoche:2018hou} and also \cite{Samary:2014tja}-\cite{Sanchez:2017gxt}. In the same way the first Ward identity \eqref{W1} has been checked to be compatible with the equation \eqref{eff4} and the melonic closed equation for the $2$-point function. However, the last condition
does not exhaust the information contained in \eqref{eff4}. Indeed, with the same level of approximation as for the computation of the flow equations \eqref{syst3}, the first Ward identity can be translated locally as a constraint between beta functions (see \cite{Lahoche:2018oeo}):
\begin{equation}
\mathcal{C}:=\beta_g+\eta\bar{g}\, \frac{\Omega(\bar{g},\bar{m}^2)}{(1+\bar{m}^2)^2}-\frac{2\pi ^2\bar{g}^2}{(1+\bar{m}^2)^3}\beta_m=0\,.\label{const}
\end{equation}
This relation hold in the deep UV limit only, i.e. for large $k$. Inessential contributions have been discarded, which play an important role in the IR sector $k\approx 0$, where one expects that Ward identity reduces to its unregularized form, i.e. for $r_k=0$. Moreover, note that in this limit, and depending on the choice of the regulator, additional inessential operators may have to be added to the original action to recover the Ward identities. \\

\noindent
Generally, the solutions of the system \eqref{syst3} do not satisfy the constraint $\mathcal{C}=0$. This is especially the case of the fixed point solution $p_1$ discussed previously. To be compatible with Ward identities, the fixed point should have crossed the red line in Figure \ref{figattractor}. We call \textit{constrained melonic phase space} and denote as $\mathcal{E}_{\mathcal{C}}$ the subspace of the melonic theory space satisfying $\mathcal{C}=0$. 
\medskip

\begin{figure}
\begin{center}
$\underset{a}{\vcenter{\hbox{\includegraphics[scale=0.27]{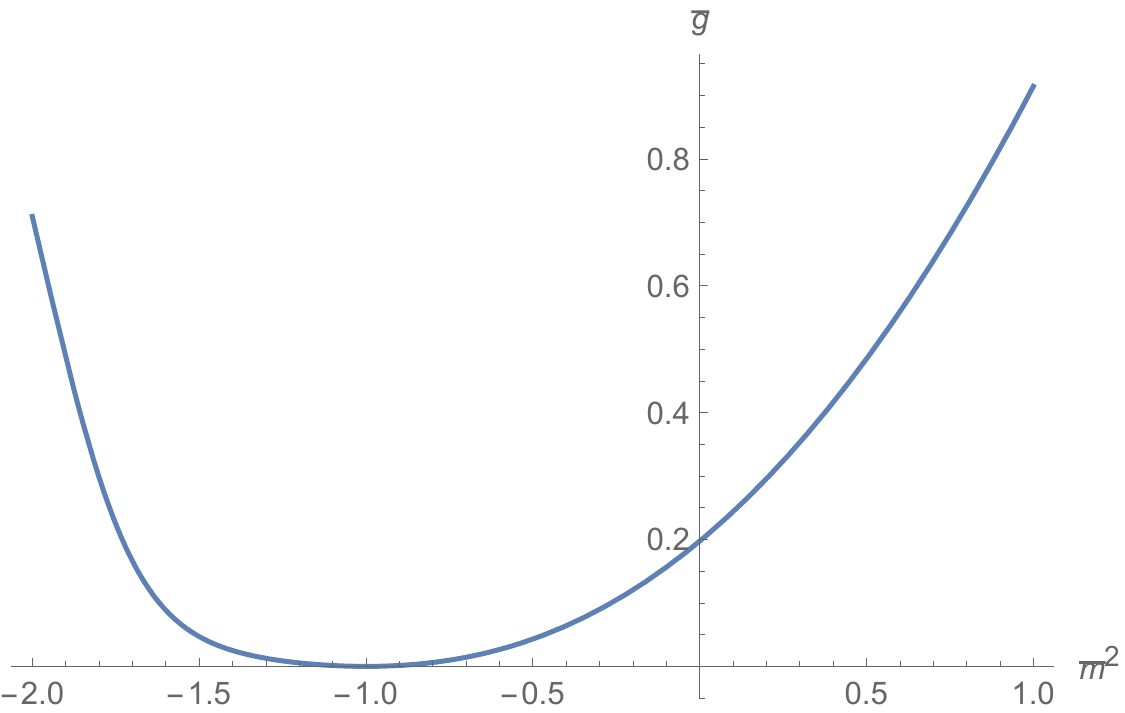} }}}$
$\underset{b}{\vcenter{\hbox{\includegraphics[scale=0.27]{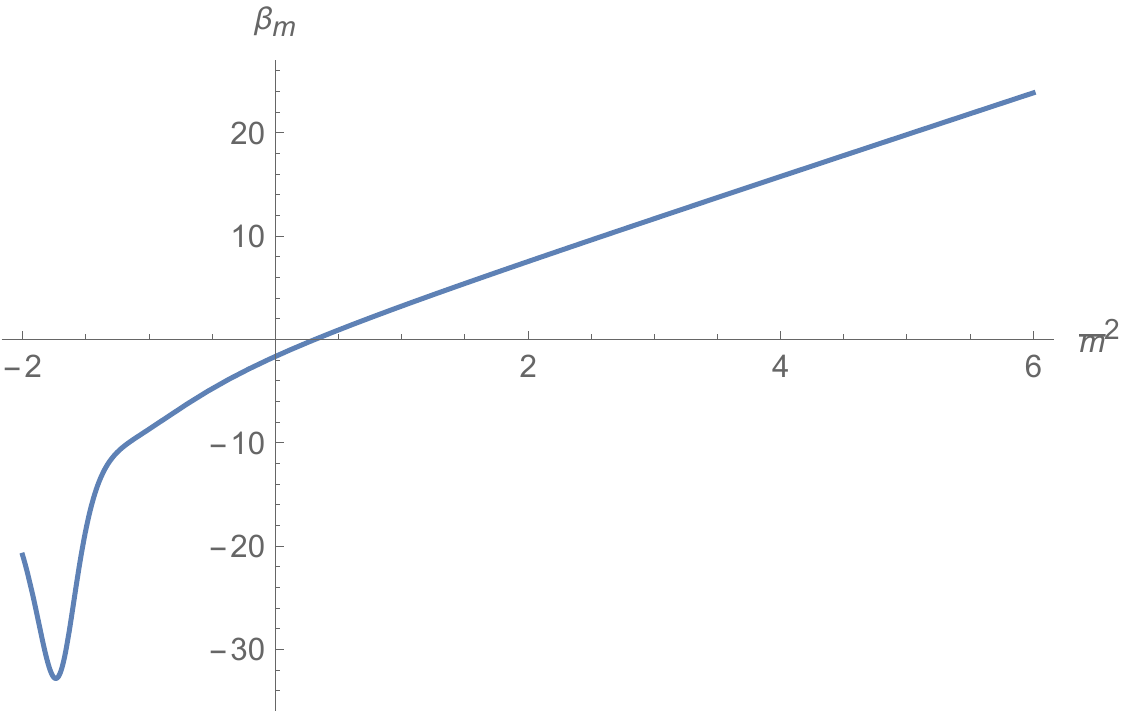} }}}$
$\underset{c}{\vcenter{\hbox{\includegraphics[scale=0.27]{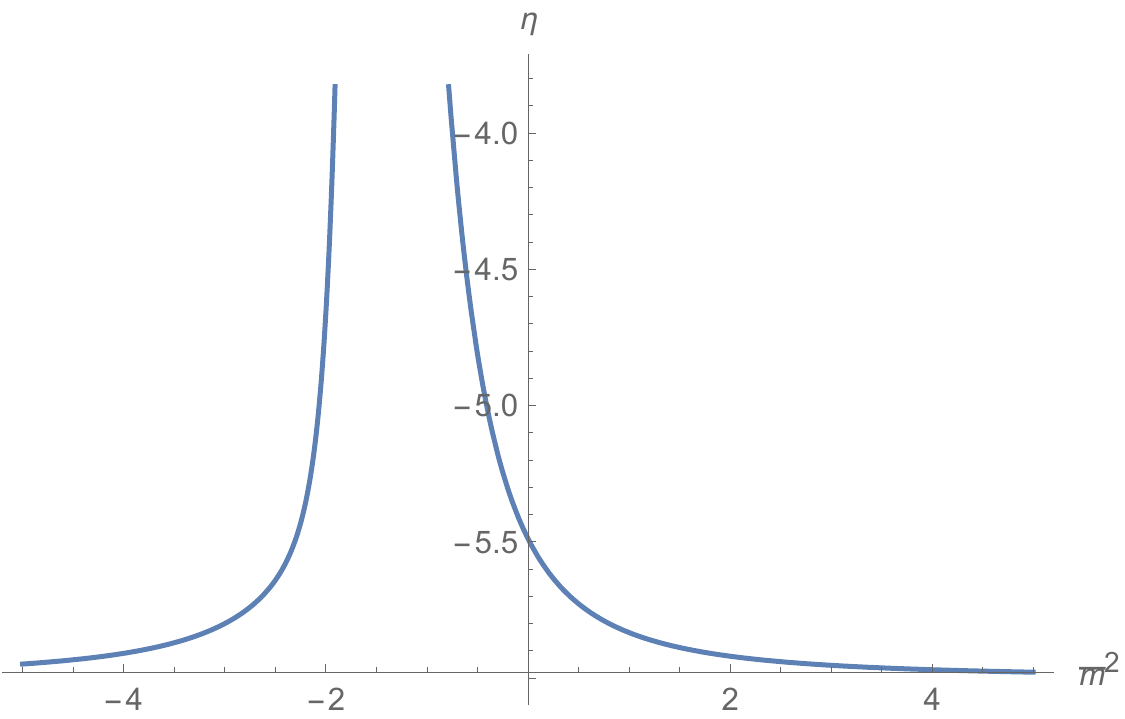} }}}$
\end{center}
\caption{The constrained melonic flow $\bar{g}=f(\bar{m}^2)$ ($\mathcal{E}_{\mathcal{C}1}$) in blue and $\bar{g}=0$ ($\mathcal{E}_{\mathcal{C}0}$) in brown (a); the corresponding beta function $\beta_m(\bar{m}^2,\bar{g}=f(\bar{m}^2))$ (b) and the anomalous dimension $\eta(\bar{m}^2,\bar{g}=f(\bar{m}^2))$ (c). The vertical dotted red line materializes the position of the UV attractive fixed point and the horizontal brown line materializes the physical bound $\eta=-2$.}\label{plotf}
\end{figure}

\noindent
In the description of the physical flow over $\mathcal{E}_{\mathcal{C}}$, we substituted the explicit expressions of $\beta_g$, $\beta_m$ and $\eta$, translating the relations between velocities as a complicated constraint on the couplings $\bar{g}$ and $\bar{m}^2$ providing a systematic projection of the RG trajectories. Explicitly, solving this constraint, we get two equations for this constrained subspace $\mathcal{E}_{\mathcal{C}}$, defining respectively $\mathcal{E}_{\mathcal{C}0}$ and $\mathcal{E}_{\mathcal{C}1}$:
\begin{equation}
\bar{g}=0 \,, \quad \text{and} \quad \bar{g}=f(\bar{m}^2)\,, \label{sol}
\end{equation}
where:
\begin{equation}
f(\bar{m}^2):=\frac{(\bar{m}^2+1)^2 (15 \bar{m}^2 (\bar{m}^2+3)+37)}{\pi ^2 (\bar{m}^2 (5\bar{m}^2+19)+19)}\,,
\end{equation}
and the non-trivial solution is pictured on Figure \ref{plotf}a below. In particular $f$ has only one real zero, $\bar{m}^2=-1$, and the coupling constant in always positive. The solution $\bar{g}=f(\bar{m}^2)$ leads to the $\beta$-function and anomalous dimension:
\begin{equation}
\beta_m=\frac{4 \bar{m}^2 (\bar{m}^2 (15 \bar{m}^2 (5 \bar{m}^2 (\bar{m}^2+6)+68)+956)+163)-740}{(15 \bar{m}^2 (5\bar{m}^2 (\bar{m}^2+6)+69)+1088)\bar{m}^2+445}\,,
\end{equation}
and:
\begin{equation}
\eta=-\frac{6 (5
\bar{m}^2 (\bar{m}^2+3)+11) (15 \bar{m}^2 (\bar{m}^2+3)+37)}{\bar{m}^2 (15 \bar{m}^2 (5 \bar{m}^2 (\bar{m}^2+6)+69)+1088)+445}\,.
\end{equation}
These functions are pictured on Figures \ref{plotf}b and \ref{plotf}c below. Interestingly it seems that the singularity occurring at $\bar{m}^2=-1$ in the unconstrained flow \eqref{florence} disappears, due to the factor $(1+\bar{m}^2)^2$ in the numerator of $f$. However the original flow remains undefined at this point, even if it seems allowed to continue it analytically. $\beta_m$ exhibits a fixed point solution for $\bar{m}_1^2\approx 0.32$. This fixed point is reminiscent of the fixed point $p_1$ discovered previously. Indeed, $\eta_*\approx -5.6$, and the fixed point solution is unphysical.

Some objections can be addressed to this method. First of all, our computation concerns only the melonic sector, and a more complete analysis should be done, using the method explained in \cite{Lahoche:2018oeo}. Secondly, the computation is based on the approximation \eqref{derivexp}, keeping only the relevant terms in the derivative expansion. This approximation is used to compute the sums involved in the flow equations, into the windows of momenta allowed by $\dot{r}_k$. Note that, because this window is the same as $\partial r_k/\partial p_1^2$, no additional approximation is used to compute the Ward constraint \eqref{const}. Indeed, the undefined term:
\begin{equation}
2g \dot{\mathcal{A}}_{2,0} \,,
\end{equation}
is fixed in term of the rate $\dot{\lambda}$, from equation \eqref{eff4}:
\begin{equation}
\dot{g}=-2g^2 \dot{\mathcal{A}}_{2,0}\,.
\end{equation}
However, the computation of the flow equations has required to compute superficially convergent sums using the approximation \eqref{derivexp}, out of the windows of momenta allowed by $\dot{r}_k$. This was concerned $\pi_3$ defined by \eqref{6pp}, and the derivative of the $4$-point vertex \eqref{vertexderiv}. We checked in \cite{Lahoche:2018oeo} the compatibility of this approximation with the second Ward identity \eqref{W2}. This is a strong constraint, in favour of the reliability of our approximation. But not a good justification, without exact computation. Such an exact computation is expected to be very hard. However, we may use the constraint, not to find where our approximation makes sense in the investigated region of the full phase space, but to fix these undefined sums. This is the strategy that we will describe now, keeping the approximation \eqref{derivexp} only for the sums in the domain $\vec{p}\,^2<k^2$. As we will see, this alternative description of the interacting sector of the theory strongly simplifies the description of the constrained space, which we call $\mathcal{E}_C$ as well, and can be easily extended for models with higher-order interactions. \\

\noindent
Another objection could concern the Ward identities themselves, or more precisely the form of Ward identities that we used to compute the constraint \eqref{const}. As discussed in this paragraph, we discarded the effective vertices with disconnected boundaries from our analysis. However, as pointed out in\cite{Lahoche:2018oeo}, the relation \eqref{W1} may be deduced from the structure of the melonic diagrams rather than a consequence of the symmetry breaking of the kinetic action. Indeed, in the melonic sector, the self-energy satisfy a closed equation see \cite{Samary:2014tja}-\cite{Sanchez:2017gxt};
\begin{proposition}
In the melonic sector, the $2$-point self energy $\Sigma(\vec{p}\,)$ decomposes as a sum over colors:
\begin{equation}
\Sigma(\vec{p}\,)=: \sum_{i=1}^d \tau(p_i)\,,
\end{equation}
where $\tau(p)$ satisfy a closed equation:
\begin{equation}
\tau(p)=-2Z_g g_r \sum_{\vec{q}}\, \delta_{q_1p}\, \frac{1}{Z_0 \vec{q}\,^2+Z_m m_b^2-\sum_i \tau(q_i)}\,. \label{closed}
\end{equation}
\end{proposition}
which combined with the melonic expansion \eqref{eff4} provides exactly the same relation than \eqref{W1}. Moreover, as explained in the previous paragraph, the second relation \eqref{W2} work as well, at least in the first order in the derivative expansion to compute the derivative of the effective vertex \eqref{vertexderiv}. Indeed, as checked in \cite{Lahoche:2018oeo}, Appendix B:
\begin{align}
\frac{d}{dp^2_1}&\pi_{2,p_1p_1}^{(1)}\big\vert_{p_1=0}=-4g^2(k) \frac{d}{dp^2_1}\mathcal{A}_{2,p_1}\big\vert_{p_1=0}=2(\pi^{(1)}_{00})^2 \left[ \sum_{\vec{p}_\bot} \left(Z(k)-Z\right)G^3_k(\vec{p}_\bot)+\mathcal{L}_{3,k}\right]
\end{align}
where we used \eqref{derivexp} for the second line and \eqref{Lsum} for the last line. Finally, from the first Ward identity \eqref{W1}, or mixing \eqref{eff4} and the closed equation \eqref{closed}, $(Z(k)-Z)=-Z\pi_{00}^{(1)}\mathcal{L}_{2,k}$; such that because \eqref{6pp} we recover exactly the Ward identity \eqref{W2}. The reliability of our approximations for the Ward constraint however cannot be extended for the RG flow equations themselves, and in the next two subsection, we investigate two heuristic ways to discuss the robustness of our conclusions.\\

A simple way of investigation for the robustness of our conclusions about the constrained phase space is to reduce the number of interactions. Indeed, configurations, as pictured in Figure \ref{figboundary}, disappears if we restrict the number of interactions to one, replacing the original model \eqref{int} by:
\begin{equation}
S_{\text{int}}[T,\bar T]= g \vcenter{\hbox{\includegraphics[scale=1.2]{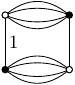} }}\,.\label{int2}
\end{equation}
In the point of view of the EVE method, equations \eqref{syst3}, the only change with respect to the fully interacting model concern the beta function $\beta_m$, the factor $10$ being in fact $2\times d$, for $d=5$. For a single color, we find two real fixed point solutions again:
\begin{equation}
p_1\approx (0.27,0.04)\,,\qquad p_2\approx (-0.009,0.08)\,.
\end{equation}
The second one has the wrong sign for the quartic coupling constant and the first one having anomalous dimension smaller than the physical bound $\eta_c$. Hence no reliable fixed point is found. 
\medskip

\begin{figure}[h!]
\begin{center}
$\underset{a}{\vcenter{\hbox{\includegraphics[scale=0.3]{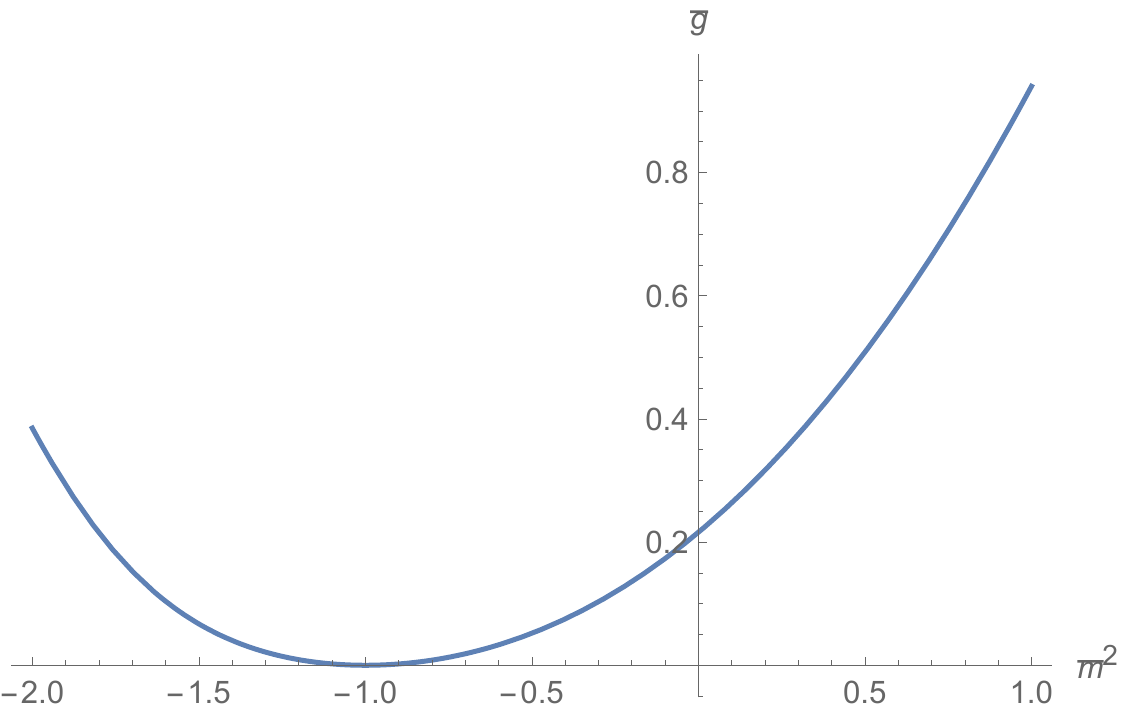} }}}$
$\vspace{0.5cm}$\qquad
$\underset{b}{\vcenter{\hbox{\includegraphics[scale=0.3]{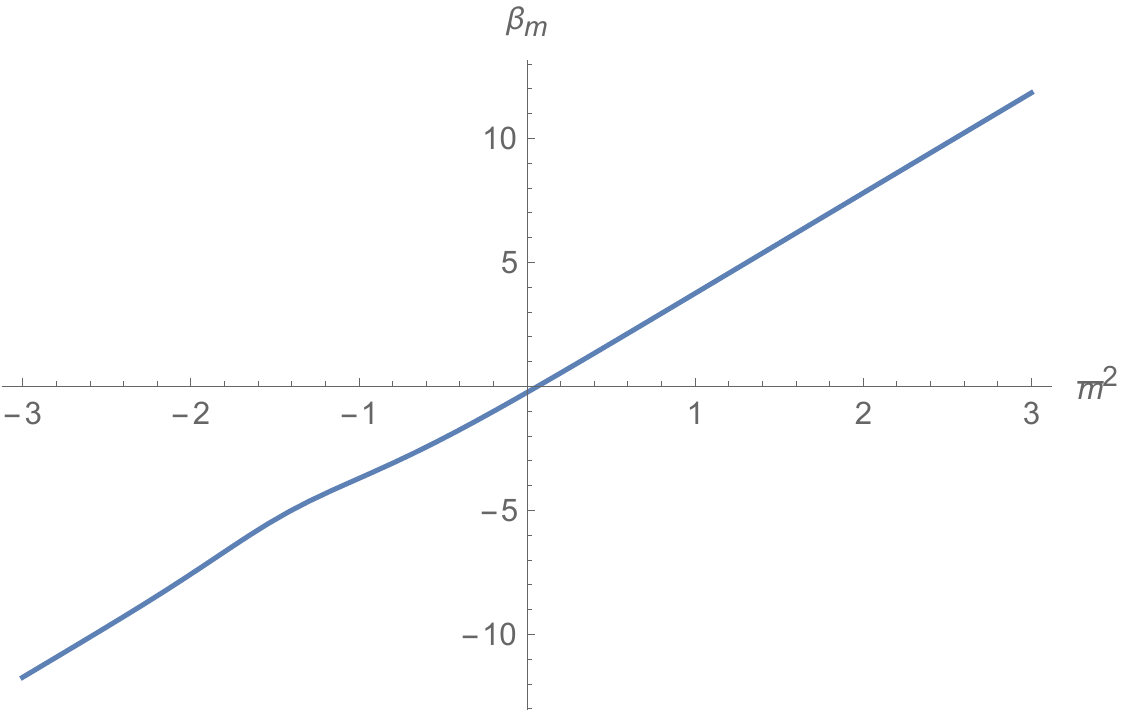} }}}$
\end{center}
\caption{The constrained melonic flow $\bar{g}=f(\bar{m}^2)$ ($\mathcal{E}_{\mathcal{C}1}$) in blue and $\bar{g}=0$ ($\mathcal{E}_{\mathcal{C}0}$) in brown for a single interaction (a); the corresponding beta function $\beta_m(\bar{m}^2,\bar{g}=f(\bar{m}^2))$ (b).}\label{fig1color}
\end{figure}

\begin{figure}
\begin{center}
\includegraphics[scale=0.4]{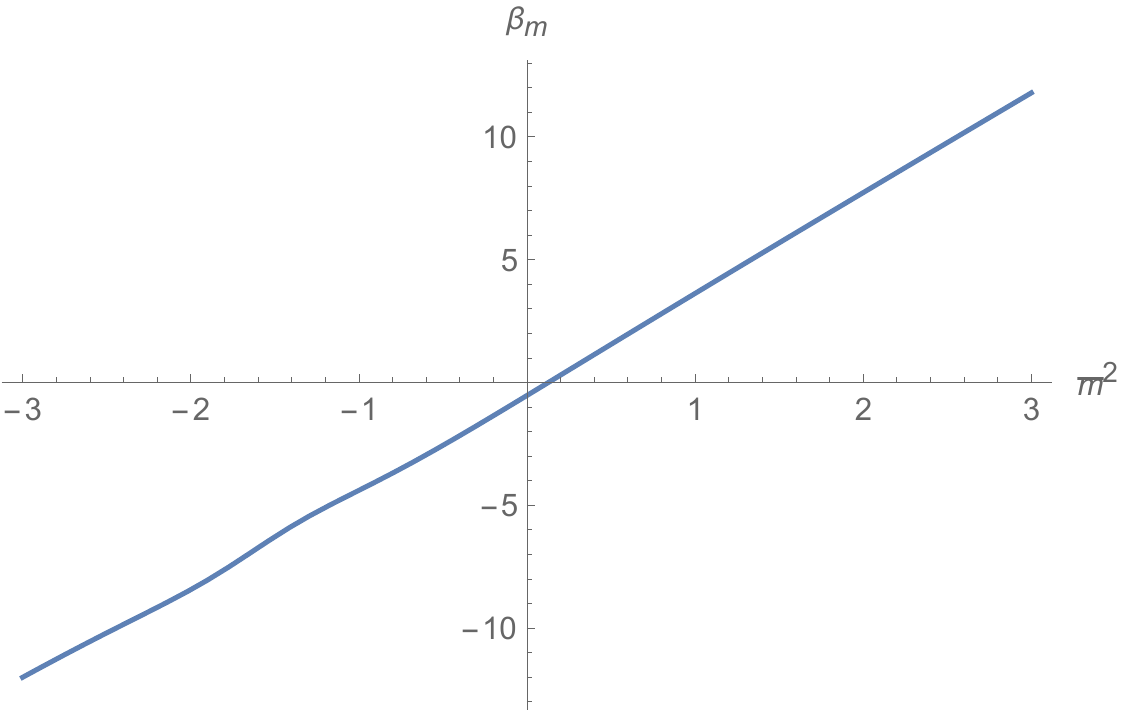}
\end{center}
\caption{The beta function $\beta_m$ over the constrained phase space for two interactions.}\label{betamd2}
\end{figure}

\begin{figure}
\begin{center}
$\underset{a}{\vcenter{\hbox{\includegraphics[scale=0.27]{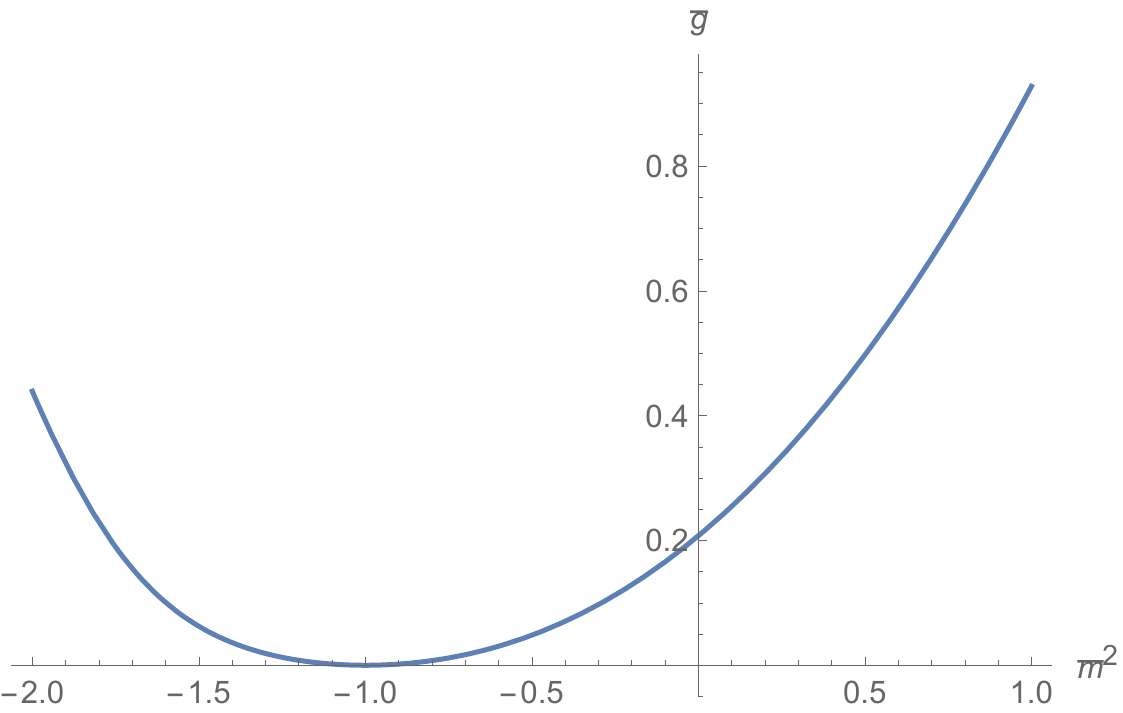} }}}$
$\underset{b}{\vcenter{\hbox{\includegraphics[scale=0.27]{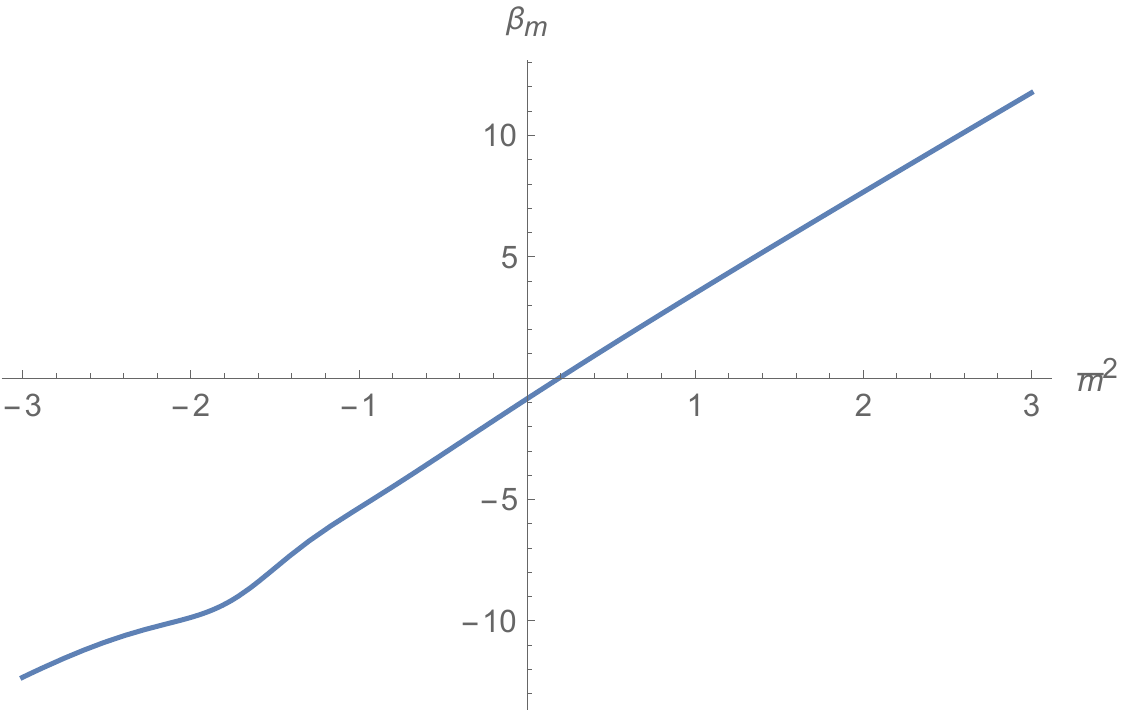} }}}$
$\underset{b}{\vcenter{\hbox{\includegraphics[scale=0.27]{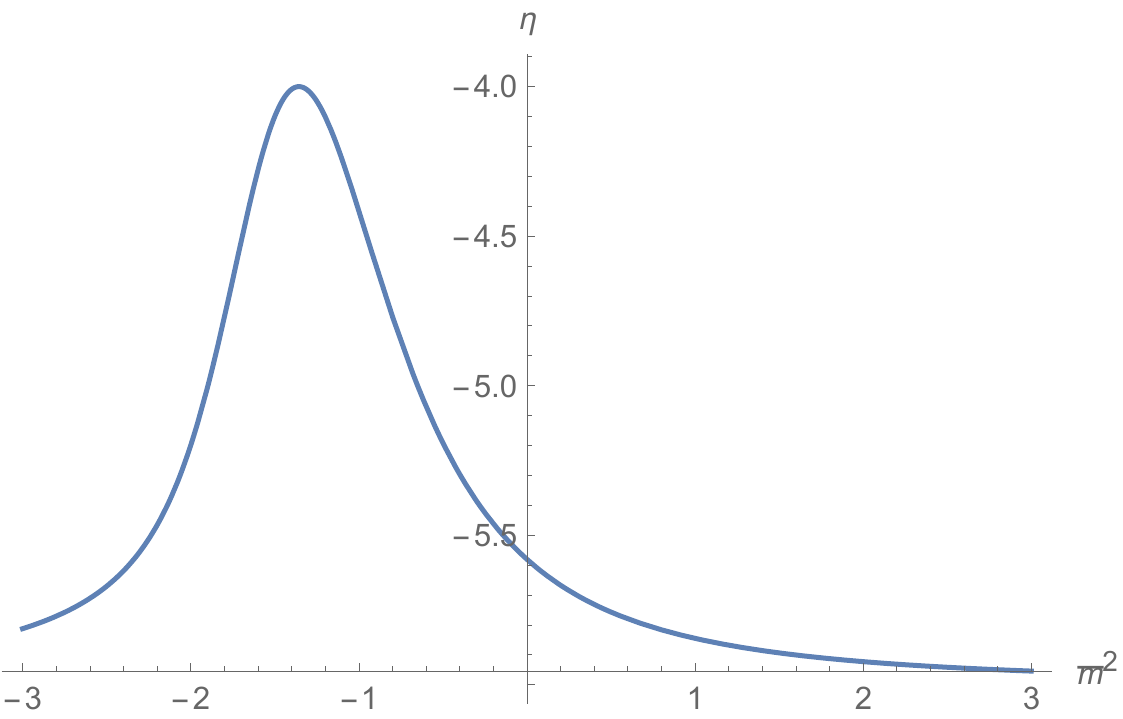} }}}$
\end{center}
\caption{The constrained melonic flow $\bar{g}=f(\bar{m}^2)$ ($\mathcal{E}_{\mathcal{C}1}$) in blue and $\bar{g}=0$ ($\mathcal{E}_{\mathcal{C}0}$) in brown for three interactions (a); the corresponding beta function $\beta_m(\bar{m}^2,\bar{g}=f(\bar{m}^2))$ (b); the anomalous dimension (c). }\label{figd3}
\end{figure}

The Ward constraint \eqref{const} remains unchanged, and inserting the flow equations following the strategy described in section \ref{sec3}, we get the solutions pictured on Figure \ref{fig1color}. This picture have to be compared with Figure \ref{plotf}. In particular, $f$ is always positive and $\beta_m$ has the same behavior in the physical domain $\bar{m}^2>-1$ as the corresponding $\beta$-function for the fully colored model. $\beta_m$ has a single real zero, for $\bar{m}^2\approx 0.06$, which is purely attractive toward IR. It is however as unphysical as the corresponding fixed point discovered for the fully colored model, the anomalous dimension so large and negative: $\eta_*\approx -5.67$ (close to the value obtained previously). Hence, these results highlight the weak dependencies of our previous conclusions in regard to the number of quartic interactions involved to span the theory space, as soon as we focus on the isotropic sector (all interactions have the same coupling constant). In particular, one expect that disconnected interactions play no significant role. 
For $2$ interactions, the results approach the full model. $f$ remains positive and the corresponding $\beta$-function for $\bar{m}^2$ is pictured on Figure \ref{betamd2}. It has the same shape as the one-colored and the full colored model, and it has a single zero for $\bar{m}^2\approx 0.13$, which is unphysical again ($\eta_* \approx -5.6$). 
For $3$ interactions finally, we find again two fixed point, one of them being unstable and the other one, reminiscent of $p_1$, has negative and large anomalous dimension $\eta_* \approx -5.8$. In the constrained space, the results are summarized on Figure \ref{figd3}: We recover the main features of the fully colored model; and no reliable fixed point. To summarize:
\begin{claim}
The relevant features of the RG in the non-branching melonic sector weakly rely on the number of interactions. In particular, the predictions of the single and fully colored models are essentially the same. For this reason, we expect that disconnected interactions play no significant role in a first approximation. 
\end{claim}

\subsection{Dynamical constrained flow}

As an alternative way of investigation, we propose to fix $\pi_3^{(i)}$, and then the sum $\mathcal{A}_{3,0}$ by the flow itself, rather than through a specific approximation for $\Gamma_k^{(2)}$, out of the windows of momenta allowed by $\dot{r}_k$. Our procedure is schematically the following: \\

\noindent
(1) We keep $\beta_m$ and fix $\beta_g$ from the equation \eqref{const}:
\begin{equation}
\left\{
\begin{array}{ll}
\beta_m&=-(2+\eta)\bar{m}^{2}-\,\frac{10\pi^2\bar{g}}{(1+\bar{m}^{2})^2}\,\left(1+\frac{\eta}{6}\right)\,,\\
\beta_g&=-\eta\bar{g}\, \frac{\Omega(\bar{g},\bar{m}^2)}{(1+\bar{m}^2)^2}+\frac{2\pi ^2\bar{g}^2}{(1+\bar{m}^2)^3}\beta_m\,.
\end{array}
\right.\label{systprime}
\end{equation}
(2) We fix $\pi_{3,000}^{(i)}$ dynamically from the flow equation \eqref{florence2}:
\begin{align}
\nonumber\beta_g=-2\eta \bar{g} &-3 \bar{\pi}_3^{(1)}\frac{\pi^2}{(1+\bar{m}^{2})^2}\left(1+\frac{\eta}{6}\right)\\
& +4\bar{g}^2 \,\frac{\pi^2}{(1+\bar{m}^{2})^3}\left(1+\frac{\eta}{6}\right)\label{pi3dyn}
\end{align}
(3) We compute $\frac{d}{dp_1^2}\pi_{2,00}^{(i)}$ from equation \eqref{W2}, and finally deduce an equation for the anomalous dimension $\eta$. The computation require the sums $\mathcal{L}_{2,k}$ and $\mathcal{L}_{3,k}$. $\mathcal{L}_{2,k}$ has a vanishing power counting, and contain the undefined sum $\mathcal{A}_{2,0}$. However, $\mathcal{L}_{2,k}$ may be expressed in term of $Z(k)$ and $g(k)$ from equation \eqref{W1}. Indeed, setting $k=0$ and fixing the renormalization condition such that $Z(k=0)=1$\footnote{This condition may be refined, see \cite{Lahoche:2018oeo}, but this point has no consequence on our discussion.}, we get that, in the continuum limit $\Lambda\to \infty$, $Z\to 0$. To summarize, in the same limit, \eqref{W1} reduces to $-2g(k)\mathcal{L}_{2,k}=Z(k)$, and from \eqref{W2}:
\begin{equation}
\frac{d}{dp_1^2}\pi_{2,00}^{(1)}=\left(Z(k)\frac{\pi_{3,00}^{(1)}}{2g(k)}+2(\pi_{2,00}^{(1)})^2\,\mathcal{L}_{3,k}\right)\,. \label{deriv2}
\end{equation}
To compute $\mathcal{L}_{3,k}$, we have to note that, because all the quantities are renormalized, only the superficial divergences survives. As a result, $\mathcal{A}_{3,p}$ does not diverge, such that in the continuum limit, $Z_{-\infty}\mathcal{A}_{3,p}$ must vanish, and we get straightforwardly:
\begin{equation}
\mathcal{L}_{3,k}=-\frac{1}{2Z^2(k)k^2}\frac{\pi^2}{(1+\bar{m}^2)^3}\,.
\end{equation}
Note that this term was computed using \eqref{derivexp} in the interior of the domain $\vec{p}\,^2 < k^2$. It is easy to check that the system \eqref{systprime} does not admits physically relevant fixed point using expression \eqref{eta1} for the anomalous dimension. Indeed, a fixed point have to satisfy $\eta=0$, i.e. $(1+\bar{m}^{2})^2-\bar{g}\pi^2(2+\bar{m}^{2})=0 $ or $(1+\bar{m}^{2})^2-\bar{g}\pi^2=0 $. Inserting the first condition in the flow equation for $\beta_m$, we get:
\begin{equation}
\beta_m=-2 \bar{m}^{2}-\frac{10}{2+\bar{m}^{2}}\,,
\end{equation}
which has no real zero. With the second condition however:
\begin{equation}
\beta_m=4 \bar{m}^{2}+\frac{2}{2+ \bar{m}^{2}}-6\,,
\end{equation}
which has two zeros, for $\bar{m}^2\approx -1.85$ and $\bar{m}^2\approx1.35$. The first one has to be discarded because it is located under the singularity $\bar{m}^2\approx -1$. The second one because it strongly violates the physical bound $\eta >-2$ (explicitly $\eta \approx -4.2$).
\medskip

From equation \eqref{systprime} and \eqref{pi3dyn}, we get (we omit the indices $0$ to simplify the notations) :
\begin{align*}
-\frac{3}{\bar{g}}\pi_3^{(1)}\frac{\pi^2}{(1+\bar{m}^{2})^2}\left(1+\frac{\eta}{6}\right)&=\eta+\eta\frac{\pi^2 \bar{g}}{(1+\bar{m}^2)^2} -\frac{2\pi^2\bar{g}\bar{m}^2}{(1+\bar{m}^2)^3}(2+\eta)\\
&-\frac{20\pi^4\bar{g}^2}{(1+\bar{m}^2)^5}\left(1+\frac{\eta}{6}\right) -4\bar{g}\frac{\pi^2}{(1+\bar{m}^2)^3}\left(1+\frac{\eta}{6}\right)\,.
\end{align*}
Interestingly, it is easy to see, using the perturbative expansion that this formula is in accordance with the formula \eqref{6pp}. Indeed, from \eqref{eta1},
\begin{equation}
\eta\approx 4\pi^2 \bar{g} (1-2\bar{m}^2)+\cdots\,,
\end{equation}
canceling all the $\bar{g}\bar{m}^2$ terms at the leading order. Then, from equations \eqref{deriv2}, \eqref{pi3dyn} and from the flow equation \eqref{florence}, it is easy to get an explicit relation, fixing $\eta$ for a given value of the pair $(\bar{m}^2,\bar{g})$ along the flow.  After some algebraic manipulations this relation takes the form:
\begin{equation}
\eta=\frac{4 \pi ^2\bar{g}  \left(\frac{\pi ^2 \bar{g} }{5 (1+\bar{m}^2)^3}+1\right)}{(1+\bar{m}^2)^2-\Omega_1(\bar{m}^2,\bar{g})}\,, \label{stateeta}
\end{equation}
where:
\begin{equation}
\Omega_1(\bar{m}^2,\bar{g}) :=\frac{6 \pi ^2\bar{g} }{5}-\frac{4 \pi ^4 \bar{g}^2}{(1+\bar{m}^2)^3}-\frac{12 \pi ^2 \bar{g}  \bar{m}^2}{5 (1+\bar{m}^2)}-\frac{4 \pi ^2 \bar{g} }{5 (1+\bar{m}^2)}\,.
\end{equation}
 Using equation \eqref{stateeta} for $\eta$, we find that no global real fixed point survives for the system \eqref{systprime}. Hence, fixing $\pi_3$ in this way, we may expect to improve the computations of the sums implied in EVE, keeping contributions which have to be treated on the same footing with respect to the power counting.
\medskip

 Alternatively, we may assume validity of $\beta_\lambda$ and $\eta$ given by system \eqref{syst3} and \eqref{eta1}. As we pointed out these equations are compatible with Ward identities \eqref{W2}, and we may assume that missing disconnected interactions, which are essentially relevant for mass is a more serious source of physical disagreement than using derivative expansion to compute the sums involved in the EVE. With this respect, we should define $\beta_m$ from equation \eqref{const} rather than from the flow equation. The resulting system satisfy Ward identities, up to less relevant effects than $\phi^6$ melonic interaction in regard to the power counting, and  no physically reliable fixed point solutions are found, in agreement with the previous results \ref{claim0}.

Equation \eqref{stateeta} determine $\eta$ in the constrained space $\mathcal{E}_{\mathcal{C}}$ excepts for $\bar{g}=0$ and   $\Omega_1=(1+\bar{m}^2)^2$, where it is undefined. In the vicinity of the Gaussian fixed point, we recover the universal one-loop asymptotic freedom:
\begin{equation}
\eta\approx 4 \pi^2\bar{g}\,, \qquad \beta_g=-4 \pi^2\bar{g}^2\,. \label{div}
\end{equation}
All these results are in agreement with the claim \ref{claim0}: \textit{No physically relevant fixed point exist for the model that we consider in the non-branching sector.}

\subsection{Extension to higher interactions and beyond melonic sector}\label{sec4}

In this section, we briefly review some results beyond the quartic model discussed in the previous sections. For more details, the reader may consult \cite{Lahoche:2018oeo},\cite{Lahoche:2019cxt} of the same authors. The EVE methods have been initially developed as a trick working well for just-renormalizable quartic models \cite{Lahoche:2018oeo}. Extension to sextic model was required from the fact that some potentially relevant TGFT for physical implications include sextic melonic interactions in the list of their interactions \cite{Carrozza:2012uv}-\cite{Lahoche:2015ola}. The rank-three model, in particular, based on the group structure $\SU(2)$ \cite{Carrozza:2013wda}, in particular, has been largely investigated for its relevance to three-dimensional quantum gravity. Unfortunately, it is not asymptotically free \cite{Carrozza:2014rba}, but potentially well defined in the UV from the existence of a deep UV attractor, compatible with an asymptotic safety scenario \cite{Carrozza:2014rya}-\cite{Carrozza:2017vkz}. In the nonperturbative analysis, the authors show the existence of a UV attractive fixed point, resistant to large truncations investigations, and with the interesting property to depend only on a subfamily of melons said \textit{non-branched}, with general structure:
\begin{equation}
\vcenter{\hbox{\includegraphics[scale=1.2]{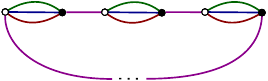} }}\,.
\end{equation}
For this subsector of the full melonic sector, a version of the EVE method has been build for an Abelian model, expected to extend the truncation results obtained in \cite{Lahoche:2016xiq}-\cite{Carrozza:2016tih}. A fixed point with one relevant direction has been found, expected to be compatible with the UV attractor. However, the details of the two investigated models being very different. In addition to the choice of the dimension and the group structure, they differ by an internal symmetry arising from loop quantum gravity and known as closure constraint, such that the physical field may satisfy:
\begin{equation}
\psi(g_1,g_2,g_3)=\phi(hg_1,hg_2,hg_3)\,,\qquad \forall\,h\in \mathbf{G}\,.
\end{equation}
Therefore, a strong statement requires to use of EVE for the non-Abelian model considered in the reference. \\

As mentioned, EVE method has been applied as well for investigations beyond the melonic sector, including a new class of observables called pseudo-melons. In rank $5$, the elementary pseudo-melon is typical:
\begin{equation}
\vcenter{\hbox{\includegraphics[scale=1]{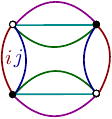} }}\,.
\end{equation}
Focusing on non-branching interactions of the form:
\begin{equation}
\vcenter{\hbox{\includegraphics[scale=1]{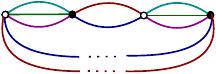} }}\,,
\end{equation}
the authors show that a renormalizable sector exists for sextic interactions, which may have zero canonical dimension, and then fix the canonical dimension for other pseudo-melonic observables. Finally, as for the melons, pseudo-melons exhibit a branched structure, which can be mapped as trees with bicolored edges, and resumed like melons. This was the first non-trivial example showing how renormalizable sectors merge, increasing drastically the number of effective graphs with the single melonic sector discussed in the previous sections. Numerically, the investigations revealed that no additional fixed points appear, in agreement with Ward identities).

\section{The nonpertubrative RG for matrix models}\label{sec4}

For TGFTs, the renormalization group flow is considered in some sense ‘‘traditionally". The non-trivial propagator fixes an abstract notion of scale, and fluctuation can be integrated out, from the deep UV, where fluctuation has large square momenta, to the deep IR, where fluctuations have small momenta. However, a more abstract renormalization process has been considered for matrix models \cite{Brezin:1992yc} to investigate double scaling limit; and extended using non-perturbative techniques in
\cite{Eichhorn:2013isa}-\cite{Eichhorn:2017xhy} for matrix and tensor models with trivial propagator. All these investigations allow recovering more and less qualitatively the exact (analytic) results. However, some subtleties appear due to the breakdown of the original $U(N)$ invariance. Because the propagator is trivial, all the quantum fluctuations play the same role. There is no canonical distinction between UV and IR scales, and the choice of integrated out degrees of freedom becomes arbitrary. The common choice has singularized a corner of the matrix, and to integrate firstly degrees of freedom with large indices as “UV". Breaking the original symmetry has an important consequence regarding the Ward-Takahashi identities, which have been started to investigate in
\cite{Eichhorn:2013isa},\cite{Lahoche:2019ocf}. In this section, we review these aspects of the RG flow for matrix models.

\subsection{Flowing in the matrix theory space}

In this section, we recall the framework allowing us to investigate nonperturbative aspects of the RG flow for matrix models. To go beyond this informal presentation the reader may consult the references \cite{Eichhorn:2013isa}, \cite{Lahoche:2019ocf}. In contrast with the model considered in the introduction, and to make contact with some reference papers \cite{DiFrancesco:1993cyw}, we focus on a quartic model:
\begin{equation}
S[\phi]=\frac{1}{2} \Tr (\phi^2)+ \frac{g}{4} \Tr (\phi^4)\,.
\end{equation}
In addition to the $U(N)$ symmetry, this model has a discrete $\mathbb{Z}_2$ symmetry $\phi\to -\phi$, and generate squarulations rather than triangulations. This distinction is unimportant for the continuum limit that we will investigate, which does not depend on the choice of elementary discrete polygons used to build random surfaces. For this model, the critical value $g_c$ and the corresponding critical exponent $\theta$ in the continuum limit have been exactly computed \cite{DiFrancesco:1993cyw}:
\begin{equation}
g_c= -\frac{1}{12}\,,\qquad \theta=\frac{4}{5}\,. \label{exact}
\end{equation}
The elementary intuition allowing to consider the renormalization group approach to investigate the continuum limit for matrix comes from the constraint \eqref{const}, freely interpreted as a fixed point with an appropriate scaling in $N$. This point of view, understanding double scaling as a fixed point of the RG flow with respect to $N$ has been firstly considered by Brezin and Zinn-Justin in \cite{Brezin:1992yc}-\cite{DiFrancesco:1993cyw}. In their elementary approach, they assume that the partition function of matrix models satisfies the Callan-Symanzik equation:
\begin{equation}
\left(N \frac{\partial}{\partial N}+\beta_g(g) \frac{\partial}{\partial g}+\eta(g)\right)\mathcal{Z}(N,g)=r(g)\,, \label{CS}
\end{equation}
where $\beta_g:= N\partial g/\partial N$ is the rate of change of the coupling constant when $N$ changes, and in contrast with just-renormalizable theories, the remainder $r(g)$ does not vanish here; the coupling constant $g$ being irrelevant with respect to its canonical scaling (see below). Reaching a fixed point $g\equiv g_*$, $\beta_g$ must vanish, and $\mathcal{Z}(N,g)$ scales as:
\begin{equation}
\mathcal{Z}(N,g)\sim (g_c-g)^{\gamma_1}f\left((g_c-g)N^{\frac{2}{\gamma_1}}\right)\,,
\end{equation}
enforcing the identification :
\begin{equation}
\gamma_1=2-\gamma=-\frac{2}{\beta^\prime(g_*)}\,, \label{string}
\end{equation}
which corresponds to the string susceptibility $\gamma$ and the critical exponent $-\beta^\prime(g_*)$ as in \cite{Brezin:1992yc}. The exact computation \eqref{exact} then requires $\beta^\prime(g_*)=4/5$. Beyond a purely scaling relation, equation \eqref{CS} can be phrased in a Wilsonian point of view; as the way that any change of matrix size $N\to N-\delta N$ may be compensated by a change $g\to g+\delta g$ of the coupling, without modify of the continuum physics. In this point of view, we introduce an arbitrary slicing in the matrix entries, and integrate out ‘‘step by step" the large $N$ degrees of freedom along lines and rows. Then, starting from $N\times N$ random matrices, we reduce them to $(N-1)\times (N-1)$ matrices after a single step, $(N-2)\times (N-2)$ matrices after two steps, and so on. To each step, $(N-i)\times (N-i)$ matrices are described by effective action, building as a sum of two distinct pieces: The classical action for $(N-i)\times (N-i)$ matrices, corresponding to the original action truncated around $(N-i)\times (N-i)$ matrices; and the fluctuations term, arising from integration of $N-i+1$ degrees of freedom. As a result, to each step, the couplings have the discrete change rule:
\begin{equation}
g_{i+1}=g_{i}+\frac{1}{N} \beta(g_i)+\mathcal{O}(1/N)\,,
\end{equation}
where the notation suggest that we consider only the large $N$ limit to define the $\beta$-function. Computing $\beta(g_i)$ from a single step, we get the one-loop beta function \cite{Brezin:1992yc}:
\begin{equation}
\beta(g)\approx g+6g^2+\mathcal{O}(g^3)\,, \label{oneloop}
\end{equation}
which vanishes for $g_*=-1/6$, in qualitative accordance with the exact result \eqref{exact}. \\

\noindent
Nonperturbative techniques have been originally introduced in \cite{Eichhorn:2013isa}-\cite{Lahoche:2019ocf} to improve this perturbative results, taking into account higher couplings and nonperturbative loop effects. In this reference, the authors introduced a FRG framework based on the Wetterich equation formalism. Recall that the Wilson RG procedure requires an appropriate slicing into modes between UV scales (when no fluctuations are integrated out) and IR scales (when all the fluctuations are integrated out), dictating how the small distance fluctuations are integrated out. Following the standard strategy in FRG formalism, we introduce a new term in the classical action,
\begin{equation}
\Delta S_N [\phi]=\frac{1}{2}\sum_{a,b,c,d} \phi_{ab} [r_N(a,b) ]_{ab;cd} \phi_{cd}\,, \label{regulator}
\end{equation}
which behaves like a scale dependence of the mass term, the specific slicing in $N$ depending on the shape of the regulator $r_N(a,b)$. Introducing this mass term into the classical action, we replace the global description given by the referent generating functional $\mathcal{Z}[J]:=\int d\phi e^{-S[\phi]+ J\cdot\phi}$, by a one-parameter set of models $\{\mathcal{Z}_N[J]\}$ defined as:
\begin{equation}
\mathcal{Z}_N[J]:=\int\, d\phi\, e^{-S[\phi]-\Delta S_N [\phi]+J\cdot\phi}\label{part2}\,,
\end{equation}
where the dot product is defined as $A\cdot B:= \sum_{mn} A_{mn}B_{mn}$. Then, because of the scale dependence of the regulator, the long distance physics effects ($(m,n)\lesssim N$) acquire a large mass and are frozen out, whereas the small distance effects ($(m,n)>N$) are integrated out. The RG flow then makes connection from $\mathcal{Z}_{N}$ to $\mathcal{Z}_{N-\delta N}$. Explicitly this relation may be re-expressed as a first order differential equation very similar to the flow equation \eqref{Wett}:
\begin{equation}
\dot{\Gamma}_N=\frac{1}{2}\Tr \left[\dot{r}_N\left(\Gamma^{(2)}_N+r_N\right)^{-1}\right]\,,\label{Wettbis}
\end{equation}
which indicates how the average \textit{effective action} $\Gamma_N$ is modified in the windows of scale $[N,N-dN]$, the dot meaning the derivative with respect to the RG parameter $t:=\ln N$: $\dot{X}=N\frac{d}{dN}X$. As in the previous sections, equation \eqref{legendre}, the average effective action is defined as slightly modified Legendre transform of the free energy $\mathcal{W}_N:=\ln\mathcal{Z}_N$ :
\begin{equation}
\Gamma_N[\Phi]+r_N[\Phi]=J\cdot \Phi-\mathcal{W}_N[J]\,,
\end{equation}
where $\Phi$ denotes the classical field:
\begin{equation}
\Phi_{mn}:= \frac{\partial \mathcal{W}_N}{\partial {J}_{mn}}\,.\label{classical}
\end{equation}
In the same way $\Gamma^{(2)}_N$ in equation \eqref{Wett} denotes the second derivative of the average effective action :
\begin{equation}
\left[\Gamma^{(2)}_{N}\right]_{mn;pq}:=\frac{\partial^2 \Gamma_k}{\partial \Phi_{mn}\partial {\Phi}_{pq}}\,.
\end{equation}
Even to close this section we have to add an important comment about the notion of \textit{canonical scaling}. Scaling, that is to say, the dependence of the quantity coming from their dimensionality play generally an important role in renormalization. In standard quantum field theory, for instance, dimensionality is closely related to renormalizability. For matrix models, as for TGFTs, the situation is quite different, because there are no referent space-time, no referent length and no canonical scaling coming from the extra-structure of the theory. However, the behavior of the RG flow with $N$ in the vicinity of the Gaussian fixed point (i.e. keeping only the part of the scaling which is independent of the couplings), provides an intrinsic notion of dimension, that we call \textit{canonical dimension}:
%\begin{definition}
%For any trace observable $g_k \Tr(\phi^k)$ in the classical action, the canonical dimension of the coupling constant $g_k$ is defined in the vicinity of the Gaussian fixed point as the part of the scaling in $N$ which is independent of $g_k$ and the other couplings.
%\end{definition}
\noindent
We denote as $d_k$ the canonical dimension of $g_k$, so that the intrinsic scaling writes as $N^{d_k+\mathcal{O}(g_1,g_2,\,\cdots)}$. To find the explicit expression of $d_k$, we then have to be investigate the behavior of the Feynman diagrams with $N$. This may be traced from the link between two dimensional quantum gravity recalled in the previous section. Up to the rescaling $\phi\to \sqrt{N} \phi$ we have stressed a relation between matrix coupling, $N$, Newton and cosmological constant. Keeping this relation implies that each Feynman diagrams scales exactly as $N^{2-2h}\equiv N^{\chi}$, where $\chi(\Delta):= V(\Delta)-E(\Delta)+F(\Delta)$ denote the Euler characteristic of the polygon decomposition $\Delta$, having $V$ vertices, $E$ edges and $F$ faces. It is not hard to see that this holds if and only if, up to the mentioned rescaling, the only $N$ dependence of the classical action comes from a global $N$ factor, enforcing the definition:
\begin{equation}
d_k=-\frac{k-2}{2}\,,
\end{equation}
in agreement with formula \eqref{defmodel}. In this paper, we consider also multi-trace interactions at the level of the effective action, and we have to extend this formula for such an interactions. In order to remain in accordance with the expected scaling $N^\chi$, we impose to cancel the additional $N$ factors coming from the additional traces. As a result, for an observable of the form $\prod_{j=1}^n\Tr (\phi^{k(j)})$, one assigns the canonical dimension $d^{(j)}_{k(1),\cdots,k(j)}$:
\begin{equation}
d^{(j)}_{k(1),\cdots,k(j)}= d_{\sum_j k(j)}-(j-1) \,.
\end{equation}
For a double trace operator for instance $\Tr (\phi^k)\Tr(\phi^l)$, one gets $d_{kl}^{(2)}=-(k+l)/2$. As pointed out in \cite{Eichhorn:2013isa}, it is interesting to note that, even for a single trace operator, the canonical dimension is negative for $k>2$, meaning that all non-Gaussian couplings are irrelevant. In this situation, the improvement of the scaling coming from radiative corrections plays an essential role in the fixed point structure. \\

\subsection{Solving RG equations in the local potential approximation}

Solving the exact flow equation \eqref{Wett} remains an open issue for a lot of interesting relevant physical models, except for very special problems.
As a result, extracting information about the nonperturbative behavior of the RG flow then requires an appropriate scheme of approximation. In this section, we review a common approach, based on a crude truncation of the full theory space. Among the existent methods, this one has a strong advantage to easily get along with the non-locality of the interactions. Note that a specific locality principle emerges from the $U(\Lambda)$ invariance of the interactions, and we focus essentially on the \textit{local potential approximation} (LPA), i.e. on single trace interactions. Enlargement to the product of traces is then discussed at the end of the section. \\

\textit{i.) Local potential approximation.} The matrix action is non-local in the usual sense in field theory. However, what allows us to say that two objects interact locally is the usual sense is precisely the interaction. The interaction then allows to define by themselves an appropriate locality principle, and we adopt the definition (which is equivalent to the definition used in the previous sections for TGFTs):
\begin{definition}
Any global trace of the form $\Tr (\phi^k)$ is said to be a local monomial interaction. In the same way, any functional of $U[\phi]$ which may be expanded as a sum of single traces are said to be a local functional.
\end{definition}\label{deflocal}
\noindent
Note that this locality principle reflects the proper invariance of the interactions concerning unitary transformations, exactly as for TGFTs\footnote{See \cite{Lahoche:2019vzy}-\cite{Lahoche:2019cxt} and references therein for an extended discussion, showing how this definition work in practical contexts, especially in the context of matrix field theory to define counter-terms for renormalization.}. \\

\noindent
The first parametrization of the theory space that we consider split the effective action $\Gamma_N(\Phi)$ as a sum of two kind of terms:
\begin{equation}
\Gamma_N(\Phi)=(\text{non-local})+U_N(\Phi)\,. \label{decomp1}
\end{equation}
The last term $U_N(\Phi)$ designates the purely local potential, expanding as a sum of single trace observables:
\begin{equation}
U_N(\Phi)= \frac{Z_N}{2} \Tr (\Phi^2)+\frac{g_{4,N}}{4} \Tr (\Phi^4)+\frac{g_{6,N}}{6} \Tr(\Phi^6)+\cdots\,.\label{potential}
\end{equation}
Following \cite{Eichhorn:2013isa}, we introduced a field strength renormalization $Z_N$ in front of the Gaussian term. The renormalized quantities are generally defined from a fixed coefficient in the Gaussian part of the original action. Rescaling the fields such that the mass term reduces to its free term $\frac{1}{2} \Tr \Phi^2$, we define the dimensionless and renormalized couplings $u_{k,N}$ as:
\begin{equation}
u_{k,N}:= N^{-d_k} Z_N^{-k/2} g_{k,N} \label{rencoupl}
\end{equation}
As pointed out in the derivation of the Ward identity, the presence of the regulator breaks the $U(\Lambda)$ invariance of the original action, and the RG flow has to generate non-invariant momentum dependent effective interactions such that, for instance:
\begin{equation}
K_N[\Phi]= \sum_{a,b} q\left(\frac{a}{N},\frac{b}{N}\right)\Phi_{ab}\Phi_{ba}\,. \label{kin}
\end{equation}
where the Taylor expansion of the function $q$ starts at the order $1$ in $a/N$ and $b/N$. The terminology ‘‘momentum dependent" simply reflects the situation in ordinary quantum field theory, the indices of the matrix field playing the role of discrete momenta. Expanding $q$ in power of $a/N$ and $b/N$ corresponds to the standard derivative expansion. As we will see from Ward identity, such a deviation from strict locality introduces relevant corrections at the leading order in $1/N$, and must be kept in the large $N$ limit. In particular, in the closure procedure around quartic interactions, the linear coupling:
\begin{equation}
q\left(\frac{a}{N},\frac{b}{N}\right)=\gamma\, \frac{a+b}{2N}\,,\label{nonlocal}
\end{equation}
plays an important role in the fixed point structure, improving strongly the local potential approximation. In a first time, in order to compare them, we keep only the strong local part of the decomposition \eqref{decomp1}:
\begin{equation}
\Gamma_N(\Phi)=U_N(\Phi)\,. \label{localpara}
\end{equation}

\noindent
The flow equations for couplings $g_n$ in the parametrization \eqref{localpara} can be deduced from the exact Wetterich equation deriving $n$ time with respect to $\Phi$ and setting $\Phi=0$ (we recall that we work in the symmetric phase). Because $\Phi$ is a Hermitian matrix, $\Phi_{ab}=\Phi^*_{ba}$, and :
\begin{equation}
\frac{\partial \Phi_{ab}}{\partial \Phi_{cd}}=\delta_{ac}\delta_{bd}\,,
\end{equation}
from which we get:
\begin{align}
[\Gamma_N^{(2)}]_{ab,cd}= \delta_{ac}\delta_{bd}Z_N\,,
\end{align}
where $g_{ab,cd}:=\delta_{ac}\delta_{bd}$ is nothing but the ‘‘bare" propagators. For the regulator function, we chose a modified version of the Litim optimized regulator \cite{Litim:2000ci}-\cite{Canet:2002gs}, allowing to make analytic computations:
\begin{equation}
\left[r_N(a,b)\right]_{ab,cd}=Z_N \delta_{cb}\delta_{ad} \left(\frac{2N}{a+b}-1\right) \Theta\left(1-\frac{a+b}{2N}\right)\,.
\end{equation}
Note that in this section we designate the Heaviside function with a big $\Theta$
to avoid confusion with critical exponents. Taking the derivative with respect to the flow parameter $t=\ln N$, we get straightforwardly:
\begin{align}
\left[\dot{r}_N(a,b)\right]_{ab,cd}= Z_N g_{ba,cd}&\frac{2N}{a+b} \Theta\left(1-\frac{a+b}{2N}\right)+ \eta_N \left[r_N(a,b)\right]_{ab,cd}\,,
\end{align}
where we introduced the anomalous dimension following the usual definition:
\begin{equation}
\eta_N:=\frac{ \dot{Z}_N}{Z_N}\,. \label{anmalous}
\end{equation}

\textit{ii.) Truncation along local theory space.} Taking successive derivative with respect to $\Phi$ of the exact flow equation \eqref{Wett}, and setting $\Phi=0$, we deduce the flow equations for all couplings involved in \eqref{localpara}. For each step, all contributions involve some powers of the effective propagator $G_N=(\Gamma_N^{(2)}+r_N)^{-1}$, evaluated for vanishing $\Phi$, and for $a+b\leq 2N$ as:
\begin{align}
(G_N)_{ab,cd}= Z_N^{-1} g_{ba,cd}\, \frac{a+b}{2N} \,. \label{eqG}
\end{align}
The one-loop sums that we will encounter in the derivation of the flow equations are all of the forms:
\begin{equation}
I_{a}^{(p)}:= \sum_b \left((G_N)_{ab,ba}\right)^p \left[\dot{r}_N(a,b)\right]_{ab,ba}\,.
\end{equation}
In the large $N$ limit, the sum can be replaced by an integration up to $1/N$ corrections. Let us introduce the continuous variable $2N x:=a+b$, running from $a/2N$ to $1$:
\begin{equation}
I_{a}^{(p)}\approx 2Z_N^{1-p} N \int_{a/2N}^{1} dx\,x^{p-1} \, \left(1+\eta_N(1-x)\right)\,, \label{Ip}
\end{equation}
leading to:
\begin{align}
I_{a}^{(p)}\approx 2Z_N^{1-p} &N \bigg[\frac{1}{p} \left(1-\left(\frac{a}{2N}\right)^p\right)(1+\eta_N)-\eta_N \frac{1}{p+1} \left(1-\left(\frac{a}{2N}\right)^{p+1}\right)\bigg]\,.\label{intapprox}
\end{align}

\begin{figure}
\begin{center}
\includegraphics[scale=0.8]{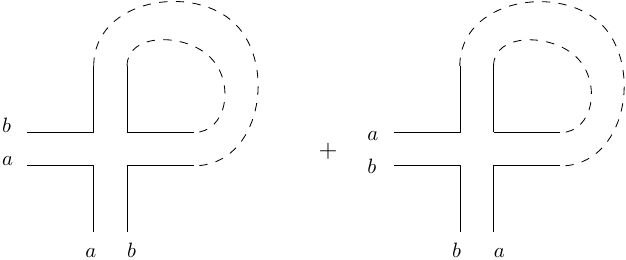}
\end{center}
\caption{Leading order contractions for 2-point graphs made of a single effective loop. } \label{figcont1}
\end{figure}

Deriving the equation \eqref{Wett} twice with respect to the $\Phi$ fields, and setting $\Phi=0$, we get:
\begin{equation}
\dot{\Gamma}_{N,ab,ba}^{(2)}=-\frac{1}{2}\,G_{N,cd,ef} \Gamma^{(4)}_{N,ef,lm,ab,ba}\tilde{G}_{N,lm,cd}\,, \label{equationwettexp1}
\end{equation}
with $\tilde{G}_{N,ab,cd}:= ({G}_{N}\,\dot{r}_N)_{ab,cd} $ and where once again we sum over repeated indices. To compute the sums, we have to take into account the symmetry structure of the external indices. From \eqref{eqG}, we get for instance
\begin{equation}
G_{N,cd,ef}({G}_{N}{r}_N)_{lm,cd}= \frac{g_{ef,lm}}{Z_N}\, \left(\frac{l+m}{2N} \right)^2f(l/N,m/N)\,,\label{eq52}
\end{equation}
where we defined:
\begin{equation}
f(a/N,b/N)=\left(\frac{2N}{a+b}-1\right) \Theta\left(1-\frac{a+b}{2N}\right)\,.
\end{equation}
For a fixed configuration of the external indices, there are two leading order contractions, both pictured on Figure \ref{figcont1}, where in this graphical representation the dotted edges correspond to the contraction with the effective propagator $G_{N,cd,ef}\tilde{G}_{N,lm,cd}$ given by equation \eqref{eq52}. \\

\begin{figure}
\begin{center}
\includegraphics[scale=0.8]{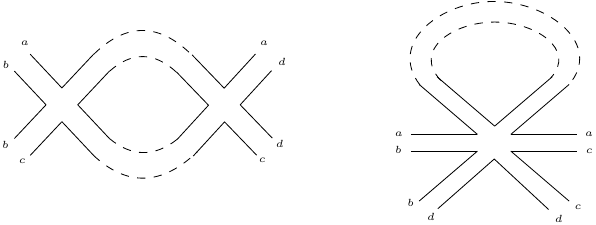}
\end{center}
\caption{Two typical leading order contractions contributing to the flow equation for $g_4$.} \label{fig2}
\end{figure}

\noindent
We now have to compute how many leading order contractions such as the one pictured in Figure \ref{figcont1} contribute. This numerical factor counts the number of perturbations for the four external edges of the effective vertex function which leads to a leading order diagram, having a single internal face. It is not hard to see that there are exactly $4\times 2$ different ways to build such a diagram: $4$ different positions for the first endpoint of the propagator edge, and two remaining positions for the second endpoint. Finally, there is an additional factor $2$ coming from the two remaining attributions for the two free external edges, sharing the momentum $(a,b)$. Then, translating the diagram into equation, and setting $a=b=0$, one gets using the integral approximation \eqref{intapprox}
\begin{equation}
\vcenter{\hbox{\includegraphics[scale=0.6]{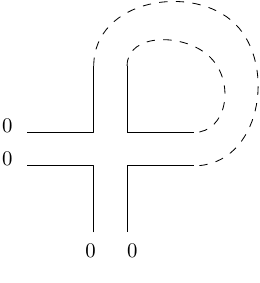} }}=4g_{4,N}I_0^{(2)}\,.
\end{equation}
Finally, computing the derivative of the left hand side of equation \label{equationwettexp} for zero external momenta, we get:
\begin{equation}
{\Gamma}_{N,00,00}^{(2,0)}=Z_N\,,
\end{equation}
from which we deduce that:
\begin{equation}
\dot{Z}_N=-\frac{4Ng_{4,N}}{Z_N}\left(\frac{1}{2}+\frac{\eta_N}{6}\right)\,.\label{eqZ}
\end{equation}
From the definitions \eqref{rencoupl}, we get finally;
\begin{equation}
\eta_N=-\dfrac{6u_{4,N}}{3+2u_{4,N}}\,.
\end{equation}
The computation of the beta function $\beta_4:=\dot{u}_{4,N}$ follows the same strategy. Deriving once again twice with respect to the fields $\Phi$, and setting $\Phi=0$ at the end of the computation, one gets, formally:
\begin{equation}
\dot{\Gamma}_{N}^{(4)}=3\Tr\,\tilde{G}\,\Gamma_N^{(4)}G\,\Gamma_N^{(4)}G-\frac{1}{2}\Tr\,\tilde{G}\,\Gamma_N^{(6)} G\,.
\end{equation}
The relevant diagrams corresponding to the two kinds of traces involved in these expressions, all including one internal face are pictured on Figure \ref{fig2}.
Each of them may be easily translated into equation like for the $2$ point diagrams. For zero external momenta we get:
\begin{equation}
\vcenter{\hbox{\includegraphics[scale=0.7]{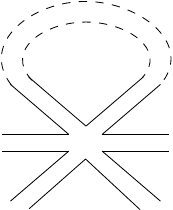} }}= 2\times 24\times \times g_{6,N} I_0^{(2)}\,,
\end{equation}
and:
\begin{equation}
\vcenter{\hbox{\includegraphics[scale=0.7]{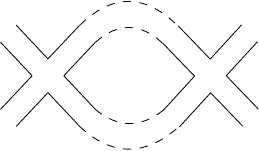} }}= 8\times g_{4,N}^2 I_0^{(3)}\,.
\end{equation}
Once again the numerical factors may be easily understood. For instance, for the diagram involving a $6$-point vertex, there are $6$ different ways to choose the first end point of the contracted edge, $2$ to choose the second one which leads to construct a leading order graph; and finally $4!$ ways to exchange the remaining external points. Because from definition ${\Gamma}_{N,00,00,00,00}^{(4)}=6g_{4,N}$, it follows that:
\begin{equation}
6\dot{g}_{4,N}=24 g_{4,N}^2 I_0^{(3)} - 24 g_{6,N} I_0^{(2)}\,,
\end{equation}
leading to:
\begin{equation}
\beta_4=(1-2\eta_N) u_{4,N} + \frac{4u_{4,N}^2}{6}(4+\eta_N)-4u_{6,N}\left(1+\frac{\eta_N}{3}\right)\,. \label{eqbeta4}
\end{equation}
\begin{remark}
Neglecting the coupling $u_{6,N}$ and expending the remaining right-hand side in the power of $u_{4,N}$, up to order $u^{3}_{4,N}$, we do not reproduce the one-loop result \eqref{oneloop}. In particular, the numerical factor in front of $u_{4,N}$ becomes $20/3$. This cannot be viewed as a defect of the approach, the one-loop beta function being non-universal for coupling with non-zero canonical dimension, as it can be easily checked.
\end{remark}

\noindent
We may proceed in the same way for $\beta_6$, $\beta_8$, $\beta_{10}$; $\beta_n$ involving $u_{n+2}$ and so one. The truncation procedure stop crudely this infinite hierarchical system, imposing $u_{n}=0$ for some $n$. For $n=10$, it is then straightforward to get the following closed system :

\begin{proposition}\label{prop1}
In the large $N$ limit, and in the local potential approximation, the truncated RG flow around $\phi^6$ interactions is described by the following closed system:
\begin{equation*}
\beta_4=(1-2\eta_N) u_{4,N} + \frac{2u_{4,N}^2}{3}(4+\eta_N)-4u_{6,N}\left(1+\frac{\eta_N}{3}\right)\,,
\end{equation*}
\begin{align*}
\nonumber \beta_6=(2-3\eta_N+2u_{4,N} (4+\eta_N)) u_{6,N}&-3u_{4,N}^3 \left( 1+\frac{\eta_N}{5}\right)-2u_{8,N}(3+\eta_N)\,.
\end{align*}
\begin{align*}
\beta_8=(3&-4\eta_N)u_{8,N}+\frac{8}{15}u_{4,N}^4 (6+\eta_N)+\frac{4}{3}u_{6,N}^2\left(4+\eta_N\right)-12u_{6,N}u_{4,N}^2\left(1+\frac{\eta_N}{5}\right)+\frac{8}{3}u_{8,N}u_{4,N} \left(4+\eta_N\right)\,.
\end{align*}
with:
\begin{equation*}
\eta_N=-\dfrac{6u_{4,N}}{3+2u_{4,N}}\,.
\end{equation*}
\end{proposition}
The fixed point solutions can be investigated for increasing truncations, from $k=6$ to $k=10$; and the results are given in Table \ref{table1} below. On this table, we tacked the only fixed point having a single relevant direction i.e. a single positive critical exponent. We recall that critical exponents $\theta_i$ are the eigenvalues of the stability matrix $\Theta_{ij}$ with entries:
\begin{equation}
\Theta_{ij}=- \frac{\partial \beta_i}{\partial u_j}\,.
\end{equation}
The results of Table \ref{table1} are in qualitative agreement with the exact statement \eqref{exact}. Moreover, It seems to improve the perturbative computation \eqref{oneloop}. Note with this respect that coupling constants are not universals, in contrast with critical exponents $\theta_i$, which provide an objective criterion to evaluate the reliability of the approximation. \\

\begin{table}[h!]
\begin{center}
\begin{tabular}{|c|c|c|c|c|c|c|c|}
\hline
\text{truncation order} ($k$) & $u_4$ & $u_6$& $u_8$ & $\theta_1$ & $\theta_2$ & $\theta_3$ & $\eta$\\
&&&&&&&
\\
\hline
6&-0.14 &-- &-- &1.09&--&--&0.3\\
\hline
8 &-0.101 &-0.007 & --& 1.06&-1.05&--&0.22 \\
\hline
10 &-0.16 &0.006 &0.002 &1.06&-1.05&-0.99&0.35\\
\hline
\end{tabular}
\caption{Numerical results for vertical truncations from $k=6$ to $k=10$. We see that increasing the number of interactions do not change the value of the positive critical exponents, the other one corresponding to irrelevant directions. Moreover, the anomalous dimension is very small in comparison to truncation with standard Litim regulator.}\label{table1}
\end{center}
\end{table}

\noindent
However, we observe that we have no significant progress passing from $k=8$ to $k=10$ truncation i.e. by adding the $k=10$ valence interactions does not modify significantly the result from $k=8$ valence. The origin of this pathology may be tracked on the hand of disconnected interactions, i.e. interactions build as a product of single traces. We may consider for instance a truncation of the form:
\begin{align}
\Gamma_N[\Phi]=&\frac{Z}{2} \Tr (\Phi^2)+\frac{g_4}{4}\Tr(\Phi^4)+ \frac{g_6}{6} \Tr(\Phi^6) + \frac{h_{2,2}}{4} (\Tr(\Phi^2))^2+\frac{h_{4,2}}{2} \Tr(\Phi^2)\Tr(\Phi^4)+\frac{h_{2,2,2}}{6} (\Tr(\Phi^2))^3+\cdots \,,
\end{align}
involving double and triple traces. Computing the beta functions in the same way as for the single trace potential, we do not obtain a significant improvement comes from these multi-trace with our previous results, at least at the order of truncation that we consider see \cite{Lahoche:2019ocf} for more detail.

\subsection{Modified Ward identities and ambiguities}

\subsubsection{Ward-Takahashi identity}

As discussed in the first part of this review, the Ward-Takahashi identities are a general feature about symmetries in quantum field theory and may be understood as a quantum version of the Noether's theorem, resulting in the translation invariance of the Lebesgue integration measure in the path integral definition of the partition function, cf. \cite{Ward:1950xp}-\cite{Takahashi:1957xn}. In the discrete quantum gravity context, they have been recently discussed in a series of papers \cite{Lahoche:2018ggd}-\cite{Lahoche:2018vun} see also \cite{Eichhorn:2014xaa} and \cite{Lahoche:2019ocf}. There is however a crude difference between TGFTs and matrix models. For TGFTs, the unitary symmetry is already broken by the Laplacian in the kinetic action; so that the regulator does not break an exact symmetry of the original action. This is not the case for matrices, whose original action is exactly unitary invariant. This is especially important regarding the solution of the RG flow that we build. For TGFT, a solution that does not respect the unitary invariance is not expected to be so pathological but appears as an artefact of the RG procedure for matrices. This feature is closely related to what happens for gauge theories, see \cite{Wetterich:2016ewc}. Following the cited references, and in the matrix context, we view the Ward identities as nontrivial functional relations, depending on the regulator like flow equations. With this respect, Ward identities and flow equations have to be treated on the same footing and solved simultaneously. \\

Without regulator term, only the source terms break the global $U(\Lambda)$ invariance (for some fundamental cut-off $\Lambda$). Therefore, infinitesimal variations provide the identification:
\begin{equation}
\Gamma_{N, \bullet\cdots (ab)(ba)}^{(n)}=\Gamma_{N, \bullet\cdots (cb)(bc)}^{(n)}\,, \label{asymptot}
\end{equation}
holding to all orders of perturbation theory. Note that for this section we restrict our investigations to the \textit{symmetric phase} again, where vanishing classical field $\Phi$ defined from equation \eqref{classical} is expected to be a good mean field, and all the odd correlation functions vanish identically. \\

\noindent
The regulator term $\frac{1}{2} \phi \,r_k \phi$ breaks explicitly the global $U(\Lambda)$ invariance, and add a new contribution to the asymptotic Ward identity \eqref{asymptot}. Let us consider an infinitesimal unitary transformation $1+\epsilon$, $\epsilon$ for some infinitesimal anti hermitian operator $\epsilon=\epsilon^\dagger$. At the leading order, the transformation rule for the matrix field $\phi$ is:
\begin{equation}
\phi \to \phi^\prime = (1+\epsilon)\phi (1+\epsilon)^{\dagger}\approx \epsilon \phi-\phi\epsilon\,.
\end{equation}
Therefore, at first order in $\epsilon$, the total variation of the generating functional $\mathcal{Z}_N$ writes as
\begin{equation}
\delta\mathcal{Z}_N= \int d\phi e^{-S_{N}[\phi,J]} \left[-\delta S[\phi]-\delta \Delta S_N[\phi]+\delta (J\cdot \phi)\right]\,, \label{var}
\end{equation}
where $S_{N}=S+\Delta S_N$. Because $S$ is a sum of traces, $\delta S[\phi]=0$. The variation of the source term is noting but :
\begin{align}
\delta (J\cdot \phi)= J\cdot \delta\phi&= \sum_{a,b,c} (J_{ab} \epsilon_{ac}\phi_{cb} -J_{ab} \phi_{ac}\epsilon_{cb} )=\sum_{a,b,c} (J_{ab} \phi_{cb} -J_{bc} \phi_{ba})\epsilon_{ac}\,.
\end{align}
The variation of the regulation term can be deduced from the same way:
\begin{align}
\delta \Delta S_N[\phi]= \sum_{a,b,c,d}\left[\delta\phi_{ab} [r_N(a,b) ]_{ab;cd} \phi_{cd} \right]\,,
\end{align}
where we assumed that $[r_N(a,b) ]_{ab;cd}=[r_N(a,b) ]_{cd;ab}$. This is exactly the same computation as for the source term, up to the replacement $J_{ab} \to \sum_{c,d}\,[r_N(a,b) ]_{ab;cd} \phi_{cd} $, leading to:
\begin{align}
\delta\Delta S_N[\phi]=\sum_{a,b,c,d,e} \phi_{de}\bigg[r_N(a,b) ]_{ab;de} \phi_{cb} -[r_N(c,b) ]_{bc;de} \phi_{ba}\bigg] \epsilon_{ac}\,.
\end{align}
Where moreover we assumed that $r_N(a,b)$ is a symmetric function with respect to $a$ and $b$. Due to the translation invariance of the Lebesgue measure, $\mathcal{Z}_N$ must be invariant up to a global reparametrization of the fields, therefore the variation of the left hand side in \eqref{var} must be vanish $\delta \mathcal{Z}_N=0$. From the identity:
\begin{equation}
\int d\phi\, \phi_{ab} e^{-S_N[\phi,J]} = \int d\phi\, \frac{\partial}{\partial J_{ab}} e^{-S_N[\phi,J]}
\end{equation}
We finally deduce the following statement:
\begin{theorem}
\textbf{Ward-Takahashi identity.} In the symmetric phase, and along the path $N=\Lambda$ to $N=0$, the following relation holds:
\begin{align}
\bigg\{ \frac{\partial}{\partial J_{de}}\bigg([r_N(a,b) ]_{ab;de} \frac{\partial}{\partial J_{cb}}-[r_N(c,b) ]_{bc;de} \frac{\partial}{\partial J_{ba}}\bigg)
- \left(J_{ab} \frac{\partial}{\partial J_{cb}} -J_{bc} \frac{\partial}{\partial J_{ba}}\right)\bigg\} e^{\mathcal{W}_N[J]}=0\,.\label{Ward}
\end{align}
where we adopted the Einstein convention for repeated indices. Note thhat there are no summation over indices $a$ and $c$.
\end{theorem}

\subsection{Influence of derivative couplings}

As we explained in section \ref{sec2}, there is no preferred notion of scale for the initial model (i.e. for the model without regulator). More precisely, for a statistical field theory formalism provides canonical notions of ‘‘what is deep UV" and ‘‘what is deep IR": the deep UV being related to the classical action $\mathcal{S}(\phi)$, without integration over statistical fluctuations and the deep IR scale, related to the effective action $\Gamma[\Phi]$, when all fluctuations are integrated out. However, for matrix models, there is no canonical way to link the deep IR region from the deep UV one. This difficulty arises because of the flatness of the kinetic spectrum. Indeed, the propagator having equals eigenvalues (all equals to $1$), all the fluctuations play the same roles, and indistinguishably ‘‘UV" or ‘‘IR". In contrast, for standard field theories, this is the spectrum of the kinetic operator which provides a canonical path from UV to IR, allowing to classify of the fluctuations following their respective energy. But for matrix models, due to the $U(\Lambda)$ invariance, all the eigenvalues of the kinetic operator are the same, and the fluctuations become indistinguishable. This highlights the role of the regulator. The regulator that we introduced broke the global $U(\Lambda)$ -- invariance at the kinetic level, providing a preferred path from UV to IR and an ordering for partial integrations over quantum fluctuations. \\

\noindent
The Ward identity derived in the previous subsection is a consequence of this symmetry breaking. It arises from the nontrivial variation of the kinetic term under infinitesimal unitary transformation, like the flow equations arise from a nontrivial variation of the kinetic term under a change of the running scale $N$. Both are consequences of the symmetry breaking and have to be treated on the same footing, as nontrivial relations between effective vertices $\Gamma^{(n+2)}$ and $\Gamma^{(n)}$. More precisely, and as it will be clearer in the rest of this section, one can say that the RG equation dictates how to move through increasing scales (from large to small $N$) whereas the Ward identity dictates how to move in the momentum space. As we will see, because of the symmetry breaking, non-local derivative-like interactions such that \eqref{nonlocal} appear even in the strictly local sector, and play an important role in the behavior of the RG flow, especially around the relevant UV fixed point. \\

\textit{i.) Explicit Ward identities. } Equation \eqref{Ward}, as the exact flow equation \eqref{Wett} is a functional relation which can be translated as an infinite hierarchical system, taking successive derivative with respect to external sources and setting $J=0$. For instance, Taking the derivative with respect to $\partial^2/\partial J_{de}\partial J_{d^\prime e^\prime}$ of the Ward identity \eqref{Ward} and setting $J=0$, we get:
\begin{align}
\nonumber Z_N\,G^{(4)}_{N, de,d^\prime e^\prime, ba, cb} \, [f(a/N,b/N)-f(c/N,b/N)]&=
\delta_{da}\delta_{eb} G^{(2)}_{N,d^\prime e^\prime,cb}+\delta_{d^\prime a}\delta_{e^\prime b} G^{(2)}_{N,de,cb}\\
&-\delta_{db}\delta_{ec} G^{(2)}_{N, d^\prime e^\prime,ba}-\delta_{d^\prime b}\delta_{e^\prime c} G^{(2)}_{N, de,ba}\,. \label{equation92}
\end{align}
Setting $d=a$, $d^\prime=e^\prime=e=c$ and $a\neq c$, and after some elementary algebraic manipulations, it is not hard to deduce the following statement, holding in the large $N$ limit:
\begin{lemma}\label{lemma1}
In the large $N$ limit, the $4$ and $2$-point functions are related by the non-trivial relation :
\begin{align}
\nonumber 4Z_N \gamma^{(4)}_{ac,cc,bc,ba} &g_{bc}^{(2)}g_{ba}^{(2)}\left[f\left(\frac{a}{N},\frac{b}{N}\right)-f\left(\frac{c}{N},\frac{b}{N}\right)\right]\\
&-Z_N\left[f\left(\frac{a}{N},\frac{c}{N}\right)-f\left(\frac{c}{N},\frac{c}{N}\right)\right]=(g_{cc}^{(2)})^{-1}-(g_{ac}^{(2)})^{-1}\,.
\end{align}
where the quantity $g_{ab}$ is defined in the relation \eqref{eqG}.
\end{lemma}
To derive this statement, we assumed that $4$-point effective vertices are built as a single trace, exactly as in the local potential approximation used to solve the flow equations in the previous section. This assumption reflects the aim to solve both Ward identities and flow equations from a single approximation scheme, treating them on the same footing. Moreover, we specify some key definitions:\\

\noindent
$\bullet$ $ \gamma^{(4)}_{A_1,A_2,A_3,A_4}$, for $A_i:=(a_i,b_i)$ designate the effective vertex with \textit{fixed index position}, with the global trace condition:
\begin{equation}
\gamma^{(4)}_{A_1,A_2,A_3,A_4} \propto \prod_i \delta_{b_ia_{i+1}}\,.
\end{equation}
and is related to the effective vertex function $\Gamma^{(4)}_{A_1,A_2,A_3,A_4}$ as:
\begin{equation}
\Gamma_{A_1,A_2,A_3,A_4}^{(4)}:= \sum_{\pi} \gamma^{(4)}_{A_{\pi(1)},A_{\pi(2)},A_{\pi(3)}, A_{\pi(4)}}\,,
\end{equation}
the sum running over the $4!$ permutations of the boundary variables $A_i$. \\

\noindent
$\bullet$ Assuming to be into the symmetric phase, the momenta conservation along the external faces ensure that $G^{(2)}_{ab,cd}\propto \delta_{ad}\delta_{bc}$, and we define $g^{(2)}_{ab}$ as the proportionality coefficient:
\begin{equation}
G^{(2)}_{ab,cd}=: g^{(2)}_{ab} \,\delta_{ad}\delta_{bc}\,.
\end{equation}

\noindent
$\bullet$ In the same notations, the $4$-point function $G^{(4)}_N$ must be decomposed into connected components as:
\begin{align}
\nonumber &G^{(4)}_{N, A_1A_2A_3A_4}=G^{(4,c)}_{N, A_1A_2A_3A_4}+\bigg(G^{(2)}_{N,A_1A_2}G^{(2)}_{N,A_3A_4} +G^{(2)}_{N,A_1A_3}G^{(2)}_{N,A_2A_4}+G^{(2)}_{N,A_1A_4}G^{(2)}_{N,A_3A_2} \bigg)\,. \label{G41}
\end{align}
the $4$-point connected component $G^{(4,c)}_{N, A_1A_2A_3A_4}$ being related to the $4$-point effective vertex as:
\begin{equation}
G^{(4,c)}_{N, A_1A_2A_3A_4}=: -\left( \prod_{i=1}^4 g^{(2)}_{A_i}\right) \Gamma^{(4)}_{N,A_1,A_2,A_3,A_4}\,.
\end{equation}
In the large $N$ limit, this relation can be turned as an integro-differential equation. Indeed, setting $a=c+1$, the finite difference between the functions $f$ may be approximated as:
\begin{equation}
f\left(\frac{a}{N},\frac{b}{N}\right)-f\left(\frac{c}{N},\frac{b}{N}\right)\approx \frac{1}{N} \frac{d}{dx} f\left(x,\frac{b}{N}\right)\bigg\vert_{x=\frac{c}{N}}\,,\label{finitediff}
\end{equation}
which can be computed explicitly for the Litim regulator, we get:
\begin{equation}
\frac{d}{dx} f\left(x,\frac{b}{N}\right)\bigg\vert_{x=\frac{c}{N}}= -2\left(\frac{N}{c+b}\right)^2 \,\Theta\left(1-\frac{c+b}{2N}\right)\,.\label{equ101}
\end{equation}
In the same way, assuming that $g^{(2)}_{ab}$ can be analytically extended by the function $g^{(2)}(x,y)$ with the continuous variables $x,y:=a/N,b/N$, we get:
\begin{equation}
(g_{ac}^{(2)})^{-1}-(g_{cc}^{(2)})^{-1}=\frac{1}{N}\,\frac{d}{dx} g\left(x,\frac{c}{N}\right)\bigg\vert_{x=\frac{c}{N}}+\mathcal{O}\left(\frac{1}{N^2}\right)\label{eq102}
\end{equation}
From equation \eqref{equ101}, it is clear that the windows of momenta allowed in the sum over $b$ from $df/dx:=f^\prime$ are the same as the one allowed by $\dot{r}_N$ in the flow equation \eqref{Wett}. Therefore, the same approximations used to solve the RG equations may be used for $g^{(2)}_{bc}$, $g^{(2)}_{ba}$ and $\gamma^{(4)}_{ac,cc,bc,ba}$. The same situation has been observed for tensor field theory \cite{Lahoche:2019vzy}, for several choices of regulator function. Then, one expects that this behavior is not due to the Litim regulator, but a general feature that the allowed windows of momenta for $\dot{r}_N$ covers the one of $f^\prime$. Moreover, equation \eqref{eq102} point out the existence of a strong relation between $4$ point functions and the momenta variations of the $2$-point functions along the path from the deep UV sector to the IR sector. Therefore, and as we will see explicitly, even in the large $N$ limit, non-local interactions such that \eqref{nonlocal} survives at the leading order in $1/N$ and cannot be discarded from any relevant parameterization of the phase space. This argument shows that strictly local potential approximations have to be enlarged with derivative-like interaction to become compatible with Ward identity. As a first improvement, we can consider the following minimal enlargement :
\begin{equation}
\Gamma[\Phi]=\gamma\,\sum_{a,b} \frac{a+b}{2N} \Phi_{ab}\Phi_{ba} +U_N[\Phi]\,, \label{improvedLPA}
\end{equation}
where $U_N[\Phi]$ expands as a single trace like in equation \eqref{potential}. We call improved LPA this parametrization allowing a small deviation from the crude LPA. From this approximation,
\begin{equation}
(g_{ab}^{(2)})^{-1}=Z_N+\gamma\,\frac{a+b}{N}+\mathcal{O}\left(a^2,b^2\right)+Z_Nf\left(\frac{a}{N},\frac{b}{N} \right) \,, \label{derivativeexp}
\end{equation}
and
\begin{equation}
\gamma^{(4)}_{ac,cc,bc,ba}\to\frac{g_4}{4}=(Z_N)^2N^{-1} \frac{u_4}{4}\,.
\end{equation}
Inserting these relations into the lemma \ref{lemma1}, we see that the second term on the left-hand side is exactly compensated with the same term on the right-hand side. Then, setting $c=0$, we get the following statement:
\begin{proposition} \label{propward1}
Up to $1/N$ corrections, and in the improved local potential approximation, the $2$ point derivative coupling $\gamma$ and the local $4$ point coupling $u_4$ are related as:
\begin{equation}
u_4 \,\bar{\mathcal{L}}_N = -\bar{\gamma}\,, \label{firstWard}
\end{equation}
where $Z_N\bar{\gamma}=:\gamma$ and ${\mathcal{L}}_N=(Z_N)^{-2}N\bar{\mathcal{L}}_N$
\begin{equation}
\mathcal{L}_N:= \sum_b \,(g_{bc}^{(2)})^2 f^\prime (c,b) \big\vert_{c=0}\,.
\end{equation}
\end{proposition}
With the truncation \eqref{improvedLPA}, $\bar{\mathcal{L}}_N$ depends only on $\bar{\gamma}$. Therefore, deriving equation \eqref{firstWard} with respect to $t$ leads to
\begin{equation}
-\beta_4 \bar{\gamma} +u_4(1+u_4 \bar{\mathcal{L}}^\prime_N)\dot{\bar{\gamma}}=0\,, \label{Wardconst}
\end{equation}
This equation that we call \textit{Ward constraint} rely on two beta functions along with the history of the RG flow since $N$ remain large. As an important consequence:
\begin{corollary}
In the large $N$ limit, any fixed point of the flow equations satisfy the Ward constraint \eqref{firstWard}.
\end{corollary}
The flow of the non local kinetic coupling $\gamma$ receives two kind of contributions. A first contribution arise from the derivative with respect to one external momentum of the loop integrals, but a direct computation show that these variations vanish identically (as we will see explicitly in the next subsection). A second contribution arises from the derivative of the effective vertex themselves. In the local potential approximation, the vertex does not depends on the external momenta. But from the Ward identity, it follows that the ultralocal information determine completely the first derivative with respect to the external momenta, like ultralocal $4$-point coupling $u_4$ determine $\gamma$ in lemma \ref{lemma1}. In order to obtain the first derivative of the $4$-point function, we need to derive the Ward identity involving $6$-point functions (i.e. provided from \eqref{Ward} by deriving four time with respect to the source $J$). Then this new Ward identity can be given by introducing the function $f_{a_1,a_2,a_3,a_4}$ as:
\begin{equation}
\gamma^{(4)}_{A_{\pi(1)},A_{\pi(2)},A_{\pi(3)}, A_{\pi(4)}}=:f_{a_1,a_2,a_3,a_4}\delta_{b_1a_2}\delta_{b_2a_3}\delta_{b_3a_4}\delta_{b_4a_1}\,. \label{def99}
\end{equation}
and the continuous function $\tilde{f}(x,c/N)$ for the continuous variable $x=a/N$, such that $\tilde{f}(a/N,c/N)\equiv f_{c,c,a,c}$; we get, after few computation:
\begin{proposition}
Up to $1/N$ corrections, and in the improved local potential approximation, the $4$-point derivative coupling and the local $4$ and $6$-point renormalized couplings $u_4$ and $u_6$ are related as:
\begin{align}
2( \,u_6\bar{\mathcal{L}}_N - u_4^2\, \bar{\mathcal{U}}_N)+\,\Xi=0\,. \label{cc}
\end{align}
where we defined:
\begin{equation}\label{ftilde}
\frac{d\tilde{f}}{dx}\left(x,0\right)=: 4(Z_N)^2 N^{-1}\Xi\,,
\end{equation}
and:
\begin{equation}
\mathcal{U}_N:=\sum_b \,(g_{bc}^{(2)})^3 f^\prime (c,b) \big\vert_{c=0} \,,\quad \mathcal{U}_N=:(Z_N)^{-3} N \bar{ \mathcal{U}}_N\,.
\end{equation}
\end{proposition}
We will explore the consequences of these relations in the next section. At this stage, one can stress that the ultralocal truncation, with $\gamma=\Xi=0$, violate strongly the Ward identities. \\

\textit{ii.) Ward identity violation and derivative couplings.} As a first improvement of the LPA, we may attempt to include the derivative coupling $\gamma$ in the truncated action; following the parametrization \eqref{improvedLPA}. In this parameterization,
a first change concerns the effective propagator \eqref{eqG}, which becomes:
\begin{align}
(G_N)_{ab,cd}= Z_N^{-1} g_{ba,cd}\, \frac{a+b}{2N} \left( \frac{1}{1+2\bar{\gamma} \left(\frac{a+b}{2N}\right)^2}\right)\,, \label{eqG2}
\end{align}
such that the integral $I_{a}^{(p)}$ as the equation \eqref{Ip} becomes:
\begin{equation}
I_{a}^{(p)}\approx 2Z_N^{1-p} N \int_{a/2N}^{1} dx\,x^{p-1} \, \frac{1+\eta_N(1-x)}{\left(1+2\bar{\gamma} x^2\right)^p}\,. \label{Ip2}
\end{equation}
To simplify our discussion, we introduce the sequence $\iota_{p,q}(y)$ for the continuous variable $y=a/2N$ such that:
\begin{equation}
\iota_{p,q}(y):=\int_y^1 dx \frac{x^q}{\left(1+2\bar{\gamma} x^2\right)^p}\,,
\end{equation}
and:
\begin{equation}
I_{a}^{(p)}\approx 2Z_N^{1-p} N\left[\iota_{p,p-1}(y)+\eta_N\left(\iota_{p,p-1}(y)-\iota_{p,p}(y)\right) \right]\,.
\end{equation}
In addition we defined the renormalized loop $\bar{I}_{a}^{(p)}:=Z_N^{p-1} N^{-1} I_{a}^{(p)}$.
Equation \eqref{eqZ} is then transformed as:
\begin{equation}
\dot{Z}_N=-\frac{4Ng_{4,N}}{Z_N}\left[\iota_{2,1}(0)+\eta_N\left(\iota_{2,1}(0)-\iota_{2,2}(0)\right) \right]\,,\label{eqZ2}
\end{equation}
solved as:
\begin{equation}\label{122}
\eta_N=-\frac{4u_4\,\iota_{2,1}}{1+4u_4\left(\iota_{2,1}-\iota_{2,2}\right) }\,,
\end{equation}
where we used of the concise notation $\iota_{p,q}\equiv \iota_{p,q}(0)$. In the same way, we get the function $\beta_4$, which is a new expression as replacement of \eqref{eqbeta4}:
\begin{align}
\beta_4=(1-2\eta) u_{4}& + 8 u_{4}^2\left[ \iota_{3,2} +\eta\left(\iota_{3,2}-\iota_{3,3}\right) \right]-8u_{6}\left[ \iota_{2,1} +\eta\left(\iota_{2,1}-\iota_{2,2}\right) \right]\,. \label{eqbeta42}
\end{align}
The flow equation for $\gamma$ can be deduced from \eqref{equationwettexp1}, like $\eta_N$. From definition:
\begin{equation}
\frac{\gamma}{N}\equiv \frac{d}{da} \Gamma^{(2)}_{ab,ba}\big\vert_{a=b=0} \,,
\end{equation}
we get ($\beta_\gamma\equiv \dot{\bar{\gamma}}$):
\begin{align}
\nonumber \beta_\gamma=-\eta\, \bar{\gamma} -4 u_4& \left[\iota_{2,1}^\prime+\eta\left(\iota_{2,1}^\prime-\iota_{2,2}^\prime \right) \right] -6 \Xi \left[\iota_{2,1}+\eta\left(\iota_{2,1}-\iota_{2,2}\right) \right]\,.
\end{align}
It is easy to check that the involves derivatives $\iota_{p,q}^\prime\equiv \iota_{p,q}^\prime(0)$ vanish identically
\begin{equation}
\iota_{p,q}^\prime= -\frac{x^q}{\left(1+2\bar{\gamma} x^2\right)^p} \bigg\vert_{x=0}=0\,,\quad \forall q\neq 0\,,
\end{equation}
such that the equation for $\dot{\bar{\gamma}}$ reduces to:
\begin{align}
\nonumber \beta_\gamma&=-\eta\, \bar{\gamma} -6 \Xi \left[\iota_{2,1}+\eta\left(\iota_{2,1}-\iota_{2,2}\right) \right] \\
&=-\eta\, \bar{\gamma} -12 \left(u_6 \iota_{2,0}-u_4^2\iota_{3,1}\right) \left[\iota_{2,1}+\eta\left(\iota_{2,1}-\iota_{2,2}\right) \right]\,,
\label{eqgamma1}
\end{align}
where $\bar{\mathcal{L}}_N$ and $\bar{\mathcal{U}}_N$ may be expressed in terms of the sequences $\iota_{p,q}$,
\begin{equation}
\bar{\mathcal{L}}_N:= - \iota_{2,0}\,,\qquad \bar{\mathcal{U}}_N:=- \iota_{3,1}\,. \label{LN}
\end{equation}
Note that the origin of the factor $6$ in front of $\Xi$ in equations \eqref{eqgamma1} counts the different localizations for the derivative $\tilde{f}^\prime$ (see equation \eqref{ftilde}) on the vertex itself, as pictured on Figure \ref{confXi} below. \\
\begin{figure}
\begin{center}
\includegraphics[scale=0.7]{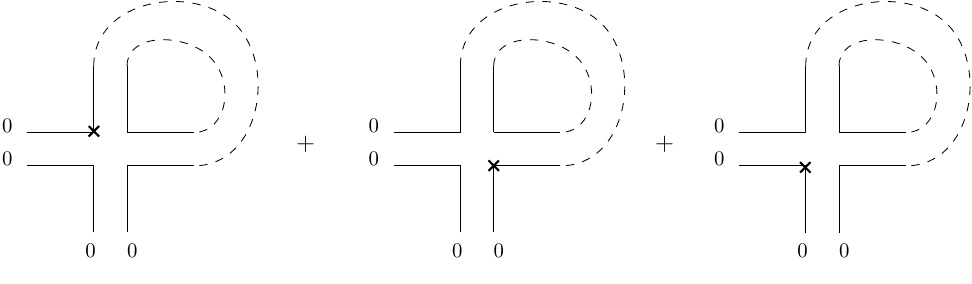}
\end{center}
\caption{The three contributions to the derivative of the effective vertex, the cross means location of the derivative $\tilde{f}^\prime$.} \label{confXi}
\end{figure}
Another expression for $\dot{\gamma}$ comes from the Ward identity, equation \eqref{Wardconst}, namely
\begin{equation}
\beta_\gamma=-\frac{ \bar{\mathcal{L}}_N}{1+u_4 \bar{\mathcal{L}}^\prime_N} \beta_4\,.
\end{equation}
Obviously $\iota^\prime_{2,0}=-4\iota_{3,2}$ (where the {\it prime} means the derivative with respect to $\bar\gamma$), so that the equation for $\dot{\gamma}$ reduces to:
\begin{equation}
\beta_\gamma=\frac{\iota_{2,0}}{1+4u_4 \iota_{3,2}} \beta_4\,. \label{WC}
\end{equation}
From the expression of $\beta_4$ given by equation \eqref{eqbeta42}, and taking into account equation \eqref{eqgamma1}, we can deduce $u_6$ in terms of $u_4$ and $\bar{\gamma}$ dynamically along the RG flow, $u_6=f(u_4,\bar{\gamma})$, with
\begin{align*}
&f(u_4,\bar{\gamma}) =-\frac{(1-\eta)\bar{\gamma}+2u_4^2\left(2\iota_{2,0} \bar{I}_0^{(3)}-3\iota_{3,1} \bar{I}_0^{(2)}(1+4u_4\iota_{3,2})\right)}{2\iota_{2,0} \bar{I}_0^{(2)}[1+12u_4\iota_{3,2}]}\,,
\end{align*}
which, from the Ward constraint \eqref{propward1} can be translated as a function on $\bar{\gamma}$ only. At this level of approximation, the problem is then completely closed. The two parameters $\bar{\gamma}$ and $u_4$ fix $u_6$, which fix $u_8$ and so one. This conclusion highlight two points. Firstly, the role played by the derivative couplings, secondly that the improved local potential parametrization \eqref{improvedLPA}, which involve an infinite number of couplings can be, in fact, reduced to a one dimensional manifold. Obviously, enlarging the theory space with more derivative and/or disconnected interactions, we lost this property. Moreover, note that we neglected the flow of the derivative coupling $\dot{\Xi}\approx 0$. \\

We now move on the essential motivation to build and improve the method discussed in the previous paragraph i.e. the investigation of the global fixed point solutions of the flow equations. Our strategy is the following. Setting $\beta_4=0$, from the linearity of this equation respect to the couplings, we fix $u_6$ \textit{uniquely} in terms of $u_4$ and $\bar{\gamma}$ through a relation of the form $u_6=F(u_4,\bar{\gamma})$. Moreover, from the first Ward identity given by proposition \eqref{propward1}, $u_4$ and $\bar{\gamma}$ are not independent, $u_4= \bar{\gamma}/\iota_{0,2}$, therefore :
\begin{equation}
u_6=F\left(\frac{\bar{\gamma}}{\iota_{0,2}},\bar{\gamma}\right)\,.
\end{equation}
Explicitly:
\begin{equation}
F(u_4,\bar{\gamma})=\frac{(1-2\eta) u_{4}+ 8 u_{4}^2\left[ \iota_{3,2} +\eta\left(\iota_{3,2}-\iota_{3,3}\right) \right]}{8\left[ \iota_{2,1} +\eta\left(\iota_{2,1}-\iota_{2,2}\right) \right]}\,.
\end{equation}
Inserting these relation into the equation \eqref{eqgamma1}, and setting $\dot{\bar{\gamma}}=0$, we deduce:
\begin{proposition} \label{fixedpointequation}
In the large $N$ limit, all the fixed points of the improved LPA have to be solution of the following fixed point equation:
\begin{align*}
0=\beta_\gamma\equiv \eta\, \bar{\gamma} +&6 \bigg(F\left(\frac{\bar{\gamma}}{\iota_{2,0}},\bar{\gamma}\right) \iota_{2,0}-\left(\frac{\bar{\gamma}}{\iota_{2,0}}\right)^2\iota_{3,1}\bigg) \bar{I}_0^{(2)}\,.
\end{align*}
\end{proposition}
This equation can be solved numerically. One may expect that the dynamical definition of $u_6$, $u_6=f(u_4,\bar{\gamma})$ breaks down at the fixed point, because both $beta_\gamma$ and $\beta_4$ vanish at this point. It is no hard however to show the following lemma:
\begin{lemma}
The effective RG flow, described by the function $f(u_4,\bar{\gamma})$ satisfy:
\begin{equation}
f(u_4,\bar{\gamma})\big\vert_{\bar{\gamma}^*}=F(u_4,\bar{\gamma})\vert_{\bar{\gamma}^*}\,,
\end{equation}
ensuring continuity at the fixed point.
\end{lemma}

\noindent
\textit{Proof.} The proof is elementary. Let us rewrite our set of flow equations as:
\begin{align}
\beta_\gamma&=-\eta \bar{\gamma}-6(a_\gamma u_6-b_\gamma u_4^2) \\
\beta_4&=(1-2\eta) u_4+b_4 u_4^2-a_4 u_6\,,
\end{align}
and the relation between them coming from Ward identity as $\beta_\gamma = A \beta_4$. Using the last one, we get the explicit expression for $f$:
\begin{equation}
u_6^\sharp= \frac{(1-2\eta) Au_4+\eta \gamma+ (b_4A-6b_\gamma)u_4^2)}{a_4 A-6a_\gamma}\,.
\end{equation}
Moreover, setting $\beta_\gamma=0$ on one hand, and $\beta_4=0$ on the second hand; we get respectively the two solutions:
\begin{align}
u_6^*&=\frac{(1-2\eta)u_4+b_4 u_4^2 }{a_4}\,,\\
u_6^{**}&=\frac{-\eta \bar{\gamma}+6b_\gamma u_4^2}{6a_\gamma}\,.
\end{align}
Inserting these two solutions into the expression of the dynamical coupling $u_6$, we get:
\begin{equation}
u_6^\sharp = \frac{A a_4 u_6^*-6a_\gamma u^{**}_6}{A a_4-6a_\gamma}\,.
\end{equation}
For a global fixed point $u_6^*=u_6^{**}$. Therefore, without singularity of the involved coefficients, we get $u_6^{\sharp}\equiv u_6^*$.
\begin{flushright}
$\square$
\end{flushright}

\begin{figure}
\begin{center}
\includegraphics[scale=0.7]{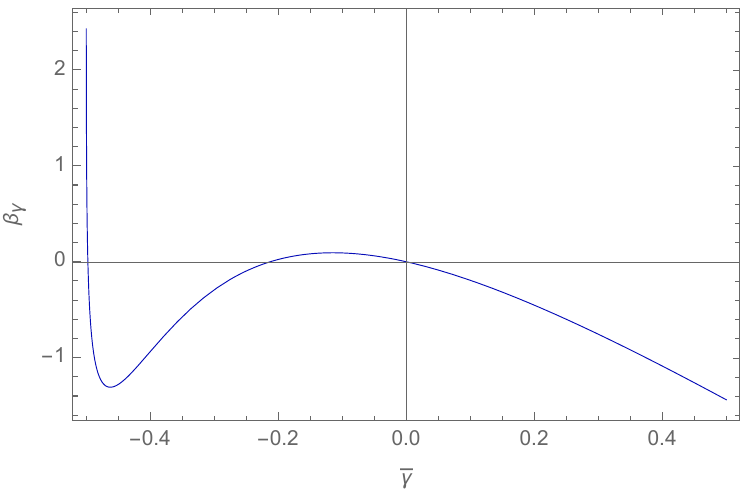}
\end{center}
\caption{Numerical plot of the beta function $\beta_\gamma$ expressed in term of $\bar{\gamma}$ only. The function has three nodes. The first one for $\bar{\gamma}=0$ corresponds to the Gaussian fixed point, and the two other ones, for $\bar{\gamma}\approx -0.21$ and $\bar{\gamma}\approx -0.49$, correspond to non-trivial interacting fixed points. } \label{figplotbetagamma}
\end{figure}

\noindent
The numerical plot of the $\beta$-function $\beta_\gamma$ is provided on Figure \ref{figplotbetagamma}, showing three zeros. The first one, for $\bar{\gamma}=0$ corresponds to the Gaussian fixed point $u_4=0$.
The others, for $\bar{\gamma}\approx -0.21$ and $\bar{\gamma}\approx -0.49$. The second one is UV-attractive, and seems to be in qualitative agreement with the UV attractive fixed point relevant for double scaling limit. Computing $u_4^*$ from Ward identity, we get:
\begin{equation}
u_4^*=\frac{\bar{\gamma}^*}{\iota_{2,0}(\bar{\gamma}^*)} \approx -0.005\,,
\end{equation}
and for the critical exponent:
\begin{equation*}
-\theta= \frac{\partial \beta_4}{\partial u_4}\bigg\vert_{\bar{\gamma}^*}+\frac{\partial \beta_4}{\partial u_6}\bigg\vert_{\bar{\gamma}^*} \frac{\partial f}{\partial \bar{\gamma}} \bigg\vert_{\bar{\gamma}^*} \frac{\partial \bar{\gamma}}{\partial u_4}\bigg\vert_{\bar{\gamma}^*}+ \frac{\partial \beta_4}{\partial \bar{\gamma}}\bigg\vert_{\bar{\gamma}^*} \frac{\partial \bar{\gamma}}{\partial u_4}\bigg\vert_{\bar{\gamma}^*}\,,
\end{equation*}
explicitly: $\theta\approx 0.97$. We then discover a relevant direction, with value less than $1$, realizing an improvement of the truncation. This result seems to indicate that our method, taking into account the Ward identity could be to go beyond the truncation wall $\theta \approx 1.06$. However, there are two conceptual difficulties with this approach. First, the coupling $\bar{\gamma}$ becomes very large at the considered fixed point (in comparison to $u_4^*$), seeming to indicate a spurious dependence on the regulator function, which is the origin of the derivative couplings, in contrast with standard assumptions neglecting the role of these regulator-dependent couplings. Second, we considered only the minimal crude truncation in the space of derivative couplings, showing the instability of the ultra-local sector due to the Ward identity. But there are no reason to stop the derivative expansion, and an infinite list of derivative couplings have to be contribute on the left hand side of the Ward identities, so that equations \eqref{firstWard} can be viewed for instance as the minimal truncation of a complete equation, involving an infinite set of couplings. \\

\textit{iii.) Fine-tuned regulator.} This last point aims to construct a well defined RG flow, following Ward identities and for which the influence of derivative couplings remains as small as possible. The Ward identities, like \eqref{firstWard} and \eqref{cc}, highlight the role played by the regulator in the emergence of the derivative couplings. We can stress a parallel between flow evolution and divergence of the flow toward the derivative sector. In both cases, this is the variation of the propagator concerning $N$ or $a/N\equiv x$ which generates the moving into the theory space, in ‘‘scales" or ‘‘momenta" directions respectively. Let us recall that: as the Wetterich equation describes the $N$ variation of $\Gamma_N$, the Ward identity describes the momentum dependence i.e. the $a/N$ variation of the same quantity. Moreover, the two transformations are not generally independents. For a regulator of the form:
\begin{equation}
r_N(x,y)=Z_N f(x,y)\,,
\end{equation}
we get:
\begin{equation}
\dot{r}_N(x,y)=\eta_N\,r_N(x,y)-Z_N\left(x\frac{\partial f}{\partial x}+y\frac{\partial f}{\partial y}\right)\,. \label{derivative}
\end{equation}
The first term is intrinsically associated with the RG flow; however, the second part involves derivatives concerning the momenta, which are the generators of the momentum displacements in Ward identities.

\begin{figure}[h!]
\begin{center}
\includegraphics[scale=0.5]{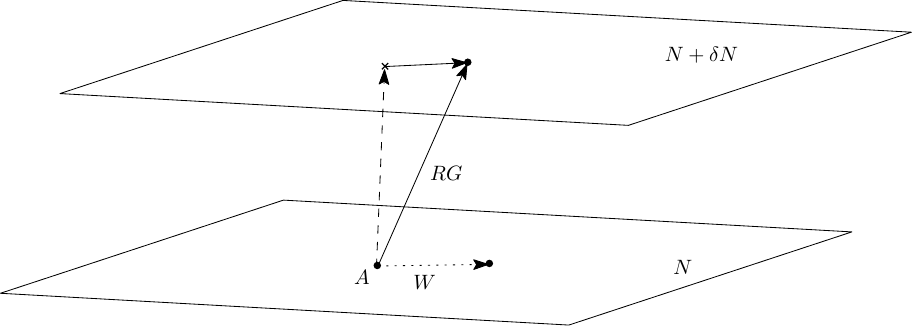}
\end{center}
\caption{Qualitative behavior of the RG map and Ward operator (W) into the full theory space. Starting from a point in the theory space (A), the Ward operator (the dotted arrow) generate horizontal moves at fixed $N$, whereas the RG map has both vertical and horizontal components.}\label{rep}
\end{figure}

Heuristically, one may picture the global dynamics as follows (see Figure \ref{rep}). Starting from a point at scale $N$ in the full theory space, the Ward operator allows moving horizontally, at fixed $N$ toward the derivative interactions world. In the same way, the RG map allows it to move vertically, from the scale $N$ to the scale $N+\delta N$, but due to the second terms of the right-hand side of \eqref{derivative}, the RG transformation generates as well a horizontal displacement. This is another way to understand the instability of the local phase space, pointed out in the previous section.
Therefore, and despite the accordance of our results with the expected ones, especially about the value of the critical exponent, and the apparent qualitatively small dependence on the regulator in a small range of values around $\alpha=1$; one cannot conclude that our results have anything to do with the original model, the explored region of the theory space being very far from one of the original ultralocal ones. \\

From the last picture, a question remains open. Can you build an RG map that is the most vertical as possible, at least for $N$ sufficiently large, in such a way that $\mathcal{L}_N$, $\mathcal{U}_N$ and their higher momenta remain small enough, such that derivative couplings can be discarded from the RG flow? This question seems to be very difficult regarding the complex hierarchical structure of the flow equations that we discussed in this paper. A heuristic attempt to solve this problem, or at least to build a flow that remains vertical for a long time is to choose a regulator such that $\mathcal{L}_N$, $\mathcal{U}_N$ vanish or become small for vanishing $\bar{\gamma}$. This can be achieved for instance with a regulator of the form:
\begin{equation}
f\left(\frac{a}{N},\frac{b}{N}\right):= \left(\frac{2N}{a+b}-1 \right)\Theta\left(\alpha-\frac{a+b}{2N} \right)\,,\label{regulnew}
\end{equation}
where the parameters $\alpha$ have to be fine-tuned such that $\mathcal{L}_N$ vanish and, $u_4^2\mathcal{U}_N$ becomes small for $\bar{\gamma}=0$. By solving $\mathcal{L}_N=0$ we get the solution $\alpha=2$. It is easy to check that this regulator satisfies the four requirements enumerated above equation \eqref{regulator}. The corresponding flow equations can be easily deduced from our previous analysis. The condition $\mathcal{L}_N\vert_{\gamma=0}=0$ and $|\mathcal{U}_N\vert_{\gamma=0}|\approx 1$ allows to keep $\Xi=\bar{\gamma}=0$ with a very good approximation along with the flow for a long time, in regards to the rapidity of the convergence of the truncated expansions. Setting $\bar{\gamma}=0$, the flow equations become the following:
\begin{proposition}
In the large $N$ limit and for the fine-tuned regulator \eqref{regulnew}, the most vertical truncated flow equations in the LPA, up to $\Phi^{10}$-interactions, write as:
\begin{align}
\nonumber \beta_4=(1-2\eta)u_4+8u_4^2[\iota^{(1)}_3 \eta+\iota^{(2)}_3+\partial \iota_3]-8u_6[\iota^{(1)}_2 \eta+\iota^{(2)}_2+\partial \iota_2]\,,
\end{align}
\begin{align}
\nonumber \beta_6&=(2-3\eta)u_6+24 u_6u_4[\iota^{(1)}_3 \eta+\iota^{(2)}_3+\partial \iota_3]-12 u_4^3[\iota^{(1)}_4 \eta+\iota^{(2)}_4+\partial \iota_4]-12 u_8 [\iota^{(1)}_2 \eta+\iota^{(2)}_2+\partial \iota_2]\,,
\end{align}
\begin{align*}
\beta_8=&(3-4\eta)u_8+16 u_4^4[\iota^{(1)}_5 \eta+\iota^{(2)}_5+\partial \iota_5]+16 u_6^2 [\iota^{(1)}_3 \eta+\iota^{(2)}_3+\partial \iota_3]-48 u_6 u_4^2 [\iota^{(1)}_4 \eta+\iota^{(2)}_4+\partial \iota_4]\\
&+32 u_8u_4 [\iota^{(1)}_3 \eta+\iota^{(2)}_3+\partial \iota_3]\,,
\end{align*}
where we used of the definitions:
\begin{equation*}
\eta:=-4u_4\frac{\iota^{(2)}_2+\partial\iota_2}{1+4u_4 \iota^{(1)}_2}
\end{equation*}
and:
\begin{equation}
\iota^{(1)}_p:=-\frac{2^{p+1}p}{2+3p+p^2}\,,
\quad
\iota^{(2)}_p:=\frac{2^p}{p}\,,
\quad
\partial \iota_p:=\alpha^{p}(1-\alpha) \,.
\end{equation}
\end{proposition}

\begin{figure}[h!]
\begin{tabular}{|c|c|c|c|c|}
\hline
truncation order $k$ & $\theta_1$ & $\theta_2$ & $\theta_3$ &anomalous dimension $\eta$\\
&&&&
\\
\hline
6&1.04&--&--&0.14\\
\hline
8 & 1.02&-1.36&--&0.08\\
\hline
10&0.98&-1.29&-2.18&0.04\\
\hline
\end{tabular}
\caption{Numerical results for vertical truncations from $k=6$ to $k=10$. We see that increasing the number of interactions does not change the value of the positive critical exponents, the other one corresponding to irrelevant directions. Moreover, the anomalous dimension is very small in comparison to truncation with the standard Litim regulator. }\label{table2}
\end{figure}

Investigating numerically the successive truncations, for $k=6$, $k=8$ and $k=10$ like in the section \ref{sec3}, we get only one fixed point with one relevant direction, the details being summarized in Table \ref{table2} below. Following the expected result seems to be not so bad. We get only one relevant direction, with a critical exponent matching with the perturbative result. Note that no significant improvement arises from the nonperturbative effects. Moreover, the value of the relevant critical exponent seems to be insensitive to the level of the truncation. However, the value of the corresponding coupling is in strong discordance with the expected one. We get a positive and very small value for $u_4$, $u_4\approx 0.016$ for $k=6$, $u_4\approx 0.01$ for $k=8$ and $u_4\approx 0.008$ for $k=10$; the values of the other couplings being of very small magnitude with respect to these values. The rapidity of the decreasing seems to decrease with the order of the truncation, evoking convergence phenomena toward a constant and nonzero value -- a conjecture which has to be confirmed with higher truncation, including disconnected interactions.\\

\section{Conclusion and outlook}\label{sec5}

In this review we provided a short introduction to a recent and active topic of investigation in quantum gravity, exploiting the nonperturbative techniques to build renormalization group flow \cite{Lahoche:2018ggd}-\cite{Lahoche:2018vun}. \\

For matrix models, we showed that taking into account the non-trivial Ward identities arising from the scale-dependent mass term reveals a strong violation along with the flow for current approximation schemes used to solve the exact RG flow equations \cite{Lahoche:2019ocf}. Keeping into account the derivative couplings generated by the regulator reveals a strong dependence of the fixed point on the choice of the regulator, which, finally questions the reliability of the results, especially regarding the original theory space. Among the possible ways discussed to solve these issues, we discussed the possibility of choosing the regulator such that the modification of the Ward identities remains ‘‘the smallest possible" along with the flow, at least for a sufficient range of time flow. This has been achieved in a first approximation from a fine-tuning of the regulator, keeping the RG flow as horizontal as possible. But the investigations toward a solid formalism, taking into account Ward identities remains to be achieved. Finally, let us mention that we can expect that these conclusions survive for higher dimensional discrete models i.e. for random tensors. A random tensor of rank $d$ is a map $\mathbb{N}^d$ to $\mathbb{C}$ or $\mathbb{R}$. Focusing on the rank $3$ complex case, the classical action for such a model describing two tensors, $T_{abc}$ and $\bar{T}_{abc}$ \cite{Geloun:2013saa}, \cite{BenGeloun:2012pu} . The tensors are assumed to be a representation of the unitary group $U(N)^{\times d}$ for $d$--independent unitary transformations, represented as $N\times N$ unitary matrices (see \cite{Gurau:2009tw}-\cite{Sasakura:1990fs} for a general reviews on tensor models), in the sense that, for any set $(U_1,U_2,U_3)\in U(N)^{\times 3}$ (for $d=3$), we must have the transformation rule:
\begin{equation}
T_{abc}\to T^{\prime}_{abc}=\sum_{a^\prime,b^\prime,c^\prime}(U_1)_{aa^\prime}(U_2)_{bb^\prime}(U_3)_{cc^\prime}T_{a^\prime b^\prime c^\prime}\,.
\end{equation}
Similarly, with matrix, interactions for tensor fields are build to remains unchanged under such a transformation. This can be achieved contracting systematically indices on the same positions between $T$ and $\bar{T}$ fields. This is the case, for instance, of the action:
\begin{equation}
\mathcal{S}[T,\bar{T}]= \sum_{abc} T_{abc}\bar{T}_{abc}+g \sum_{a,b,c,a^\prime,b^\prime,c^\prime} \bar{T}_{abc} T_{abc^\prime} \bar{T}_{a^\prime b^\prime c^\prime} T_{a^\prime b^\prime c}\,,
\end{equation}
interaction building following these rules being called \textit{tensorial}. The dual interpretation follows the matrix models as well. In the point of view of the Feynman diagrams, each $T_{abc}$ sites can be pictured as a $d$-simplex, each of the three indices, $a$, $b$ and $c$ being associated to $d-1$ faces, the remaining face corresponding to the propagator edge hooked to $T$ from the Wick contraction. Once again, the nonperturbative renormalization group approach requires a regulator function $r_N$ which breaks the explicit invariance. For instance, we can consider the generalization of the regulator \eqref{regulator} as:
\begin{equation}
[r_N(a,b,c)]_{abc,a^\prime b^\prime c^\prime}:=\delta_{a a^\prime}\delta_{ b b^\prime}\delta_{c c^\prime}\left(\frac{3N}{a+b+c}-1\right)\Theta\left(\frac{3N}{a+b+c}-1\right)\,.
\end{equation}
We expect that the same kind of difficulties as for matrix models. Breaking the $U(N)^{\times d}$ symmetry, the regulator will generate modified Ward identities, ensuring the proliferation of derivative couplings and finally the large deviation effect of the RG flow, so far from the ultra-local (i.e. tensorial) theory space. Investigations in this direction, however, remain to be addressed as a continuation of the recent work \cite{Lahoche:2020pjo}.\\

For TGFTs, we showed that taking into account the constraint arising from Ward identities strongly modify the landscape of the RG flow. This improvement has been supported by another approximation scheme, the EVE method, allowing to keep entire sectors rather than a finite-dimensional subspace of the full theory space. This technique has been extended from quartic to sextic interactions, as well as for sectors beyond the melonic one, including for instance pseudo-melons \cite{Lahoche:2018oeo}. Once again, this domain remains active, and some complementary approximations are being considered to explore ever-larger areas of the tensor theory space, and test the reliability of the results. As we have emphasized above, the solution of the closed equations of the correlation functions will allow us to calculate the true flow equations and the conclusions which will result from them will be the right solution without approximation. Major advances have already been made in the case of large dimensions \cite{Lahoche:2019ehm}. For arbitrary dimensions, the solution of the closed equation remains a work in progress.


\begin{thebibliography}{99}
 
 %\cite{Rovelli:2008brv}
\bibitem{Rovelli:2008brv} 
  C.~Rovelli,
  ``Quantum gravity,''
  Scholarpedia {\bf 3}, no. 5, 7117 (2008).
  doi:10.4249/scholarpedia.7117.
  %%CITATION = doi:10.4249/scholarpedia.7117;%%

%\cite{Donoghue:1994dn}
\bibitem{Donoghue:1994dn} 
  J.~F.~Donoghue,
  ``General relativity as an effective field theory: The leading quantum corrections,''
  Phys.\ Rev.\ D {\bf 50}, 3874 (1994)
  doi:10.1103/PhysRevD.50.3874
  [gr-qc/9405057].
  %%CITATION = doi:10.1103/PhysRevD.50.3874;%%
  %730 citations counted in INSPIRE as of 12 Nov 2019
  
%\cite{Reuter:1996cp}
\bibitem{Reuter:1996cp} 
  M.~Reuter,
  ``Nonperturbative evolution equation for quantum gravity,''
  Phys.\ Rev.\ D {\bf 57}, 971 (1998)
  doi:10.1103/PhysRevD.57.971
  [hep-th/9605030].
  %%CITATION = doi:10.1103/PhysRevD.57.971;%%
  %716 citations counted in INSPIRE as of 12 Nov 2019 

%\cite{Feynman:1963uxa}
\bibitem{Feynman:1963uxa} 
  R.~P.~Feynman, R.~B.~Leighton and M.~Sands,
  ``The Feynman Lectures on Physics,''
  %%CITATION = INSPIRE-1283666;%%
  %20 citations counted in INSPIRE as of 24 Dec 2019
 
 
%\small
%
%
%\cite{Schwarz:1982jn}
\bibitem{Schwarz:1982jn} 
  J.~H.~Schwarz,
  ``Superstring Theory,''
  Phys.\ Rept.\  {\bf 89}, 223 (1982).
  doi:10.1016/0370-1573(82)90087-4
  %%CITATION = doi:10.1016/0370-1573(82)90087-4;%%
  %1238 citations counted in INSPIRE as of 17 Mar 2020
  %
  %



%\cite{Dirac:1974dc}
\bibitem{Dirac:1974dc} 
  P.~A.~M.~Dirac,
  ``The Geometrical Nature of Space and Time,''
  Stud.\ Nat.\ Sci.\  {\bf 5}, 1 (1974).
  doi:10.1007/978-1-4684-2913-81
  %%CITATION = doi:10.1007/978-1-4684-2913-8_1;%%



%\cite{DeWitt:1960fc}
\bibitem{DeWitt:1960fc} 
  B.~S.~DeWitt and R.~W.~Brehme,
  ``Radiation damping in a gravitational field,''
  Annals Phys.\  {\bf 9}, 220 (1960).
  doi:10.1016/0003-4916(60)90030-0
  %%CITATION = doi:10.1016/0003-4916(60)90030-0;%%
  %553 citations counted in INSPIRE as of 19 Oct 2019



%\cite{DeWitt:1964dp}
\bibitem{DeWitt:1964dp} 
  B.~S.~DeWitt,
  ``Gravity,''
  Adv.\ Space Sci.\ Technol.\  {\bf 6}, 1 (1964).
  %%CITATION = AVSTA,6,1;%%
  %%
 

%\cite{Rovelli:1996ti}
\bibitem{Rovelli:1996ti} 
  C.~Rovelli,
  ``Loop quantum gravity and black hole physics,''
  Helv.\ Phys.\ Acta {\bf 69}, 582 (1996)
  [gr-qc/9608032].
  %%CITATION = GR-QC/9608032;%%
  %64 citations counted in INSPIRE as of 22 Dec 2019


%\cite{Oriti:2009nd}
\bibitem{Oriti:2009nd} 
  D.~Oriti,
  ``Group field theory and simplicial quantum gravity,''
  Class.\ Quant.\ Grav.\  {\bf 27}, 145017 (2010)
  doi:10.1088/0264-9381/27/14/145017
  [arXiv:0902.3903 [gr-qc]].
  %%CITATION = doi:10.1088/0264-9381/27/14/145017;%%
  %20 citations counted in INSPIRE as of 11 Aug 2018
  
  
 
  %\cite{Oriti:2018dsg}
\bibitem{Oriti:2018dsg} 
  D.~Oriti,
  ``Levels of spacetime emergence in quantum gravity,''
  arXiv:1807.04875 [physics.hist-ph].
  %%CITATION = ARXIV:1807.04875;%%
  %5 citations counted in INSPIRE as of 17 Jun 2019


%\cite{Baratin:2011aa}
\bibitem{Baratin:2011aa} 
  A.~Baratin and D.~Oriti,
  ``Ten questions on Group Field Theory (and their tentative answers),''
  J.\ Phys.\ Conf.\ Ser.\  {\bf 360}, 012002 (2012)
  doi:10.1088/1742-6596/360/1/012002
  [arXiv:1112.3270 [gr-qc]].
  %%CITATION = doi:10.1088/1742-6596/360/1/012002;%%
  %56 citations counted in INSPIRE as of 17 Jun 2019

%\cite{Oriti:2009wn}
\bibitem{Oriti:2009wn} 
  D.~Oriti,
  ``The Group field theory approach to quantum gravity: Some recent results,''
  AIP Conf.\ Proc.\  {\bf 1196}, no. 1, 209 (2009)
  doi:10.1063/1.3284386
  [arXiv:0912.2441 [hep-th]].
  %%CITATION = doi:10.1063/1.3284386;%%
  %61 citations counted in INSPIRE as of 17 Jun 2019
  
  
  %\cite{Penrose:1972ia}
\bibitem{Penrose:1972ia} 
  R.~Penrose and M.~A.~H.~MacCallum,
  ``Twistor theory: An Approach to the quantization of fields and space-time,''
  Phys.\ Rept.\  {\bf 6}, 241 (1972).
  doi:10.1016/0370-1573(73)90008-2
  %%CITATION = doi:10.1016/0370-1573(73)90008-2;%%
  %420 citations counted in INSPIRE as of 17 Mar 2020

%\cite{Ambjorn:1992eh}
\bibitem{Ambjorn:1992eh} 
  J.~Ambjorn, Z.~Burda, J.~Jurkiewicz and C.~F.~Kristjansen,
  ``Quantum gravity represented as dynamical triangulations,''
  Acta Phys.\ Polon.\ B {\bf 23}, 991 (1992).
  %%CITATION = APPOA,B23,991;%%
  %33 citations counted in INSPIRE as of 11 Aug 2018



%\cite{Ambjorn:1995jt}
\bibitem{Ambjorn:1995jt} 
  J.~Ambjorn,
  ``Quantum gravity represented as dynamical triangulations,''
  Class.\ Quant.\ Grav.\  {\bf 12}, 2079 (1995).
  doi:10.1088/0264-9381/12/9/002.
  %%CITATION = doi:10.1088/0264-9381/12/9/002;%%
  %24 citations counted in INSPIRE as of 11 Aug 2018


%\cite{Ambjorn:2013tki}
\bibitem{Ambjorn:2013tki} 
  J.~Ambjørn, A.~Görlich, J.~Jurkiewicz and R.~Loll,
  ``Quantum Gravity via Causal Dynamical Triangulations,''
  doi:10.1007/978-3-642-41992-8-34
  arXiv:1302.2173 [hep-th].
  %%CITATION = doi:10.1007/978-3-642-41992-8_34;%%
  %30 citations counted in INSPIRE as of 11 Aug 2018







%\cite{Eichhorn:2018yfc}
\bibitem{Eichhorn:2018yfc} 
  A.~Eichhorn,
  ``An asymptotically safe guide to quantum gravity and matter,''
  Front.\ Astron.\ Space Sci.\  {\bf 5}, 47 (2019)
  doi:10.3389/fspas.2018.00047
  [arXiv:1810.07615 [hep-th]].
  %%CITATION = doi:10.3389/fspas.2018.00047;%%
  %42 citations counted in INSPIRE as of 12 Nov 2019  



%\cite{Rovelli:1995ac}
\bibitem{Rovelli:1995ac} 
  C.~Rovelli and L.~Smolin,
  ``Spin networks and quantum gravity,''
  Phys.\ Rev.\ D {\bf 52}, 5743 (1995)
  doi:10.1103/PhysRevD.52.5743
  [gr-qc/9505006].
  %%CITATION = doi:10.1103/PhysRevD.52.5743;%%
  %433 citations counted in INSPIRE as of 22 Dec 2019


%\cite{Witten:1981nf}
\bibitem{Witten:1981nf} 
  E.~Witten,
  ``Dynamical Breaking of Supersymmetry,''
  Nucl.\ Phys.\ B {\bf 188}, 513 (1981).
  doi:10.1016/0550-3213(81)90006-7
  %%CITATION = doi:10.1016/0550-3213(81)90006-7;%%
  %3119 citations counted in INSPIRE as of 22 Dec 2019


%%\cite{Witten:1981my}
%\bibitem{Witten:1981my} 
%  E.~Witten,
%  ``Lecture Notes On Supersymmetry,''
%  Preprint - WITTEN, E. (REC.SEP. 81) 64p

%\cite{Witten:1985cc}
\bibitem{Witten:1985cc} 
  E.~Witten,
  ``Noncommutative Geometry and String Field Theory,''
  Nucl.\ Phys.\ B {\bf 268}, 253 (1986).
  doi:10.1016/0550-3213(86)90155-0
  %%CITATION = doi:10.1016/0550-3213(86)90155-0;%%
  %1786 citations counted in INSPIRE as of 22 Dec 2019


  

%\cite{Percacci:2017fkn}
\bibitem{Percacci:2017fkn} 
  R.~Percacci,
  ``An Introduction to Covariant Quantum Gravity and Asymptotic Safety,''
  doi:10.1142/10369.
  %%CITATION = doi:10.1142/10369;%%
  %22 citations counted in INSPIRE as of 12 Nov 2019

%\cite{Reuter:2019byg}
\bibitem{Reuter:2019byg} 
  M.~Reuter and F.~Saueressig,
  ``Quantum Gravity and the Functional Renormalization Group : The Road towards Asymptotic Safety,''
  %%CITATION = INSPIRE-1716753;%%
  %5 citations counted in INSPIRE as of 12 Nov 2019

%\cite{Anselmi:2018ibi}
\bibitem{Anselmi:2018ibi} 
  D.~Anselmi and M.~Piva,
  ``The Ultraviolet Behavior of Quantum Gravity,''
  JHEP {\bf 1805}, 027 (2018)
  doi:10.1007/JHEP05(2018)027
  [arXiv:1803.07777 [hep-th]].
  %%CITATION = doi:10.1007/JHEP05(2018)027;%%
  %19 citations counted in INSPIRE as of 12 Nov 2019

{\color{blue}
%\cite{Basile:2020dzh}
\bibitem{Basile:2020dzh}
I.~Basile and A.~Platania,
``Cosmological $\alpha'$-corrections from the functional renormalization group,''
JHEP \textbf{21} (2020), 045
doi:10.1007/JHEP06(2021)045
[arXiv:2101.02226 [hep-th]].
%7 citations counted in INSPIRE as of 16 Jul 2021
%
}

%\cite{Hamber:2009zz}
\bibitem{Hamber:2009zz} 
  H.~W.~Hamber,
  ``Quantum gravitation: The Feynman path integral approach,''
  doi:10.1007/978-3-540-85293-3.
  %%CITATION = doi:10.1007/978-3-540-85293-3;%%
  %26 citations counted in INSPIRE as of 12 Nov 2019
 

  
  
  

  %\cite{Chirco:2019dlx}
\bibitem{Chirco:2019dlx} 
  G.~Chirco, A.~Goeßmann, D.~Oriti and M.~Zhang,
  ``Group Field Theory and Holographic Tensor Networks: Dynamical Corrections to the Ryu-Takayanagi formula,''
  arXiv:1903.07344 [hep-th].
  %%CITATION = ARXIV:1903.07344;%%
  %1 citations counted in INSPIRE as of 17 Jun 2019



%\cite{Ashtekar:2014kba}
\bibitem{Ashtekar:2014kba} 
  A.~Ashtekar, M.~Reuter and C.~Rovelli,
  ``From General Relativity to Quantum Gravity,''
  arXiv:1408.4336 [gr-qc].
  %%CITATION = ARXIV:1408.4336;%%
  %45 citations counted in INSPIRE as of 17 Jun 2019

  %\cite{Rovelli:1997yv}
\bibitem{Rovelli:1997yv} 
  C.~Rovelli,
  ``Loop quantum gravity,''
  Living Rev.\ Rel.\  {\bf 1}, 1 (1998)
  doi:10.12942/lrr-1998-1
  [gr-qc/9710008].
  %%CITATION = doi:10.12942/lrr-1998-1;%%
  %544 citations counted in INSPIRE as of 11 Aug 2018

%\cite{Rovelli:1998gg}
\bibitem{Rovelli:1998gg} 
  C.~Rovelli and P.~Upadhya,
  ``Loop quantum gravity and quanta of space: A Primer,''
  gr-qc/9806079.
  %%CITATION = GR-QC/9806079;%%
  %34 citations counted in INSPIRE as of 11 Aug 2018


%\cite{Rovelli:2011eq}
\bibitem{Rovelli:2011eq} 
  C.~Rovelli,
  ``Zakopane lectures on loop gravity,''
  PoS QGQGS {\bf 2011}, 003 (2011)
  [arXiv:1102.3660 [gr-qc]].
  %%CITATION = ARXIV:1102.3660;%%
  %161 citations counted in INSPIRE as of 25 Mar 2016


%\cite{Rovelli:2010bf}
\bibitem{Rovelli:2010bf} 
  C.~Rovelli,
  ``Loop quantum gravity: the first twenty five years,''
  Class.\ Quant.\ Grav.\  {\bf 28}, 153002 (2011)
  doi:10.1088/0264-9381/28/15/153002
  [arXiv:1012.4707 [gr-qc]].
  %%CITATION = doi:10.1088/0264-9381/28/15/153002;%%
  %42 citations counted in INSPIRE as of 25 Mar 2016





%\cite{Connes:1990qp}
\bibitem{Connes:1990qp} 
  A.~Connes and J.~Lott,
``Particle Models and Noncommutative Geometry (Expanded
Version),''
  Nucl.\ Phys.\ Proc.\ Suppl.\  {\bf 18B}, 29 (1991).
  doi:10.1016/0920-5632(91)90120-4
  %%CITATION = doi:10.1016/0920-5632(91)90120-4;%%
  %262 citations counted in INSPIRE as of 11 Aug 2018



%\cite{Aastrup:2006ib}
\bibitem{Aastrup:2006ib} 
  J.~Aastrup and J.~M.~Grimstrup,
``Intersecting connes noncommutative geometry with quantum
gravity,''
  Int.\ J.\ Mod.\ Phys.\ A {\bf 22}, 1589 (2007)
  doi:10.1142/S0217751X07035306
  [hep-th/0601127].
  %%CITATION = doi:10.1142/S0217751X07035306;%%
  %19 citations counted in INSPIRE as of 11 Aug 2018


%\cite{Perez:2002vg}
\bibitem{Perez:2002vg} 
  A.~Perez,
  ``Spin foam quantization of SO(4) Plebanski's action,''
  Adv.\ Theor.\ Math.\ Phys.\  {\bf 5}, 947 (2002)
  Erratum: [Adv.\ Theor.\ Math.\ Phys.\  {\bf 6}, 593 (2003)]
  doi:10.4310/ATMP.2001.v5.n5.a4, 10.4310/ATMP.2002.v6.n3.e1
  [gr-qc/0203058].
  %%CITATION = doi:10.4310/ATMP.2001.v5.n5.a4, 10.4310/ATMP.2002.v6.n3.e1;%%
  %29 citations counted in INSPIRE as of 11 Apr 2019





%%%%%%%%%%%%%%%%%%%%%%%
%%%%%%%%%%%%%%%%%%%%%%% 
%\cite{Oriti:2005jr}
\bibitem{Oriti:2005jr} 
  D.~Oriti,
  ``Generalised group field theories and quantum gravity transition amplitudes,''
  Phys.\ Rev.\ D {\bf 73}, 061502 (2006)
  doi:10.1103/PhysRevD.73.061502
  [gr-qc/0512069].
  %%CITATION = doi:10.1103/PhysRevD.73.061502;%%
  %21 citations counted in INSPIRE as of 24 Mar 2020



%\cite{Freidel:2005cg}
\bibitem{Freidel:2005cg} 
  L.~Freidel, D.~Oriti and J.~Ryan,
  ``A Group field theory for 3-D quantum gravity coupled to a scalar field,''
  gr-qc/0506067.
  %%CITATION = GR-QC/0506067;%%
  %30 citations counted in INSPIRE as of 24 Mar 2020

%%%%%%%%%%%%%%%%%%%%%%%
%\cite{Oriti:2006ar}
\bibitem{Oriti:2006ar} 
  D.~Oriti,
``A Quantum field theory of simplicial geometry and the
emergence of spacetime,''
  J.\ Phys.\ Conf.\ Ser.\  {\bf 67}, 012052 (2007)
  doi:10.1088/1742-6596/67/1/012052
  [hep-th/0612301].
  %%CITATION = doi:10.1088/1742-6596/67/1/012052;%%
  %23 citations counted in INSPIRE as of 25 Mar 2016



%\cite{Oriti:2006se}
\bibitem{Oriti:2006se} 
  D.~Oriti,
  ``The Group field theory approach to quantum gravity,''
  In *Oriti, D. (ed.): Approaches to quantum gravity* 310-331
  [gr-qc/0607032].
  %%CITATION = GR-QC/0607032;%%
  %183 citations counted in INSPIRE as of 24 Mar 2020


%\cite{deCesare:2016rsf}
\bibitem{deCesare:2016rsf} 
  M.~de Cesare, A.~G.~A.~Pithis and M.~Sakellariadou,
``Cosmological implications of interacting Group Field Theory
models: cyclic Universe and accelerated expansion,''
  Phys.\ Rev.\ D {\bf 94}, no. 6, 064051 (2016)
  doi:10.1103/PhysRevD.94.064051
  [arXiv:1606.00352 [gr-qc]].
  %%CITATION = doi:10.1103/PhysRevD.94.064051;%%
  %18 citations counted in INSPIRE as of 19 Mar 2018



%\cite{Gielen:2016dss}
\bibitem{Gielen:2016dss} 
  S.~Gielen and L.~Sindoni,
``Quantum Cosmology from Group Field Theory Condensates: a
Review,''
  SIGMA {\bf 12}, 082 (2016)
  doi:10.3842/SIGMA.2016.082
  [arXiv:1602.08104 [gr-qc]].
  %%CITATION = doi:10.3842/SIGMA.2016.082;%%
  %31 citations counted in INSPIRE as of 19 Mar 2018



%\cite{Gielen:2017eco}
\bibitem{Gielen:2017eco} 
  S.~Gielen and D.~Oriti,
  ``Cosmological perturbations from full quantum gravity,''
  arXiv:1709.01095 [gr-qc].
  %%CITATION = ARXIV:1709.01095;%%
  %4 citations counted in INSPIRE as of 19 Mar 2018

%\cite{Gielen:2014uga}
\bibitem{Gielen:2014uga} 
  S.~Gielen and D.~Oriti,
  ``Quantum cosmology from quantum gravity condensates: cosmological variables and lattice-refined dynamics,''
  New J.\ Phys.\  {\bf 16}, no. 12, 123004 (2014)
  doi:10.1088/1367-2630/16/12/123004
  [arXiv:1407.8167 [gr-qc]].
  %%CITATION = doi:10.1088/1367-2630/16/12/123004;%%
  %50 citations counted in INSPIRE as of 27 Jul 2019


%\cite{Gielen:2013naa}
\bibitem{Gielen:2013naa} 
  S.~Gielen, D.~Oriti and L.~Sindoni,
  ``Homogeneous cosmologies as group field theory condensates,''
  JHEP {\bf 1406}, 013 (2014)
  doi:10.1007/JHEP06(2014)013
  [arXiv:1311.1238 [gr-qc]].
  %%CITATION = doi:10.1007/JHEP06(2014)013;%%
  %99 citations counted in INSPIRE as of 27 Jul 2019

    %\cite{Gielen:2018xph}
\bibitem{Gielen:2018xph} 
  S.~Gielen,
  ``Inhomogeneous universe from group field theory condensate,''
  JCAP {\bf 1902}, 013 (2019)
  doi:10.1088/1475-7516/2019/02/013
  [arXiv:1811.10639 [gr-qc]].
  %%CITATION = doi:10.1088/1475-7516/2019/02/013;%%
  %3 citations counted in INSPIRE as of 27 Jul 2019


%\cite{Mandrysz:2018sle}
\bibitem{Mandrysz:2018sle} 
  M.~L.~Mandrysz and J.~Mielczarek,
  ``Ultralocal nature of geometrogenesis,''
  Class.\ Quant.\ Grav.\  {\bf 36}, no. 1, 015004 (2019)
  doi:10.1088/1361-6382/aaef71
  [arXiv:1804.10793 [gr-qc]].
  %%CITATION = doi:10.1088/1361-6382/aaef71;%%
  %3 citations counted in INSPIRE as of 25 Dec 2019
  %
  
  %
  %%\cite{Oriti:2013jga}
\bibitem{Oriti:2013jga} 
  D.~Oriti,
  ``Disappearance and emergence of space and time in quantum gravity,''
  Stud.\ Hist.\ Phil.\ Sci.\ B {\bf 46}, 186 (2014)
  doi:10.1016/j.shpsb.2013.10.006
  [arXiv:1302.2849 [physics.hist-ph]].
  %%CITATION = doi:10.1016/j.shpsb.2013.10.006;%%
  %63 citations counted in INSPIRE as of 25 Dec 2019


    


%\cite{Bardeen:1973gs}
\bibitem{Bardeen:1973gs} 
  J.~M.~Bardeen, B.~Carter and S.~W.~Hawking,
  ``The Four laws of black hole mechanics,''
  Commun.\ Math.\ Phys.\  {\bf 31}, 161 (1973).
  doi:10.1007/BF01645742
  %%CITATION = doi:10.1007/BF01645742;%%
  %2127 citations counted in INSPIRE as of 24 Mar 2020

%\cite{Hawking:1974sw}
\bibitem{Hawking:1974sw} 
  S.~W.~Hawking,
  ``Particle Creation by Black Holes,''
  Commun.\ Math.\ Phys.\  {\bf 43}, 199 (1975)
  Erratum: [Commun.\ Math.\ Phys.\  {\bf 46}, 206 (1976)].
  doi:10.1007/BF02345020, 10.1007/BF01608497
  %%CITATION = doi:10.1007/BF02345020, 10.1007/BF01608497;%%
  %8156 citations counted in INSPIRE as of 24 Mar 2020
  %
  %
  %%\cite{Gibbons:1977mu}
\bibitem{Gibbons:1977mu} 
  G.~W.~Gibbons and S.~W.~Hawking,
  ``Cosmological Event Horizons, Thermodynamics, and Particle Creation,''
  Phys.\ Rev.\ D {\bf 15}, 2738 (1977).
  doi:10.1103/PhysRevD.15.2738
  %%CITATION = doi:10.1103/PhysRevD.15.2738;%%
  %2232 citations counted in INSPIRE as of 24 Mar 2020



  %\cite{Colosi:2004vw}
\bibitem{Colosi:2004vw} 
  D.~Colosi and C.~Rovelli,
  ``What is a particle?,''
  Class.\ Quant.\ Grav.\  {\bf 26}, 025002 (2009)
  doi:10.1088/0264-9381/26/2/025002
  [gr-qc/0409054].
  %%CITATION = doi:10.1088/0264-9381/26/2/025002;%%
  %31 citations counted in INSPIRE as of 18 Jun 2019
  


%%%%%%%%%%%%


  %\cite{Brezin:1992yc}
\bibitem{Brezin:1992yc} 
  E.~Brezin and J.~Zinn-Justin,
  ``Renormalization group approach to matrix models,''
  Phys.\ Lett.\ B {\bf 288}, 54 (1992)
  doi:10.1016/0370-2693(92)91953-7
  [hep-th/9206035].
  %%CITATION = doi:10.1016/0370-2693(92)91953-7;%%
  %99 citations counted in INSPIRE as of 17 Aug 2019
 

 %\cite{DiFrancesco:1993cyw}
\bibitem{DiFrancesco:1993cyw} 
  P.~Di Francesco, P.~H.~Ginsparg and J.~Zinn-Justin,
  ``2-D Gravity and random matrices,''
  Phys.\ Rept.\  {\bf 254}, 1 (1995)
  doi:10.1016/0370-1573(94)00084-G
  [hep-th/9306153].
  %%CITATION = doi:10.1016/0370-1573(94)00084-G;%%
  %733 citations counted in INSPIRE as of 17 Aug 2019
  
  %\cite{Higuchi:1993pu}
\bibitem{Higuchi:1993pu} 
  S.~Higuchi, C.~Itoi and N.~Sakai,
  ``Renormalization group approach to matrix models and vector models,''
  Prog.\ Theor.\ Phys.\ Suppl.\  {\bf 114}, 53 (1993)
  doi:10.1143/PTPS.114.53
  [hep-th/9307154].
  %%CITATION = doi:10.1143/PTPS.114.53;%%
  %13 citations counted in INSPIRE as of 24 Aug 2019




  %\cite{Zinn-Justin:2014wva}
\bibitem{Zinn-Justin:2014wva} 
  J.~Zinn-Justin,
  ``Random vector and matrix and vector theories: a renormalization group approach,''
  J.\ Statist.\ Phys.\  {\bf 157}, 990 (2014)
  doi:10.1007/s10955-014-1103-y
  [arXiv:1410.1635 [math-ph]].
  %%CITATION = doi:10.1007/s10955-014-1103-y;%%
  %3 citations counted in INSPIRE as of 17 Aug 2019
  
  
  %\cite{Klebanov:1994kv}
\bibitem{Klebanov:1994kv} 
  I.~R.~Klebanov and A.~Hashimoto,
  ``Nonperturbative solution of matrix models modified by trace squared terms,''
  Nucl.\ Phys.\ B {\bf 434}, 264 (1995)
  doi:10.1016/0550-3213(94)00518-J
  [hep-th/9409064].
  %%CITATION = doi:10.1016/0550-3213(94)00518-J;%%
  %56 citations counted in INSPIRE as of 24 Dec 2019


%\cite{Duplantier:1989sx}
\bibitem{Duplantier:1989sx} 
  B.~Duplantier and I.~K.~Kostov,
  ``Geometrical Critical Phenomena on a Random Surface of Arbitrary Genus,''
  Nucl.\ Phys.\ B {\bf 340}, 491 (1990).
  doi:10.1016/0550-3213(90)90456-N.
  %%CITATION = doi:10.1016/0550-3213(90)90456-N;%%
  %55 citations counted in INSPIRE as of 21 Dec 2019

%\cite{Ambjorn:1990pt}
\bibitem{Ambjorn:1990pt} 
  J.~Ambjorn and J.~Greensite,
  ``Nonperturbative calculation of correlators in 2-D quantum gravity,''
  Phys.\ Lett.\ B {\bf 254}, 66 (1991).
  doi:10.1016/0370-2693(91)90397-9.
  %%CITATION = doi:10.1016/0370-2693(91)90397-9;%%
  %36 citations counted in INSPIRE as of 21 Dec 2019

%\cite{Gross:1989aw}
\bibitem{Gross:1989aw} 
  D.~J.~Gross and A.~A.~Migdal,
  ``A Nonperturbative Treatment of Two-dimensional Quantum Gravity,''
  Nucl.\ Phys.\ B {\bf 340}, 333 (1990).
  doi:10.1016/0550-3213(90)90450-R.
  %%CITATION = doi:10.1016/0550-3213(90)90450-R;%%
  %441 citations counted in INSPIRE as of 22 Dec 2019

%\cite{Ginsparg:1993is}
\bibitem{Ginsparg:1993is} 
  P.~H.~Ginsparg and G.~W.~Moore,
  ``Lectures on 2-D gravity and 2-D string theory,''
  Yale Univ. New Haven - YCTP-P23-92 (92,rec.Apr.93) 197 p. Los Alamos Nat. Lab. - LA-UR-92-3479 (92,rec.Apr.93) 197 p. e: LANL hep-th/9304011
  [hep-th/9304011].
  %%CITATION = HEP-TH/9304011;%%
  %500 citations counted in INSPIRE as of 22 Dec 2019




%\cite{Marino:2004eq}
\bibitem{Marino:2004eq} 
  M.~Marino,
  ``Les Houches lectures on matrix models and topological strings,''
  hep-th/0410165.
  %%CITATION = HEP-TH/0410165;%%
  %164 citations counted in INSPIRE as of 22 Dec 2019
  
%\cite{Gervais:1990be}
\bibitem{Gervais:1990be} 
  J.~L.~Gervais,
  ``Solving the strongly coupled 2-D gravity: 1. Unitary truncation and quantum group structure,''
  Commun.\ Math.\ Phys.\  {\bf 138}, 301 (1991).
  doi:10.1007/BF02099495
  %%CITATION = doi:10.1007/BF02099495;%%
  %66 citations counted in INSPIRE as of 21 Dec 2019


%\cite{Bowick:1991ky}
\bibitem{Bowick:1991ky} 
  M.~J.~Bowick and E.~Brezin,
  ``Universal scaling of the tail of the density of eigenvalues in random matrix models,''
  Phys.\ Lett.\ B {\bf 268}, 21 (1991).
  doi:10.1016/0370-2693(91)90916-E
  %%CITATION = doi:10.1016/0370-2693(91)90916-E;%%
  %49 citations counted in INSPIRE as of 21 Dec 2019


%\cite{Dalley:1991qg}
\bibitem{Dalley:1991qg} 
  S.~Dalley, C.~V.~Johnson and T.~R.~Morris,
  ``Multicritical complex matrix models and nonperturbative 2-D quantum gravity,''
  Nucl.\ Phys.\ B {\bf 368}, 625 (1992).
  doi:10.1016/0550-3213(92)90217-Y
  %%CITATION = doi:10.1016/0550-3213(92)90217-Y;%%
  %48 citations counted in INSPIRE as of 21 Dec 2019

%\cite{Ambjorn:1992gw}
\bibitem{Ambjorn:1992gw} 
  J.~Ambjorn, L.~Chekhov, C.~F.~Kristjansen and Y.~Makeenko,
  ``Matrix model calculations beyond the spherical limit,''
  Nucl.\ Phys.\ B {\bf 404}, 127 (1993)
  Erratum: [Nucl.\ Phys.\ B {\bf 449}, 681 (1995)]
  doi:10.1016/0550-3213(93)90476-6, 10.1016/0550-3213(95)00391-5
  [hep-th/9302014].
  %%CITATION = doi:10.1016/0550-3213(93)90476-6, 10.1016/0550-3213(95)00391-5;%%
  %226 citations counted in INSPIRE as of 17 Aug 2019
  
  
   %\cite{Ambjorn:1992aw}
\bibitem{Ambjorn:1992aw} 
  J.~Ambjorn, J.~Jurkiewicz and C.~F.~Kristjansen,
  ``Quantum gravity, dynamical triangulations and higher derivative regularization,''
  Nucl.\ Phys.\ B {\bf 393}, 601 (1993)
  doi:10.1016/0550-3213(93)90075-Z
  [hep-th/9208032].
  %%CITATION = doi:10.1016/0550-3213(93)90075-Z;%%
  %71 citations counted in INSPIRE as of 17 Aug 2019
  
  

%\cite{Higuchi:1994rv}
\bibitem{Higuchi:1994rv} 
  S.~Higuchi, C.~Itoi, S.~Nishigaki and N.~Sakai,
  ``Renormalization group flow in one and two matrix models,''
  Nucl.\ Phys.\ B {\bf 434}, 283 (1995)
  Erratum: [Nucl.\ Phys.\ B {\bf 441}, 405 (1995)]
  doi:10.1016/0550-3213(95)00119-D, 10.1016/0550-3213(94)00437-J
  [hep-th/9409009].
  %%CITATION = doi:10.1016/0550-3213(95)00119-D, 10.1016/0550-3213(94)00437-J;%%
  %33 citations counted in INSPIRE as of 17 Aug 2019
  
  %\cite{Canet:2003qd}
\bibitem{Canet:2003qd} 
  L.~Canet, B.~Delamotte, D.~Mouhanna and J.~Vidal,
  ``Nonperturbative renormalization group approach to the Ising model: A Derivative expansion at order partial**4,''
  Phys.\ Rev.\ B {\bf 68}, 064421 (2003)
  doi:10.1103/PhysRevB.68.064421
  [hep-th/0302227].
  %%CITATION = doi:10.1103/PhysRevB.68.064421;%%
  %122 citations counted in INSPIRE as of 17 Aug 2019

%\cite{Ambjorn:1991cs}
\bibitem{Ambjorn:1991cs} 
  J.~Ambjorn, J.~Jurkiewicz, S.~Varsted, A.~Irback and B.~Petersson,
  ``Critical properties of the dynamical random surface with extrinsic curvature,''
  Phys.\ Lett.\ B {\bf 275}, 295 (1992).
  doi:10.1016/0370-2693(92)91593-X
  %%CITATION = doi:10.1016/0370-2693(92)91593-X;%%
  %37 citations counted in INSPIRE as of 17 Aug 2019


%\cite{Alfaro:1992nq}
\bibitem{Alfaro:1992nq} 
  J.~Alfaro and P.~H.~Damgaard,
  ``The D = 1 matrix model and the renormalization group,''
  Phys.\ Lett.\ B {\bf 289}, 342 (1992)
  doi:10.1016/0370-2693(92)91229-3
  [hep-th/9206099].
  %%CITATION = doi:10.1016/0370-2693(92)91229-3;%%
  %16 citations counted in INSPIRE as of 17 Aug 2019


  
  
%\cite{Itoh:2017huk}
\bibitem{Itoh:2017huk} 
  K.~Itoh,
  ``Gauge symmetry and the functional renormalization group,''
  Int.\ J.\ Mod.\ Phys.\ A {\bf 32}, no. 35, 1747011 (2017).
  doi:10.1142/S0217751X1747011X
  %%CITATION = doi:10.1142/S0217751X1747011X;%%
  
  
  %\cite{Gao:1995vv}
\bibitem{Gao:1995vv} 
  H.~B.~Gao,
  ``On renormalization group flow in matrix model,''
  hep-th/9209089.
  %%CITATION = HEP-TH/9209089;%%
  
  %\cite{Ayala:1993fj}
\bibitem{Ayala:1993fj} 
  C.~Ayala,
  ``Renormalization group approach to matrix models in two-dimensional quantum gravity,''
  Phys.\ Lett.\ B {\bf 311}, 55 (1993)
  doi:10.1016/0370-2693(93)90533-N
  [hep-th/9304090].
  %%CITATION = doi:10.1016/0370-2693(93)90533-N;%%
  %9 citations counted in INSPIRE as of 17 Aug 2019
  
  %\cite{Higuchi:1993pu}
\bibitem{Higuchi:1993pu} 
  S.~Higuchi, C.~Itoi and N.~Sakai,
  ``Renormalization group approach to matrix models and vector models,''
  Prog.\ Theor.\ Phys.\ Suppl.\  {\bf 114}, 53 (1993)
  doi:10.1143/PTPS.114.53
  [hep-th/9307154].
  %%CITATION = doi:10.1143/PTPS.114.53;%%
  %13 citations counted in INSPIRE as of 17 Aug 2019

  %\cite{Eichhorn:2013isa}
\bibitem{Eichhorn:2013isa} 
  A.~Eichhorn and T.~Koslowski,
  ``Continuum limit in matrix models for quantum gravity from the Functional Renormalization Group,''
  Phys.\ Rev.\ D {\bf 88}, 084016 (2013)
  doi:10.1103/PhysRevD.88.084016
  [arXiv:1309.1690 [gr-qc]].
  %%CITATION = doi:10.1103/PhysRevD.88.084016;%%
  %43 citations counted in INSPIRE as of 25 Dec 2019

  %\cite{Eichhorn:2014xaa}
\bibitem{Eichhorn:2014xaa} 
  A.~Eichhorn and T.~Koslowski,
  ``Towards phase transitions between discrete and continuum quantum spacetime from the Renormalization Group,''
  Phys.\ Rev.\ D {\bf 90}, no. 10, 104039 (2014)
  doi:10.1103/PhysRevD.90.104039
  [arXiv:1408.4127 [gr-qc]].
  %%CITATION = doi:10.1103/PhysRevD.90.104039;%%
  %29 citations counted in INSPIRE as of 17 Aug 2019



 %\cite{Eichhorn:2019hsa}
\bibitem{Eichhorn:2019hsa} 
  A.~Eichhorn, J.~Lumma, A.~D.~Pereira and A.~Sikandar,
  ``Universal critical behavior in tensor models for four-dimensional quantum gravity,''
  arXiv:1912.05314 [gr-qc].
  %%CITATION = ARXIV:1912.05314;%%





  %\cite{Eichhorn:2018phj}
\bibitem{Eichhorn:2018phj} 
  A.~Eichhorn, T.~Koslowski and A.~D.~Pereira,
  ``Status of background-independent coarse-graining in tensor models for quantum gravity,''
  Universe {\bf 5}, no. 2, 53 (2019)
  doi:10.3390/universe5020053
  [arXiv:1811.12909 [gr-qc]].
  %%CITATION = doi:10.3390/universe5020053;%%
  %9 citations counted in INSPIRE as of 17 Aug 2019

%\cite{Eichhorn:2017xhy}
\bibitem{Eichhorn:2017xhy} 
  A.~Eichhorn and T.~Koslowski,
  ``Flowing to the continuum in discrete tensor models for quantum gravity,''
  Ann.\ Inst.\ H.\ Poincare Comb.\ Phys.\ Interact.\  {\bf 5}, no. 2, 173 (2018)
  doi:10.4171/AIHPD/52
  [arXiv:1701.03029 [gr-qc]].
  %%CITATION = doi:10.4171/AIHPD/52;%%
  %15 citations counted in INSPIRE as of 22 Dec 2019  

 {\color{blue}

 %\cite{Eichhorn:2020sla}
\bibitem{Eichhorn:2020sla}
A.~Eichhorn, A.~D.~Pereira and A.~G.~A.~Pithis,
``The phase diagram of the multi-matrix model with ABAB-interaction from functional renormalization,''
JHEP \textbf{12} (2020), 131
doi:10.1007/JHEP12(2020)131
[arXiv:2009.05111 [gr-qc]].
%3 citations counted in INSPIRE as of 16 Jul 2021



   %\cite{Perez-Sanchez:2020ngw}
\bibitem{Perez-Sanchez:2020ngw}
C.~I.~Perez-Sanchez,
``Comment on \textquotedblleft{}The phase diagram of the multi-matrix model with ABAB-interaction from functional renormalization\textquotedblright{},''
JHEP \textbf{21} (2020), 042
doi:10.1007/JHEP07(2021)042
[arXiv:2102.06999 [hep-th]].
%1 citations counted in INSPIRE as of 16 Jul 2021
}
%\cite{Lahoche:2019ocf}
\bibitem{Lahoche:2019ocf} 
  V.~Lahoche and D.~Ousmane Samary,
  ``Revisited functional renormalization group approach for random matrices in the large-$N$ limit,''
  Phys.\ Rev.\ D {\bf 101}, no. 10, 106015 (2020)
  doi:10.1103/PhysRevD.101.106015
  [arXiv:1909.03327 [hep-th]].
  %%CITATION = doi:10.1103/PhysRevD.101.106015;%%
  %4 citations counted in INSPIRE as of 31 May 2020
  
  
  
  %\cite{Lahoche:2020pjo}
\bibitem{Lahoche:2020pjo}
V.~Lahoche and D.~O.~Samary,
``Reliability of the local truncations for the random tensor models renormalization group flow,''
Phys. Rev. D \textbf{102} (2020) no.5, 056002
doi:10.1103/PhysRevD.102.056002
[arXiv:2005.11846 [hep-th]].
%3 citations counted in INSPIRE as of 16 Jul 2021
%\cite{Stanford:2019vob}
\bibitem{Stanford:2019vob} 
  D.~Stanford and E.~Witten,
  ``JT Gravity and the Ensembles of Random Matrix Theory,''
  arXiv:1907.03363 [hep-th].
  %%CITATION = ARXIV:1907.03363;%%
  %6 citations counted in INSPIRE as of 17 Aug 2019

 
  
  %\cite{Gurau:2009tw}
\bibitem{Gurau:2009tw} 
  R.~Gurau,
  ``Colored Group Field Theory,''
  Commun.\ Math.\ Phys.\  {\bf 304}, 69 (2011)
  doi:10.1007/s00220-011-1226-9
  [arXiv:0907.2582 [hep-th]].
  %%CITATION = doi:10.1007/s00220-011-1226-9;%%
  %228 citations counted in INSPIRE as of 19 Oct 2019
  

  %\cite{Delporte:2018iyf}
\bibitem{Delporte:2018iyf} 
  N.~Delporte and V.~Rivasseau,
  ``The Tensor Track V: Holographic Tensors,''
  arXiv:1804.11101 [hep-th].
  %%CITATION = ARXIV:1804.11101;%%
  %19 citations counted in INSPIRE as of 19 Oct 2019

%\cite{Rivasseau:2013uca}
\bibitem{Rivasseau:2013uca} 
  V.~Rivasseau,
  ``The Tensor Track, III,''
  Fortsch.\ Phys.\  {\bf 62}, 81 (2014)
  doi:10.1002/prop.201300032
  [arXiv:1311.1461 [hep-th]].
  %%CITATION = doi:10.1002/prop.201300032;%%
  %74 citations counted in INSPIRE as of 22 Dec 2019

%\cite{Rivasseau:2012yp}
\bibitem{Rivasseau:2012yp} 
  V.~Rivasseau,
  ``The Tensor Track: an Update,''
  arXiv:1209.5284 [hep-th].
  %%CITATION = ARXIV:1209.5284;%%
  %61 citations counted in INSPIRE as of 22 Dec 2019



%\cite{Rivasseau:2014ima}
\bibitem{Rivasseau:2014ima} 
  V.~Rivasseau,
  ``The Tensor Theory Space,''
  Fortsch.\ Phys.\  {\bf 62}, 835 (2014)
  doi:10.1002/prop.201400057
  [arXiv:1407.0284 [hep-th]].
  %%CITATION = doi:10.1002/prop.201400057;%%
  %33 citations counted in INSPIRE as of 22 Dec 2019




%\cite{Gurau:2019qag}
\bibitem{Gurau:2019qag} 
  R.~Gurau,
  ``Notes on Tensor Models and Tensor Field Theories,''
  arXiv:1907.03531 [hep-th].
  %%CITATION = ARXIV:1907.03531;%%

%\cite{Gurau:2011aq}
\bibitem{Gurau:2011aq} 
  R.~Gurau and V.~Rivasseau,
  ``The 1/N expansion of colored tensor models in arbitrary dimension,''
  EPL {\bf 95}, no. 5, 50004 (2011)
  doi:10.1209/0295-5075/95/50004
  [arXiv:1101.4182 [gr-qc]].
  %%CITATION = doi:10.1209/0295-5075/95/50004;%%
  %167 citations counted in INSPIRE as of 17 Nov 2019
 
%\cite{Gurau:2013pca}
\bibitem{Gurau:2013pca} 
  R.~Gurau,
  ``The 1/N Expansion of Tensor Models Beyond Perturbation Theory,''
  Commun.\ Math.\ Phys.\  {\bf 330}, 973 (2014)
  doi:10.1007/s00220-014-1907-2
  [arXiv:1304.2666 [math-ph]].
  %%CITATION = doi:10.1007/s00220-014-1907-2;%%
  %71 citations counted in INSPIRE as of 25 Dec 2019

%\cite{Gurau:2011xp}
\bibitem{Gurau:2011xp} 
  R.~Gurau and J.~P.~Ryan,
  ``Colored Tensor Models - a review,''
  SIGMA {\bf 8}, 020 (2012)
  doi:10.3842/SIGMA.2012.020
  [arXiv:1109.4812 [hep-th]].
  %%CITATION = doi:10.3842/SIGMA.2012.020;%%
  %243 citations counted in INSPIRE as of 25 Dec 2019


 %\cite{Gurau:2011xq}
\bibitem{Gurau:2011xq} 
  R.~Gurau,
  ``The complete 1/N expansion of colored tensor models in arbitrary dimension,''
  Annales Henri Poincare {\bf 13}, 399 (2012)
  doi:10.1007/s00023-011-0118-z
  [arXiv:1102.5759 [gr-qc]].
  %%CITATION = doi:10.1007/s00023-011-0118-z;%%
  %194 citations counted in INSPIRE as of 19 Oct 2019

%\cite{Calcagni:2013dna}
\bibitem{Calcagni:2013dna} 
  G.~Calcagni, D.~Oriti and J.~Thürigen,
  ``Spectral dimension of quantum geometries,''
  Class.\ Quant.\ Grav.\  {\bf 31}, 135014 (2014)
  doi:10.1088/0264-9381/31/13/135014
  [arXiv:1311.3340 [hep-th]].
  %%CITATION = doi:10.1088/0264-9381/31/13/135014;%%
  %30 citations counted in INSPIRE as of 25 Dec 2019

%\cite{Dartois:2013sra}
\bibitem{Dartois:2013sra} 
  S.~Dartois, R.~Gurau and V.~Rivasseau,
  ``Double Scaling in Tensor Models with a Quartic Interaction,''
  JHEP {\bf 1309}, 088 (2013)
  doi:10.1007/JHEP09(2013)088
  [arXiv:1307.5281 [hep-th]].
  %%CITATION = doi:10.1007/JHEP09(2013)088;%%
  %61 citations counted in INSPIRE as of 25 Dec 2019

%\cite{Dartois:2018kfy}
\bibitem{Dartois:2018kfy} 
  S.~Dartois, O.~Evnin, L.~Lionni, V.~Rivasseau and G.~Valette,
  ``Melonic Turbulence,''
  arXiv:1810.01848 [math-ph].
  %%CITATION = ARXIV:1810.01848;%%
  %7 citations counted in INSPIRE as of 25 Dec 2019



  %\cite{Benedetti:2019rja}
\bibitem{Benedetti:2019rja} 
  D.~Benedetti, N.~Delporte, S.~Harribey and R.~Sinha,
  ``Sextic tensor field theories in rank $3$ and $5$,''
  arXiv:1912.06641 [hep-th].
  %%CITATION = ARXIV:1912.06641;%%
  %3 citations counted in INSPIRE as of 24 Mar 2020

%\cite{Benedetti:2019ikb}
\bibitem{Benedetti:2019ikb} 
  D.~Benedetti, R.~Gurau, S.~Harribey and K.~Suzuki,
  ``Hints of unitarity at large $N$ in the $O(N)^3$ tensor field theory,''
  JHEP {\bf 2002}, 072 (2020)
  doi:10.1007/JHEP02(2020)072
  [arXiv:1909.07767 [hep-th]].
  %%CITATION = doi:10.1007/JHEP02(2020)072;%%
  %6 citations counted in INSPIRE as of 24 Mar 2020

{\color{blue}
 %\cite{Diaz:2020wtr}
\bibitem{Diaz:2020wtr}
P.~Diaz,
``Backgrounds from tensor models: A proposal,''
Phys. Rev. D \textbf{103} (2021) no.6, 066010
doi:10.1103/PhysRevD.103.066010
[arXiv:2009.00623 [hep-th]].
%1 citations counted in INSPIRE as of 16 Jul 2021
    }

%\cite{Benedetti:2019eyl}
\bibitem{Benedetti:2019eyl} 
  D.~Benedetti, R.~Gurau and S.~Harribey,
  ``Line of fixed points in a bosonic tensor model,''
  JHEP {\bf 1906}, 053 (2019)
  doi:10.1007/JHEP06(2019)053
  [arXiv:1903.03578 [hep-th]].
  %%CITATION = doi:10.1007/JHEP06(2019)053;%%
  %14 citations counted in INSPIRE as of 24 Mar 2020



%\cite{Ooguri:1991ib}
\bibitem{Ooguri:1991ib} 
  H.~Ooguri and N.~Sasakura,
  ``Discrete and continuum approaches to three-dimensional quantum gravity,''
  Mod.\ Phys.\ Lett.\ A {\bf 6}, 3591 (1991)
  doi:10.1142/S0217732391004140
  [hep-th/9108006].
  %%CITATION = doi:10.1142/S0217732391004140;%%
  %60 citations counted in INSPIRE as of 17 Nov 2019

%\cite{Sasakura:1990fs}
\bibitem{Sasakura:1990fs} 
  N.~Sasakura,
  ``Tensor model for gravity and orientability of manifold,''
  Mod.\ Phys.\ Lett.\ A {\bf 6}, 2613 (1991).
  doi:10.1142/S0217732391003055
  %%CITATION = doi:10.1142/S0217732391003055;%%
  %169 citations counted in INSPIRE as of 17 Nov 2019

%\cite{Bonzom:2011zz}
\bibitem{Bonzom:2011zz} 
  V.~Bonzom, R.~Gurau, A.~Riello and V.~Rivasseau,
  ``Critical behavior of colored tensor models in the large N limit,''
  Nucl.\ Phys.\ B {\bf 853}, 174 (2011)
  doi:10.1016/j.nuclphysb.2011.07.022
  [arXiv:1105.3122 [hep-th]].
  %%CITATION = doi:10.1016/j.nuclphysb.2011.07.022;%%
  %212 citations counted in INSPIRE as of 22 Dec 2019


%\cite{Ooguri:1992tw}
\bibitem{Ooguri:1992tw} 
  H.~Ooguri,
  ``Schwinger-Dyson equation in three-dimensional simplicial quantum gravity,''
  Prog.\ Theor.\ Phys.\  {\bf 89}, 1 (1993)
  doi:10.1143/PTP.89.1
  [hep-th/9210028].
  %%CITATION = doi:10.1143/PTP.89.1;%%
  %15 citations counted in INSPIRE as of 17 Nov 2019
  

  %\cite{Boulatov:1992vp}
\bibitem{Boulatov:1992vp} 
  D.~V.~Boulatov,
  ``A Model of three-dimensional lattice gravity,''
  Mod.\ Phys.\ Lett.\ A {\bf 7}, 1629 (1992)
  doi:10.1142/S0217732392001324
  [hep-th/9202074].
  %%CITATION = doi:10.1142/S0217732392001324;%%
  %222 citations counted in INSPIRE as of 17 Nov 2019


%\cite{Witten:1990hr}
\bibitem{Witten:1990hr} 
  E.~Witten,
  ``Two-dimensional gravity and intersection theory on moduli space,''
  Surveys Diff.\ Geom.\  {\bf 1}, 243 (1991).
  doi:10.4310/SDG.1990.v1.n1.a5
  %%CITATION = doi:10.4310/SDG.1990.v1.n1.a5;%%
  %332 citations counted in INSPIRE as of 17 Nov 2019
  %
  %
  %
  %%\cite{Geloun:2011cy}
\bibitem{Geloun:2011cy} 
  J.~Ben Geloun and V.~Bonzom,
  ``Radiative corrections in the Boulatov-Ooguri tensor model: The 2-point function,''
  Int.\ J.\ Theor.\ Phys.\  {\bf 50}, 2819 (2011)
  doi:10.1007/s10773-011-0782-2
  [arXiv:1101.4294 [hep-th]].
  %%CITATION = doi:10.1007/s10773-011-0782-2;%%
  %87 citations counted in INSPIRE as of 25 Dec 2019







%\cite{Sfondrini:2010zm}
\bibitem{Sfondrini:2010zm} 
  A.~Sfondrini and T.~A.~Koslowski,
  ``Functional Renormalization of Noncommutative Scalar Field Theory,''
  Int.\ J.\ Mod.\ Phys.\ A {\bf 26}, 4009 (2011)
  doi:10.1142/S0217751X11054048
  [arXiv:1006.5145 [hep-th]].
  %%CITATION = doi:10.1142/S0217751X11054048;%%
  %18 citations counted in INSPIRE as of 22 Dec 2019



  
  


  
 




%\cite{Safari:2015dva}
\bibitem{Safari:2015dva} 
  M.~Safari,
  ``Splitting Ward identity,''
  Eur.\ Phys.\ J.\ C {\bf 76}, no. 4, 201 (2016)
  doi:10.1140/epjc/s10052-016-4036-6
  [arXiv:1508.06244 [hep-th]].
  %%CITATION = doi:10.1140/epjc/s10052-016-4036-6;%%
  %16 citations counted in INSPIRE as of 26 Aug 2019

  
%\cite{Morris:2016spn}
\bibitem{Morris:2016spn} 
  T.~R.~Morris,
  ``Large curvature and background scale independence in single-metric approximations to asymptotic safety,''
  JHEP {\bf 1611}, 160 (2016)
  doi:10.1007/JHEP11(2016)160
  [arXiv:1610.03081 [hep-th]].
  %%CITATION = doi:10.1007/JHEP11(2016)160;%%
  %34 citations counted in INSPIRE as of 17 Aug 2019
  
  
  %\cite{Wetterich:1991be}
\bibitem{Wetterich:1991be} 
  C.~Wetterich,
 ``The Average action for scalar fields near phase transitions,''
  Z.\ Phys.\ C {\bf 57}, 451 (1993).
  doi:10.1007/BF01474340
  %%CITATION = doi:10.1007/BF01474340;%%
  %139 citations counted in INSPIRE as of 09 Apr 2019


 %\cite{Wetterich:1992yh}
\bibitem{Wetterich:1992yh} 
  C.~Wetterich,
  ``Exact evolution equation for the effective potential,''
  Phys.\ Lett.\ B {\bf 301}, 90 (1993)
  doi:10.1016/0370-2693(93)90726-X
  [arXiv:1710.05815 [hep-th]].
  %%CITATION = doi:10.1016/0370-2693(93)90726-X;%%
  %1305 citations counted in INSPIRE as of 09 Apr 2019

%\cite{Litim:2000ci}
\bibitem{Litim:2000ci} 
  D.~F.~Litim,
  ``Optimization of the exact renormalization group,''
  Phys.\ Lett.\ B {\bf 486}, 92 (2000)
 doi:10.1016/S0370-2693(00)00748-6
  [hep-th/0005245].
  


%\cite{Litim:2001dt}
\bibitem{Litim:2001dt} 
  D.~F.~Litim,
  ``Derivative expansion and renormalization group flows,''
  JHEP {\bf 0111}, 059 (2001)
  doi:10.1088/1126-6708/2001/11/059
  [hep-th/0111159].
  %%CITATION = doi:10.1088/1126-6708/2001/11/059;%%
  %73 citations counted in INSPIRE as of 23 Mar 2018
  
  
%\cite{Canet:2002gs}
\bibitem{Canet:2002gs} 
  L.~Canet, B.~Delamotte, D.~Mouhanna and J.~Vidal,
  ``Optimization of the derivative expansion in the nonperturbative renormalization group,''
  Phys.\ Rev.\ D {\bf 67}, 065004 (2003)
  doi:10.1103/PhysRevD.67.065004
  [hep-th/0211055].
  %%CITATION = doi:10.1103/PhysRevD.67.065004;%%
  %113 citations counted in INSPIRE as of 25 Aug 2019
  
  %\cite{Delamotte:2007pf}
\bibitem{Delamotte:2007pf} 
  B.~Delamotte,
  ``An Introduction to the nonperturbative renormalization group,''
  Lect.\ Notes Phys.\  {\bf 852}, 49 (2012)
  doi:10.1007/978-3-642-27320-$9_2$
  [cond-mat/0702365 [cond-mat.stat-mech]].
  %%CITATION = doi:10.1007/978-3-642-27320-9_2;%%
  %268 citations counted in INSPIRE as of 25 Aug 2019


%\cite{Berges:2000ew}
\bibitem{Berges:2000ew} 
  J.~Berges, N.~Tetradis and C.~Wetterich,
  ``Nonperturbative renormalization flow in quantum field theory and statistical physics,''
  Phys.\ Rept.\  {\bf 363}, 223 (2002)
  doi:10.1016/S0370-1573(01)00098-9
  [hep-ph/0005122].
  %%CITATION = doi:10.1016/S0370-1573(01)00098-9;%%
  %990 citations counted in INSPIRE as of 17 Aug 2019
  
  %\cite{Gies:2001nw}
\bibitem{Gies:2001nw} 
  H.~Gies and C.~Wetterich,
  ``Renormalization flow of bound states,''
  Phys.\ Rev.\ D {\bf 65}, 065001 (2002)
  doi:10.1103/PhysRevD.65.065001
  [hep-th/0107221].
  %%CITATION = doi:10.1103/PhysRevD.65.065001;%%
  %149 citations counted in INSPIRE as of 17 Aug 2019
  %
  %
  %

  %%\cite{Wetterich:2016ewc}
\bibitem{Wetterich:2016ewc} 
  C.~Wetterich,
  %``Gauge invariant flow equation,''
  Nucl.\ Phys.\ B {\bf 931}, 262 (2018)
  doi:10.1016/j.nuclphysb.2018.04.020
  [arXiv:1607.02989 [hep-th]].
  %%CITATION = doi:10.1016/j.nuclphysb.2018.04.020;%%
  %28 citations counted in INSPIRE as of 25 Dec 2019
  
  %\cite{Ferrari:2019ogc}
\bibitem{Ferrari:2019ogc} 
  F.~Ferrari and F.~I.~Schaposnik Massolo,
  ``Phases Of Melonic Quantum Mechanics,''
  Phys.\ Rev.\ D {\bf 100}, no. 2, 026007 (2019)
  doi:10.1103/PhysRevD.100.026007
  [arXiv:1903.06633 [hep-th]].
  %%CITATION = doi:10.1103/PhysRevD.100.026007;%%
  %7 citations counted in INSPIRE as of 17 Aug 2019

  
  %\cite{Delepouve:2015nia}
\bibitem{Delepouve:2015nia} 
  T.~Delepouve and R.~Gurau,
  ``Phase Transition in Tensor Models,''
  JHEP {\bf 1506}, 178 (2015)
  doi:10.1007/JHEP06(2015)178
  [arXiv:1504.05745 [hep-th]].
  %%CITATION = doi:10.1007/JHEP06(2015)178;%%
  %20 citations counted in INSPIRE as of 17 Aug 2019







%\cite{Duplantier:2009np}
\bibitem{Duplantier:2009np} 
  B.~Duplantier and S.~Sheffield,
  ``Duality and KPZ in Liouville Quantum Gravity,''
  Phys.\ Rev.\ Lett.\  {\bf 102}, 150603 (2009)
  doi:10.1103/PhysRevLett.102.150603
  [arXiv:0901.0277 [math-ph]].
  %%CITATION = doi:10.1103/PhysRevLett.102.150603;%%
  %21 citations counted in INSPIRE as of 24 Aug 2019
  
  %\cite{Duplantier:2009np2}
\bibitem{Duplantier:2009np2} 
  B.~Duplantier and S.~Sheffield,
  `` Liouville Quantum Gravity and KPZ,''
  [arXiv:0808.1560 [math-ph]].
  %%CITATION = doi:10.1103/PhysRevLett.102.150603;%%
  %21 citations counted in INSPIRE as of 24 Aug 2019
  

%\cite{Pawlowski:2015mlf}
\bibitem{Pawlowski:2015mlf} 
  J.~M.~Pawlowski, M.~M.~Scherer, R.~Schmidt and S.~J.~Wetzel,
  ``Physics and the choice of regulators in functional renormalisation group flows,''
  Annals Phys.\  {\bf 384}, 165 (2017)
  doi:10.1016/j.aop.2017.06.017
  [arXiv:1512.03598 [hep-th]].
  %%CITATION = doi:10.1016/j.aop.2017.06.017;%%
  %27 citations counted in INSPIRE as of 24 Aug 2019




%\cite{Freidel:2009hd}
\bibitem{Freidel:2009hd} 
  L.~Freidel, R.~Gurau and D.~Oriti,
``Group field theory renormalization - the 3d case: Power
counting of divergences,''
  Phys.\ Rev.\ D {\bf 80}, 044007 (2009)
  doi:10.1103/PhysRevD.80.044007
  [arXiv:0905.3772 [hep-th]].
  %%CITATION = doi:10.1103/PhysRevD.80.044007;%%
  %88 citations counted in INSPIRE as of 11 Aug 2018


%\cite{Carrozza:2012uv}
\bibitem{Carrozza:2012uv} 
  S.~Carrozza, D.~Oriti and V.~Rivasseau,
``Renormalization of Tensorial Group Field Theories: Abelian
U(1) Models in Four Dimensions,''
  Commun.\ Math.\ Phys.\  {\bf 327}, 603 (2014)
  doi:10.1007/s00220-014-1954-8
  [arXiv:1207.6734 [hep-th]].
  %%CITATION = doi:10.1007/s00220-014-1954-8;%%
  %75 citations counted in INSPIRE as of 25 Mar 2016


%\cite{Carrozza:2013mna}
\bibitem{Carrozza:2013mna} 
  S.~Carrozza,
``Tensorial methods and renormalization in Group Field
Theories,''
  doi:10.1007/978-3-319-05867-2
  arXiv:1310.3736 [hep-th].
  %%CITATION = doi:10.1007/978-3-319-05867-2;%%
  %28 citations counted in INSPIRE as of 25 Mar 2016



%\cite{Carrozza:2013wda}
\bibitem{Carrozza:2013wda} 
  S.~Carrozza, D.~Oriti and V.~Rivasseau,
``Renormalization of a SU(2) Tensorial Group Field Theory in
Three Dimensions,''
  Commun.\ Math.\ Phys.\  {\bf 330}, 581 (2014)
  doi:10.1007/s00220-014-1928-x
  [arXiv:1303.6772 [hep-th]].
  %%CITATION = doi:10.1007/s00220-014-1928-x;%%
  %69 citations counted in INSPIRE as of 25 Mar 2016


%\cite{Geloun:2013saa}
\bibitem{Geloun:2013saa} 
  J.~Ben Geloun,
``Renormalizable Models in Rank $d\geq 2$ Tensorial Group Field
Theory,''
  Commun.\ Math.\ Phys.\  {\bf 332}, 117 (2014)
  doi:10.1007/s00220-014-2142-6
  [arXiv:1306.1201 [hep-th]].
  %%CITATION = doi:10.1007/s00220-014-2142-6;%%
  %43 citations counted in INSPIRE as of 25 Mar 2016


%\cite{Lahoche:2015tqa}
\bibitem{Lahoche:2015tqa} 
  V.~Lahoche and D.~Oriti,
  ``Renormalization of a tensorial field theory on the homogeneous space SU(2)/U(1),''
  J.\ Phys.\ A {\bf 50}, no. 2, 025201 (2017)
  doi:10.1088/1751-8113/50/2/025201
  [arXiv:1506.08393 [hep-th]].
  %%CITATION = doi:10.1088/1751-8113/50/2/025201;%%
  %21 citations counted in INSPIRE as of 25 Dec 2019



%\cite{Lahoche:2015ola}
\bibitem{Lahoche:2015ola} 
  V.~Lahoche, D.~Oriti and V.~Rivasseau,
``Renormalization of an Abelian Tensor Group Field Theory:
Solution at Leading Order,''
  JHEP {\bf 1504}, 095 (2015)
  doi:10.1007/JHEP04(2015)095
  [arXiv:1501.02086 [hep-th]].
  %%CITATION = doi:10.1007/JHEP04(2015)095;%%
  %17 citations counted in INSPIRE as of 25 Mar 2016


%\cite{Geloun:2012bz}
\bibitem{Geloun:2012bz} 
  J.~Ben Geloun and E.~R.~Livine,
  ``Some classes of renormalizable tensor models,''
  J.\ Math.\ Phys.\  {\bf 54}, 082303 (2013)
  doi:10.1063/1.4818797
  [arXiv:1207.0416 [hep-th]].
  %%CITATION = doi:10.1063/1.4818797;%%
  %42 citations counted in INSPIRE as of 15 Jun 2016


%\cite{Samary:2012bw}
\bibitem{Samary:2012bw} 
  D.~Ousmane~Samary and F.~Vignes-Tourneret,
``Just Renormalizable TGFT's on $U(1)^{d}$ with Gauge
Invariance,''
  Commun.\ Math.\ Phys.\  {\bf 329}, 545 (2014)
  doi:10.1007/s00220-014-1930-3
  [arXiv:1211.2618 [hep-th]].
  %%CITATION = doi:10.1007/s00220-014-1930-3;%%
  %44 citations counted in INSPIRE as of 25 Mar 2016


%\cite{Samary:2013xla}
\bibitem{Samary:2013xla} 
  D.~Ousmane Samary,
  ``Beta functions of  $U(1)^d$ gauge invariant just renormalizable tensor models,''
  Phys.\ Rev.\ D {\bf 88}, no. 10, 105003 (2013)
  doi:10.1103/PhysRevD.88.105003
  [arXiv:1303.7256 [hep-th]].
  %%CITATION = doi:10.1103/PhysRevD.88.105003;%%
  %37 citations counted in INSPIRE as of 25 Dec 2019

%\cite{BenGeloun:2012pu}
\bibitem{BenGeloun:2012pu} 
  J.~Ben Geloun and D.~Ousmane.~Samary,
``3D Tensor Field Theory: Renormalization and One-loop
$\beta$-functions,''
  Annales Henri Poincare {\bf 14}, 1599 (2013)
  doi:10.1007/s00023-012-0225-5
  [arXiv:1201.0176 [hep-th]].
  %%CITATION = doi:10.1007/s00023-012-0225-5;%%
  %62 citations counted in INSPIRE as of 25 Mar 2016

%\cite{BenGeloun:2011rc}
\bibitem{BenGeloun:2011rc} 
  J.~Ben Geloun and V.~Rivasseau,
  ``A Renormalizable 4-Dimensional Tensor Field Theory,''
  Commun.\ Math.\ Phys.\  {\bf 318}, 69 (2013)
  doi:10.1007/s00220-012-1549-1
  [arXiv:1111.4997 [hep-th]].
  %%CITATION = doi:10.1007/s00220-012-1549-1;%%
  %88 citations counted in INSPIRE as of 25 Mar 2016



%\cite{Lahoche:2015ola}
\bibitem{Lahoche:2015ola} 
  V.~Lahoche, D.~Oriti and V.~Rivasseau,
``Renormalization of an Abelian Tensor Group Field Theory:
Solution at Leading Order,''
  JHEP {\bf 1504}, 095 (2015)
  doi:10.1007/JHEP04(2015)095
  [arXiv:1501.02086 [hep-th]].
  %%CITATION = doi:10.1007/JHEP04(2015)095;%%
  %24 citations counted in INSPIRE as of 23 Mar 2018

%\cite{Carrozza:2014rba}
\bibitem{Carrozza:2014rba} 
  S.~Carrozza,
  ``Discrete Renormalization Group for SU(2) Tensorial Group Field Theory,''
  Ann.\ Inst.\ H.\ Poincare Comb.\ Phys.\ Interact.\  {\bf 2}, 49 (2015)
  doi:10.4171/AIHPD/15
  [arXiv:1407.4615 [hep-th]].
  %%CITATION = doi:10.4171/AIHPD/15;%%
  %55 citations counted in INSPIRE as of 01 Jan 2020


%\cite{Carrozza:2014rya}
\bibitem{Carrozza:2014rya} 
  S.~Carrozza,
  ``Group field theory in dimension $4-\epsilon$,''
  Phys.\ Rev.\ D {\bf 91}, no. 6, 065023 (2015)
  doi:10.1103/PhysRevD.91.065023
  [arXiv:1411.5385 [hep-th]].
  %%CITATION = doi:10.1103/PhysRevD.91.065023;%%
  %25 citations counted in INSPIRE as of 01 Jan 2020
  %
  
  %\cite{Geloun:2016qyb}
\bibitem{Geloun:2016qyb} 
  J.~Ben Geloun, R.~Martini and D.~Oriti,
  ``Functional Renormalisation Group analysis of Tensorial Group Field Theories on $\mathbb{R}^d$,''
  Phys.\ Rev.\ D {\bf 94}, no. 2, 024017 (2016)
  doi:10.1103/PhysRevD.94.024017
  [arXiv:1601.08211 [hep-th]].
  %%CITATION = doi:10.1103/PhysRevD.94.024017;%%
  %47 citations counted in INSPIRE as of 23 Mar 2020

  
  %\cite{Carrozza:2017vkz}
\bibitem{Carrozza:2017vkz} 
  S.~Carrozza, V.~Lahoche and D.~Oriti,
  ``Renormalizable Group Field Theory beyond melonic diagrams: an example in rank four,''
  Phys.\ Rev.\ D {\bf 96}, no. 6, 066007 (2017)
  doi:10.1103/PhysRevD.96.066007
  [arXiv:1703.06729 [gr-qc]].
  %%CITATION = doi:10.1103/PhysRevD.96.066007;%%
  %29 citations counted in INSPIRE as of 01 Jan 2020


%\cite{Lahoche:2016xiq}
\bibitem{Lahoche:2016xiq} 
  V.~Lahoche and D.~Ousmane Samary,
  ``Functional renormalization group for the U(1)-T$_5^6$ tensorial group field theory with closure constraint,''
  Phys.\ Rev.\ D {\bf 95}, no. 4, 045013 (2017)
  doi:10.1103/PhysRevD.95.045013
  [arXiv:1608.00379 [hep-th]].
  %%CITATION = doi:10.1103/PhysRevD.95.045013;%%
  %21 citations counted in INSPIRE as of 01 Jan 2020


%\cite{Carrozza:2016tih}
\bibitem{Carrozza:2016tih} 
  S.~Carrozza and V.~Lahoche,
  ``Asymptotic safety in three-dimensional SU(2) Group Field Theory: evidence in the local potential approximation,''
  Class.\ Quant.\ Grav.\  {\bf 34}, no. 11, 115004 (2017)
  doi:10.1088/1361-6382/aa6d90
  [arXiv:1612.02452 [hep-th]].
  %%CITATION = doi:10.1088/1361-6382/aa6d90;%%
  %29 citations counted in INSPIRE as of 01 Jan 2020


%\cite{Rivasseau:2017xbk}
\bibitem{Rivasseau:2017xbk} 
  V.~Rivasseau and F.~Vignes-Tourneret,
  ``Constructive tensor field theory: The $T^{4}_{4}$ model,''
  arXiv:1703.06510 [math-ph].
  %%CITATION = ARXIV:1703.06510;%%
  %6 citations counted in INSPIRE as of 23 Mar 2018


%\cite{Delporte:2019tof}
\bibitem{Delporte:2019tof} 
  N.~Delporte and V.~Rivasseau,
  ``Perturbative Quantum Field Theory on Random Trees,''
  arXiv:1905.12783 [hep-th].
  %%CITATION = ARXIV:1905.12783;%%
  %2 citations counted in INSPIRE as of 25 Dec 2019


%\cite{Delepouve:2014hfa}
\bibitem{Delepouve:2014hfa} 
  T.~Delepouve and V.~Rivasseau,
  %``Constructive Tensor Field Theory: The $T^4_3$ Model,''
  Commun.\ Math.\ Phys.\  {\bf 345}, no. 2, 477 (2016)
  doi:10.1007/s00220-016-2680-1
  [arXiv:1412.5091 [math-ph]].
  %%CITATION = doi:10.1007/s00220-016-2680-1;%%
  %22 citations counted in INSPIRE as of 25 Dec 2019



%\cite{Lionni:2016ush}
\bibitem{Lionni:2016ush} 
  L.~Lionni and V.~Rivasseau,
``Intermediate Field Representation for Positive Matrix and
Tensor Interactions,''
  arXiv:1609.05018 [math-ph].
  %%CITATION = ARXIV:1609.05018;%%
  %4 citations counted in INSPIRE as of 23 Mar 2018


  %\cite{Lahoche:2018ggd}
\bibitem{Lahoche:2018ggd} 
  V.~Lahoche and D.~Ousmane Samary,
  ``Ward identity violation for melonic $T^4$-truncation,''
  Nucl.\ Phys.\ B {\bf 940}, 190 (2019)
  doi:10.1016/j.nuclphysb.2019.01.005
  [arXiv:1809.06081 [hep-th]].
  %%CITATION = doi:10.1016/j.nuclphysb.2019.01.005;%%
  %7 citations counted in INSPIRE as of 17 Aug 2019
  
  %\cite{Lahoche:2018oeo}
\bibitem{Lahoche:2018oeo} 
  V.~Lahoche and D.~Ousmane Samary,
  ``Nonperturbative renormalization group beyond the melonic sector: The effective vertex expansion method for group fields theories,''
  Phys.\ Rev.\ D {\bf 98}, no. 12, 126010 (2018)
  doi:10.1103/PhysRevD.98.126010
  [arXiv:1809.00247 [hep-th]].
  %%CITATION = doi:10.1103/PhysRevD.98.126010;%%
  %7 citations counted in INSPIRE as of 17 Aug 2019
  
 
  %\cite{Lahoche:2019vzy}
\bibitem{Lahoche:2019vzy} 
  V.~Lahoche and D.~Ousmane.~Samary,
  ``Pedagogical comments about nonperturbative Ward-constrained melonic renormalization group flow,''  Phys.\ Rev.\ D,
  {\bf 101}, 024001 (2020) doi.org/10.1103/PhysRevD.2.1541.
  %%CITATION = ARXIV:1904.05655;%%
  %2 citations counted in INSPIRE as of 24 Aug 2019
  
%\cite{Lahoche:2019orv}
\bibitem{Lahoche:2019orv} 
  V.~Lahoche, D.~Ousmane Samary and A.~D.~Pereira,
  ``Renormalization group flow of coupled tensorial group field theories: Towards the Ising model on random lattices,''
  Phys.\ Rev.\ D {\bf 101}, no. 6, 064014 (2020)
  doi:10.1103/PhysRevD.101.064014
  [arXiv:1911.05173 [hep-th]].
  %%CITATION = doi:10.1103/PhysRevD.101.064014;%%
  %4 citations counted in INSPIRE as of 31 May 2020


%\cite{Lahoche:2019ehm}
\bibitem{Lahoche:2019ehm} 
  V.~Lahoche and D.~Ousmane.~Samary,
  ``Large-$d$ behavior of the Feynman amplitudes for a just-renormalizable tensorial group field theory,''
  arXiv:1911.08601 [hep-th].
  %%CITATION = ARXIV:1911.08601;%%



 %\cite{Lahoche:2019vzy}
\bibitem{Lahoche:2019vzy} 
  V.~Lahoche and D.~Ousmane Samary,
  ``Ward-constrained melonic renormalization group flow,''
  Phys.\ Lett.\ B {\bf 802}, 135173 (2020)
  doi:10.1016/j.physletb.2019.135173
  [arXiv:1904.05655 [hep-th]].
  %%CITATION = doi:10.1016/j.physletb.2019.135173;%%
  %9 citations counted in INSPIRE as of 31 May 2020
  
 
 %\cite{Lahoche:2019cxt}
\bibitem{Lahoche:2019cxt} 
  V.~Lahoche and D.~Ousmane.~Samary,
  ``Ward-constrained melonic renormalization group flow for the rank-four $\phi^6$ tensorial group field theory,''
  Phys.\ Rev.\ D {\bf 100}, no. 8, 086009 (2019)
  doi:10.1103/PhysRevD.100.086009
  [arXiv:1908.03910 [hep-th]].
  %%CITATION = doi:10.1103/PhysRevD.100.086009;%%
  %4 citations counted in INSPIRE as of 21 Dec 2019

 
 %\cite{Lahoche:2018hou}
\bibitem{Lahoche:2018hou} 
  V.~Lahoche and D.~Ousmane~Samary,
  ``Progress in the solving nonperturbative renormalization group for tensorial group field theory,''
  Universe {\bf 5}, 86 (2019)
  doi:10.3390/universe5030086
  [arXiv:1812.00905 [hep-th]].
  %%CITATION = doi:10.3390/universe5030086;%%
  %3 citations counted in INSPIRE as of 24 Aug 2019
 
 
 

  %\cite{Lahoche:2018vun}
\bibitem{Lahoche:2018vun} 
  V.~Lahoche and D.~Ousmane Samary,
  ``Unitary symmetry constraints on tensorial group field theory renormalization group flow,''
  Class.\ Quant.\ Grav.\  {\bf 35}, no. 19, 195006 (2018)
  doi:10.1088/1361-6382/aad83f
  [arXiv:1803.09902 [hep-th]].
  %%CITATION = doi:10.1088/1361-6382/aad83f;%%
  %8 citations counted in INSPIRE as of 17 Aug 2019
  %


%\cite{BenGeloun:2011xu}
\bibitem{BenGeloun:2011xu} 
  J.~Ben Geloun,
  ``Ward-Takahashi identities for the colored Boulatov model,''
  J.\ Phys.\ A {\bf 44}, 415402 (2011)
  doi:10.1088/1751-8113/44/41/415402
  [arXiv:1106.1847 [hep-th]].
  %%CITATION = doi:10.1088/1751-8113/44/41/415402;%%
  %17 citations counted in INSPIRE as of 24 Mar 2020


%\cite{Samary:2014tja}
\bibitem{Samary:2014tja} 
  D.~Ousmane~Samary,
  ``Closed equations of the two-point functions for tensorial group field theory,''
  Class.\ Quant.\ Grav.\  {\bf 31}, 185005 (2014)
  doi:10.1088/0264-9381/31/18/185005
  [arXiv:1401.2096 [hep-th]].
  %%CITATION = doi:10.1088/0264-9381/31/18/185005;%%
  %36 citations counted in INSPIRE as of 25 Dec 2019
  
  
  %\cite{Samary:2014oya}
\bibitem{Samary:2014oya} 
  D.~Ousmane Samary, C.~I.~Pérez-Sánchez, F.~Vignes-Tourneret and R.~Wulkenhaar,
  ``Correlation functions of a just renormalizable tensorial group field theory: the melonic approximation,''
  Class.\ Quant.\ Grav.\  {\bf 32}, no. 17, 175012 (2015)
  doi:10.1088/0264-9381/32/17/175012
  [arXiv:1411.7213 [hep-th]].
  %%CITATION = doi:10.1088/0264-9381/32/17/175012;%%
  %31 citations counted in INSPIRE as of 25 Dec 2019


%\cite{Sanchez:2017gxt}
\bibitem{Sanchez:2017gxt} 
  R.~Pascalie, C.~I.~P.~Sánchez and R.~Wulkenhaar,
  ``Correlation functions of $\mathrm{U}(N)$-tensor models and their Schwinger-Dyson equations,''
  arXiv:1706.07358 [math-ph].
  %%CITATION = ARXIV:1706.07358;%%
  %9 citations counted in INSPIRE as of 25 Dec 2019

  
%\cite{Itoyama:2017xid}
\bibitem{Itoyama:2017xid} 
  H.~Itoyama, A.~Mironov and A.~Morozov,
  ``Ward identities and combinatorics of rainbow tensor models,''
  JHEP {\bf 1706}, 115 (2017)
  doi:10.1007/JHEP06(2017)115
  [arXiv:1704.08648 [hep-th]].
  %%CITATION = doi:10.1007/JHEP06(2017)115;%%
  %37 citations counted in INSPIRE as of 21 Dec 2019
  
  
%\cite{Patkos:2012ex}
\bibitem{Patkos:2012ex} 
  A.~Patkos,
  ``Invariant formulation of the Functional Renormalisation Group method for $U(n)\times U(n)$ symmetric matrix models,''
  Mod.\ Phys.\ Lett.\ A {\bf 27}, 1250212 (2012)
  doi:10.1142/S0217732312502124
  [arXiv:1210.6490 [hep-ph]].
  %%CITATION = doi:10.1142/S0217732312502124;%%
  %7 citations counted in INSPIRE as of 21 Dec 2019


%\cite{Ward:1950xp}
\bibitem{Ward:1950xp} 
  J.~C.~Ward,
  ``An Identity in Quantum Electrodynamics,''
  Phys.\ Rev.\  {\bf 78}, 182 (1950).
  doi:10.1103/PhysRev.78.182
  %%CITATION = doi:10.1103/PhysRev.78.182;%%
  %435 citations counted in INSPIRE as of 15 Aug 2019


%\cite{Takahashi:1957xn}
\bibitem{Takahashi:1957xn} 
  Y.~Takahashi,
  ``On the generalized Ward identity,''
  Nuovo Cim.\  {\bf 6}, 371 (1957).
  doi:10.1007/BF02832514
  %%CITATION = doi:10.1007/BF02832514;%%
  %418 citations counted in INSPIRE as of 15 Aug 2019







  
 
\end{thebibliography}
\end{document}